\documentclass[prc,aps,floats,onecolumn,floatfix,showpacs,nofootinbib,%
superscriptaddress]{revtex4}

\usepackage[T1]{fontenc}
\usepackage{amsmath}
\usepackage{amssymb}
\usepackage{amsbsy}
\usepackage{amsfonts}
\usepackage{graphicx}
\usepackage{tabularx}
\usepackage{multirow}
\usepackage{rotating}   
\usepackage{pstricks}
\usepackage{appendix}
\usepackage{mathrsfs}
\usepackage{enumitem}
\usepackage{verbatim} 

\usepackage{xcolor}
\usepackage[
    colorlinks=true,
    linkcolor=blue,
    citecolor=blue,
    urlcolor=red
]{hyperref}
\def\be{\begin{equation}}
\def\ee{\end{equation}}
\def\bea{\begin{eqnarray}}
\def\eea{\end{eqnarray}}

\begin{document}

\title{Solution of the generalised Bohr Hamiltonian with Gogny interactions}


\author{C. Azam}
\email{clementine.azam@cea.fr}
\affiliation{ CEA, DES, IRESNE, DER, SPRC, F-13108 Saint Paul Lez Durance, France}

\author{D. Davesne}
\email{davesne@ipnl.in2p3.fr}
\affiliation{Institut de Physique des deux  Infinis, CNRS-IN2P3, UMR 5822, Universit\'e Lyon 1 \\
             43 Bd. du 11 Novembre 1918, F-69622 Villeurbanne cedex, France}

\author{ Y. Lallouet}
\affiliation{Lycée Malherbe, 14, Avenue Albert Sorel, 14000 Caen, France}

\author{L. Pr\'ochniak }
\email{prochniak@slcj.uw.edu.pl}
\affiliation{ Heavy Ion Laboratory, University of Warsaw, PL-02-093 Warsaw, Poland}

\author{M. Frosini}
\email{mikael.frosini@cea.fr}
\affiliation{ CEA, DES, IRESNE, DER, SPRC, F-13108 Saint Paul Lez Durance, France}

\author{A. Pastore}
\email{alessandro.pastore@cea.fr}
\affiliation{ CEA, DES, IRESNE, DER, SPRC, F-13108 Saint Paul Lez Durance, France}

\begin{abstract}
We present a comprehensive derivation of the generalized Bohr Hamiltonian formalism for quadrupole collective degrees of freedom, together with a detailed description of the numerical framework employed to solve the associated eigenvalue problem through a basis-expansion method. The present formulation provides a unified and transparent framework that is particularly well suited for future methodological extensions and applications. In addition, the numerical implementation is benchmarked against analytical test cases in order to validate the accuracy and stability of the adopted computational approach.

Using fully microscopic constrained Hartree–Fock–Bogoliubov calculations based on four different Gogny effective interactions, we investigate the excited spectra of Gd isotopes and assess the sensitivity of the predicted spectroscopic properties to the underlying interaction. While all parametrizations reproduce the main features of the experimental spectra, noticeable differences are observed in the electromagnetic transition probabilities, indicating that these observables are significantly more sensitive to the details of the underlying interaction than the excitation energies themselves.
\end{abstract}


 
\date{\today}


\maketitle

\section{Introduction}
Mean-field (MF) methods based on effective nucleon--nucleon (NN) interactions are nowadays capable of achieving a remarkable level of accuracy in the description of nuclear ground-state properties \cite{goriely2009first,ryssens2022skyrme}.
The main advantage of such approaches, compared with more fine-tuned microscopic-macroscopic models \cite{pomorski2003nuclear,moller2012new,aza26}, is that the same NN interaction can also be used to calculate excited states in combination with other many-body methods, such as the quasiparticle random-phase approximation (QRPA), the rigid-rotor model, or the generator coordinate method (GCM) \cite{ring2004nuclear,peru2014mean}.

Each of these many-body methods has its own advantages and limitations, depending on the observable of interest. For example, the QRPA is better suited to describing the structure of giant resonances \cite{bortignon2019giant}, whereas the rigid-rotor model is more appropriate for describing rotational bands \cite{boh76}.
Ideally, the GCM could be employed in all cases, including situations in which the vibrational or rotational character of the states is not strongly pronounced \cite{bender2006global,peru2014mean}. However, the computational cost of GCM calculations can become very large, depending on the symmetries that are broken and restored \cite{sheikh2021symmetry}. As a consequence, their use in systematic calculations remains challenging \cite{bender2006global}, especially in combination with finite-range effective interactions \cite{decharge1980hartree,nakada2008mean}.

One way to overcome this difficulty without significantly deteriorating the description of the nuclear spectrum is to employ the generalized Bohr Hamiltonian (BH). Such an approach was developed in Refs.~\cite{Kumar_1967,roh77,lib82,Libert_1999,pro99b} (and references therein) and later applied systematically in Refs.~\cite{delaroche2010structure,delaroche2024investigations} to investigate the properties of low-lying states using the Gogny D1S interaction \cite{decharge1980hartree}.
In recent years, the BH framework has also been successfully employed in conjunction with zero-range interactions \cite{pro09,mui20,mui21} and relativistic Lagrangians \cite{nikvsic2009beyond,nik11}, further demonstrating the versatility of the approach.

Although the BH method has been extensively studied in connection with analytical potentials \cite{rowe,Fortunato_2005}, recent years have witnessed renewed interest in applying the BH framework to potentials derived from microscopic mean-field models. In parallel, significant progress in the development of fast and efficient algorithms for calculating low-lying vibrational states \cite{car12,avo13,hin13} has made it possible to improve the standard approximations commonly used to evaluate collective mass parameters \cite{bar11}, leading to remarkably accurate results \cite{was24}. Finally, we note that substantial efforts are currently being devoted to formulating the Bohr Hamiltonian within the framework of effective field theory \cite{per24}.
The vast majority of studies based on the BH framework are restricted to quadrupole collective degrees of freedom. However, in recent years, several extensions have been proposed \cite{dob16,zhao20,xia25} to include couplings between quadrupole motion and pairing or octupole degrees of freedom. These developments further enhance the descriptive power of the model and pave the way for systematic studies of additional observables, such as nuclear level densities \cite{bloch1954theory}.

All these studies provide a strong motivation to revisit some of the fundamental aspects of the quadrupole BH model, in particular the numerical methods required to determine the excited spectra. As discussed later, the traditional approach used to solve the corresponding equations is based on finite-difference techniques \cite{Kumar_1967,roh77,smi76}. However, such methods are limited by the resolution of the underlying microscopic calculations and by the rapidly increasing number of coupled differential equations with increasing angular momentum $I$ of the excited states. To overcome these limitations, an alternative approach based on the diagonalization of the generalized Hamiltonian over a suitable basis has been developed. The main advantage of this method is its flexibility, which allows for straightforward extensions to the description of yrast bands in super-deformed nuclei \cite{Libert_1999}.

In the present article, we provide a detailed and unified derivation of the Bohr Hamiltonian  equations, together with the ingredients required to construct the corresponding basis of wave functions. Although several elements of the formalism have previously appeared in the literature \cite{Libert_1999,pro09}, the relevant equations are often scattered across different works and formulated using different conventions and phase choices. Here, we bring these elements together into a consistent formulation and systematically analyse the construction of the BH basis, with particular emphasis on the role of the free parameters entering its definition and their influence on the resulting wave functions and observables.

The main novelty of this article resides in the investigation of the predictive capabilities of the BH when combined with modern effective interactions whose fitting protocols incorporate the BH framework \cite{goriely2009first,batail2025three}. In particular, we focus on the newly adjusted D3G3M Gogny interaction~\cite{batail2025three} and assess its performance within the BH framework. In contrast to previous calculations performed using the Gogny D1S interaction \cite{delaroche2010structure}, this interaction has been directly adjusted by accounting for the additional correlation energy arising from the BH, as was done for D1M \cite{goriely2009first}. However, unlike D1M, it also includes an additional long-range term in its central component, which significantly improves its isovector properties \cite{davesne2025isovector}.
 To facilitate comparison with existing theoretical studies and experimental data, we analyse the low-lying spectra and electromagnetic transitions of selected even-even gadolinium isotopes.

Finally, we perform a systematic sensitivity analysis of the different ingredients entering the BH calculations. We examine, in particular, the dependence on the underlying NN interaction, the role of pairing correlations, and the impact of the scaling parameter and the zero-point energy. Since there is currently no consensus in the literature on how these ingredients should be treated within the BH framework, this analysis provides a systematic assessment of their impact and, more generally, of the uncertainties associated with BH predictions.

The article is organised as follows: in Sec.~\ref{sec:ref:frame}, we introduce the intrinsic and laboratory frames, discuss the transformations between them, and present the basic tools required to construct the collective Hamiltonian. In Sec.~\ref{basis:wf}, we describe the construction of the basis used to diagonalise the collective Hamiltonian, while Sec.~\ref{sec:analytical} presents the numerical tests performed to characterise its properties against selected analytical cases. The results obtained from microscopic potentials are presented in Sec.~\ref{sec:micro}. We first analyse the spectra of Gd isotopes using the D3G3M Gogny interaction and then discuss the sensitivity of the results to the different ingredients of the BH calculations. Finally, our conclusions are presented in Sec.~\ref{sec:conclusion}.

\section{Quadrupole Hamiltonian}\label{sec:ref:frame}

In the original Bohr model \cite{Bohr_1952}, collective excitations are described in terms of dynamic deformations of the nuclear surface.
Assuming volume conservation and invariance of the center of mass, the nuclear surface is described in terms of spherical harmonics, $Y_{\lambda\mu}$, as

\begin{equation} \label{eq:R}
    R(\theta,\phi)=R_0\left[1 +\sum_{\lambda=2}^\infty\sum_{\mu=-\lambda}^\lambda \alpha_{\lambda\mu}Y^*_{\lambda\mu}(\theta,\phi) \right]\,,
\end{equation}

\noindent where $\alpha_{\lambda\mu}$ are the collective coordinates describing the deformation. Since the nuclear surface is real, the parameters satisfy the following relation $\alpha^*_{\lambda \mu}=(-)^\mu \alpha_{\lambda -\mu}$. In the following, we truncate the above expansion at the quadrupole level, but recent work has extended it to include octupole deformations \cite{dob16,xia25}. For simplicity, we adopt the decomposition into real $a_{2\mu}$ and imaginary $b_{2\mu}$ parts, as in \cite{pro09}.

\begin{equation}
\begin{split}
     \alpha_{2 0} = a_{2 0} & \quad \quad \alpha_{2\mu} = \frac{1}{\sqrt{2}}  \left(a_{2\mu} +i b_{2\mu}\right) \text{  for   } \mu>0 \,.
\end{split}
\end{equation}

In most practical cases, it is more convenient to work in the intrinsic frame of the nucleus. For quadrupole deformations, this concept has a clear physical meaning: it corresponds to choosing the coordinate axes along the principal axes of the nucleus. This is achieved by applying an appropriate rotation, defined by the Euler angles $\Omega=(\phi, \theta, \psi)$, to the reference axes. The collective coordinates in the intrinsic frame, $\alpha_{\lambda\nu}^{\mathrm{(in)}}$, are related to those in the laboratory frame through the Wigner functions $\mathcal{D}^{\lambda}_{\mu\nu}$ by the following transformation.

\begin{equation}
\begin{aligned}
    \alpha_{\lambda\nu}^{\mathrm{(in)}} &= \sum_{\mu=-\lambda}^{\lambda} \mathcal{D}^{\lambda *}_{\mu\nu}(\Omega) \, \alpha_{\lambda\mu}^{(\mathrm{lab})} \,, &
    \alpha_{\lambda\mu}^{(\mathrm{lab})} &= \sum_{\nu=-\lambda}^{\lambda} \mathcal{D}^{\lambda}_{\mu\nu}(\Omega) \, \alpha_{\lambda\nu}^{\mathrm{(in)}} \,.
\end{aligned}
\label{eq:coord_combined}
\end{equation}

Various authors use different phase conventions for the Wigner matrices; here, we adopt the convention used by Bohr \cite{Bohr_1952, BohrMotelson_1953}.

\begin{figure}[h]
    \centering
    \includegraphics[width=8cm]{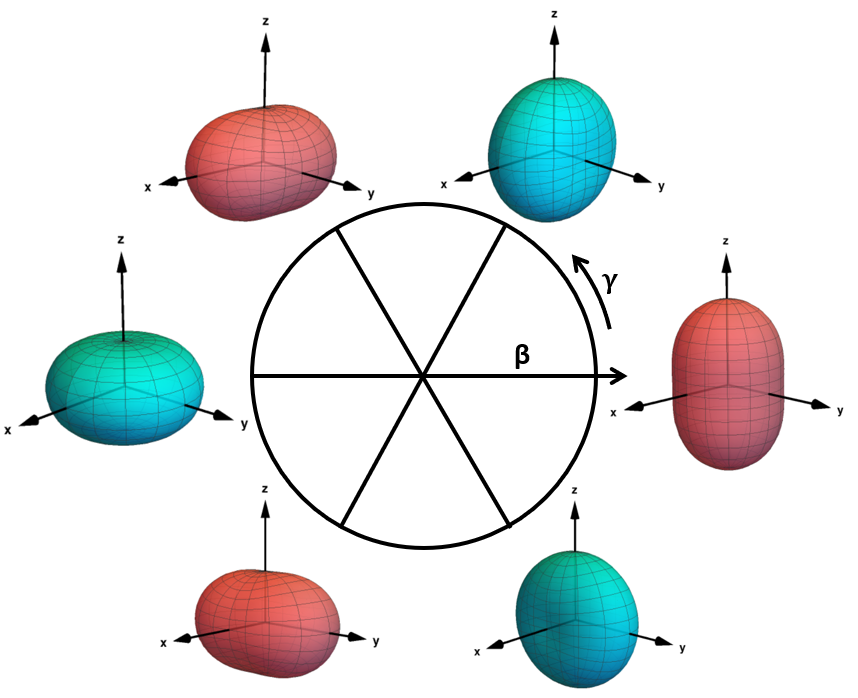}
    \caption{Schematic representation of shape deformation in the $\left( \beta, \gamma \right)$ plane.}
    \label{fig:beta_gamma_plane}
\end{figure}

In this new frame, the coefficients $a_{21}$, $b_{21}$, and $b_{22}$ vanish.
Defining $\mathcal{D}^{\lambda(\pm)}_{\mu \nu}=[2(1+\delta_{\nu0})]^{-1/2}[\mathcal{D}^{\lambda}_{\mu \nu}\pm(-1)^\nu\mathcal{D}^{\lambda}_{\mu- \nu}]$, we characterise the collective coordinates using two deformation parameters, $a_0$ and $a_2$, and three Euler angles, as
\begin{equation}
    \alpha_{2\mu}^{(\mathrm{lab})} =  \mathcal{D}^{2(+)}_{\mu0}(\Omega) \,  a_0 + \mathcal{D}^{2(+)}_{\mu2}(\Omega) \, a_2 \, .
\end{equation}
The parameters $a_0$ and $a_2$ are usually expressed in terms of Bohr's deformation parameters $\beta$ and $\gamma$ via the relations $a_0 = \beta \cos \gamma$ and $a_2 = \beta \sin \gamma$. In the MF approach, the latter are related to the expectation values of the intrinsic quadrupole operators (expressed in Cartesian coordinates).
\begin{eqnarray}
   \hat{Q}^{(in)}_{20}=\frac{1}{2}\sqrt{\frac{5}{4\pi}}\sum_i(2z_i^2-x_i^2-y_i^2)\;\;, \;\;  \hat{Q}^{(in)}_{22}=\frac{1}{2}\sqrt{\frac{5}{4\pi}}\sum_i\sqrt{3}(x_i^2-y_i^2)\,,
 \end{eqnarray}
 via the relations
\begin{eqnarray}
\beta \cos\gamma = \frac{4\pi}{5} \frac{1}{A \langle r^2 \rangle}  \langle \hat{Q}^{(in)}_{20} \rangle \;\;, \;\;  
\beta \sin\gamma = \frac{4\pi}{5} \frac{1}{A \langle r^2 \rangle}  \langle \hat{Q}^{(in)}_{22} \rangle ,
\end{eqnarray}

\noindent where A is the number of nucleons and $\langle r^2 \rangle$ is the mean-square radius estimated as extracted from the liquid drop model \cite{Libert_1999}.
The parameter $\beta$ describes the elongation of the nucleus along the $z$-axis, while $\gamma$ is an angle characterising nuclear triaxiality. As an illustration, Fig.~\ref{fig:beta_gamma_plane} shows how the nuclear shape evolves for a fixed value of $\beta$ and a few selected values of $\gamma$. For simplicity, we represent only axially symmetric shapes here, but this transformation is general and also applies to triaxial shapes.

We note that this rotation allows the three axes of the reference frame to be aligned with the principal axes of the nucleus, but it does not specify their ordering or orientation. Therefore, there are $3! \cdot 2^3 = 48$ possible ways to define this intrinsic frame, or only 24 if a right-handed frame is considered. All quantities (wave functions, moments of inertia, \dots) must therefore be invariant under these transformations, which form the octahedral group $\mathcal{O}$ (see Appendix \ref{app:rotations} for more details).

This symmetry reduces the range of the $\gamma$ parameter to $0 \le \gamma \le \pi/3$. Indeed, as shown in Fig.~\ref{fig:beta_gamma_plane}, the shapes corresponding to $\gamma=k\pi/3$, with even $k$, are identical up to a rotation of the reference frame; these shapes are called prolate. Similarly, the shapes corresponding to odd values of $k$ are also equivalent up to a rotation and are referred to as oblate.
\subsection{Collective Hamiltonian}

In order to obtain the form of the quantized BH, we start from the general classical form of the Hamiltonian in the laboratory frame \cite{pro09} written as

\begin{equation}\label{eq:ham}
    H_\text{coll}=\frac{1}{2} \left( \dot{\alpha}_{2}^{(\mathrm{lab})} \cdot  B_{22}^\mathrm{(lab)}(\alpha_{2}) \cdot \dot{\alpha}_{2}^{(\mathrm{lab})} \right) +V_\text{coll}(\alpha_{2}) \, ,
\end{equation}

where $V_\text{coll}(\alpha_{2})$ is the real potential and $B_{22}^\text{lab}(\alpha_{2})$ is the mass bitensor. We refer to Ref.~\cite{pro09} for details about its construction.
In order to evaluate the kinetic terms, we introduce the time derivative of the coordinates as

\begin{eqnarray}
\dot{\alpha}_{2\mu}^{(\mathrm{lab})} = \mathcal{D}^{2(+)}_{\mu0} \dot{a_0} + \mathcal{D}^{2(+)}_{\mu2} \dot{a_2} - 2 i a_x \mathcal{D}^{2(-)}_{\mu1} \omega_x  - 2 a_y \mathcal{D}^{2(+)}_{\mu1} \omega_y + 2 i a_z \mathcal{D}^{2(-)}_{\mu2} \omega_z \, , 
\end{eqnarray}

\noindent where $\omega_x, \omega_y, \omega_z$ are the intrinsic components of the angular velocity in the $x$, $y$, and $z$ axes, related to the Euler angles, and their time derivatives, by 
\begin{eqnarray}
     \omega_x = \dot{\theta} \sin \psi - \dot{\phi} \sin \theta \cos \psi \, , \quad
    &\omega_y = \dot{\theta} \cos \psi + \dot{\phi} \sin \theta \sin \psi \, ,  \quad
    & \omega_z = \dot{\psi} + \dot{\phi} \cos \theta \,.
\end{eqnarray}

\noindent The explicit expressions for the terms $a_x,a_y,a_z$ as functions of $a_0$ and $a_2$ are given in Eq.~\eqref{eq:axayaz}.

\noindent Using the definition of the scalar product between tensors and bitensors $\left(\alpha_2 \cdot B_{2\mu2\nu} \cdot \alpha_2 \right) = \sum_{\mu\nu} (-1)^{\mu+\nu} \alpha_{2\mu} B_{2-\mu2-\nu} \alpha_{2\nu}$, its transformation under a given rotation $ B_{2\mu'2\nu'}^{(in)} = \sum_{\mu \nu} \mathcal{D}_{\mu\mu'}^{2*} B_{2\mu2\nu}^{(lab)} \mathcal{D}_{\nu\nu'}^{2*} \, 
$, and the following property of Wigner functions $\mathcal{D}^{2(\pm)*}_{\mu\nu} = \pm (-1)^{\mu} \mathcal{D}^{2(\pm)}_{-\mu\nu}$, we express the scalar product in Eq.~\eqref{eq:ham} in terms of intrinsic collective coordinates. This expression contains 25 terms, but only 6 are non-zero. They read



\begin{eqnarray}
    \sum_{\mu\nu}  (-1)^{\mu+\nu}\dot{a}_0 \mathcal{D}^{2(+)}_{\mu0} B^{\mathrm{(lab)}}_{2-\mu2-\nu} \dot{a}_0 \mathcal{D}^{2(+)}_{\nu0}  & = & \dot{a}_0^2  B_{00},\label{eq:B00}  \\
    \sum_{\mu\nu}  (-1)^{\mu+\nu} \dot{a}_0 \mathcal{D}^{2(+)}_{\mu0} B^{\mathrm{(lab)}}_{2-\mu2-\nu} \dot{a}_2 \mathcal{D}^{2(+)}_{\nu2} &= & \dot{a}_0 \dot{a}_2   B_{02}\label{eq:B02} , \\ 
     \sum_{\mu\nu}  (-1)^{\mu+\nu} \dot{a}_2 \mathcal{D}^{2(+)}_{\mu2} B^{\mathrm{(lab)}}_{2-\mu2-\nu} \dot{a}_2 \mathcal{D}^{2(+)}_{\nu2}  &= & \dot{a}_2^2 B_{22} ,\label{eq:B22}\\
    \sum_{\mu\nu}  (-1)^{\mu+\nu} (-i2\omega_x a_x) \mathcal{D}^{2(-)}_{\mu1} B^{\mathrm{(lab)}}_{2-\mu2-\nu} (-i2\omega_x a_x) \mathcal{D}^{2(-)}_{\nu1}  
    &= &  4 \omega_x^2 a_x^2 B_x, \\
        \sum_{\mu\nu}  (-1)^{\mu+\nu} (-2\omega_y a_y) \mathcal{D}^{2(+)}_{\mu1} B^{\mathrm{(lab)}}_{2-\mu2-\nu}  (-2\omega_y a_y) \mathcal{D}^{2(+)}_{\nu1} &= & 4 \omega_y^2 a_y^2 B_y ,\\
      \sum_{\mu\nu}  (-1)^{\mu+\nu} (i2\omega_z a_z) \mathcal{D}^{2(-)}_{\mu2} B^{\mathrm{(lab)}}_{2-\mu2-\nu} (i2\omega_z a_z) \mathcal{D}^{2(-)}_{\nu2}   & = & 4\omega_z^2 a_z^2 B_z .
\end{eqnarray}

\noindent The expressions for $B_{00}$, $B_{02}$, $B_{22}$, $B_x$, $B_y$, and $B_z$ are given by

\begin{equation}\label{eq:appendix:B:B0B1B2BxByBz}
\begin{aligned}
B_x &= - B^{\mathrm{(in)}}_{21,21} - B^{\mathrm{(in)}}_{21,2-1}\,, & \qquad B_{00} &= B^{\mathrm{(in)}}_{20,20}\,, \\
B_y &= \phantom{-}B^{\mathrm{(in)}}_{21,21} - B^{\mathrm{(in)}}_{21,2-1}\,, & \qquad B_{02} &= \sqrt{2}\, B^{\mathrm{(in)}}_{20,22}\,, \\
B_z &= - B^{\mathrm{(in)}}_{22,22} + B^{\mathrm{(in)}}_{22,2-2} \,, & \qquad B_{22} &= B^{\mathrm{(in)}}_{22,22} + B^{\mathrm{(in)}}_{22,2-2}\,.
\end{aligned}
\end{equation}

\noindent Putting all the terms together, the classical collective Hamiltonian in the intrinsic frame is then formed by three parts as follows
\begin{equation}\label{eq:classic:H}
{H}_\text{coll} = {T}_\text{vib} + {T}_\text{rot} + V_\text{coll} \,.
\end{equation}

\noindent The vibrational part describes changes in the nuclear shape at fixed orientation, while the rotational kinetic part corresponds to rotations of the nucleus at fixed deformation \cite{Bohr_1952}. These two terms read, respectively,
\begin{eqnarray}\label{eq:Hvibrot_cl}
  T_\text{vib} &=&
 \frac{1}{2} \left( B_{00} \dot{a}_0^2 + 2B_{02} \dot{a}_0 \dot{a}_2 + B_{22}\dot{a}_2^2 \right)\,, \\
 &=&    \frac{1}{2} \left( B_{\beta\beta}\dot{\beta}^2 + 2B_{\beta\gamma}\beta \dot{\beta} \dot{\gamma} + B_{\gamma\gamma} \beta^2 \dot{\gamma}^2 \right)\,,\\
    {T}_\text{rot} &=& \frac{1}{2} \left(\mathcal{I}_x{\omega_x}^2 +\mathcal{I}_y{\omega_y}^2+\mathcal{I}_z{\omega_z}^2 \right)\,,
\end{eqnarray}

\noindent where we introduced the transformation

\begin{equation}
\begin{split}\label{transformation:B}
    B_{\beta\beta} = & B_{00} \cos^2 \gamma + 2 B_{02} \sin \gamma \cos \gamma + B_{22} \sin^2 \gamma\,, \\
    B_{\beta\gamma} = & \left( B_{22} - B_{00} \right) \sin \gamma \cos \gamma + B_{02} \left( \cos^2 \gamma - \sin^2 \gamma \right)\,, \\
    B_{\gamma\gamma} = & B_{00} \sin^2 \gamma - 2 B_{02} \sin \gamma \cos \gamma + B_{22} \cos^2 \gamma\,,
\end{split}
\end{equation}

\noindent and 
\begin{equation}
\mathcal{I}_k = 4B_ka_k^2  = 4B_k\beta^2 \sin^2\left(\gamma-2 k \pi/3 \right) \,,
\label{eq:Ixyz}
\end{equation}
with $k=\{ 1,2,3 \}=\{ x,y,z \}$ depending on whether $k$ appears as an integer index or as a coordinate label.
The Hamiltonian in Eq.~\eqref{eq:classic:H}  is re-expressed as follows
\begin{equation}
    H_\text{coll} = \frac{1}{2} \sum_{\mu'\nu'} B_{2\mu',2\nu'} \dot{q}_{\mu'}\dot{q}_{\nu'}+ V_\text{coll}\,,
\end{equation}
where $\dot{q}_{\mu'} = \{\dot\beta, \dot\gamma, \omega_x, \omega_y, \omega_z\}$. Although the angular velocity components $\omega_i$ are not time derivatives of true coordinate positions, they describe the instantaneous rotational dynamics of the intrinsic frame. Consequently, upon quantization, the rotational contribution is naturally expressed in terms of the angular momentum operators $\hat{L}_i$, which act as the generators of infinitesimal rotations around the intrinsic axes. Using the standard Podolsky-Pauli procedure \cite{pod28,pauli} and adopting the generalized derivative notation $\{\frac{\partial}{\partial q_{\mu'} }\} = \{  \frac{\partial}{\partial \beta}, \frac{\partial}{\partial \gamma}, \frac{i}{\hbar}\hat{L}_x, \frac{i}{\hbar}\hat{L}_y, \frac{i}{\hbar}\hat{L}_z\}$ \footnote{Notice we made the derivative dimensionless by scaling the intrinsic angular momentum  operators in the intrinsic frame by $i/\hbar$. See Appendix \ref{app:curv} for details.}, the quantized collective Hamiltonian reads
\begin{equation}
    \hat{H}_\text{coll} =-\frac{\hbar^2}{2\sqrt{\vert \text{det}(B) \vert}} \sum_{\mu',\nu'} \frac{\partial}{\partial q_{\mu'} } \sqrt{\vert \text{det}(B) \vert} (B^{-1})_{2\mu',2\nu'} \frac{\partial}{\partial q_{\nu'} } + \hat{V}_\text{coll}\,,
    \label{H:general}
\end{equation}
where
\begin{equation}
    B =  
    \begin{pmatrix}
        B_\text{vib} & 0 \\
        0 & B_\text{rot} 
    \end{pmatrix} =
    \begin{pmatrix}
        B_{\beta\beta} & \beta B_{\beta\gamma} & 0 & 0 & 0 \\
        \beta B_{\beta\gamma} & \beta^2 B_{\gamma\gamma} & 0 & 0 & 0 \\
        0 & 0 & \mathcal{I}_x & 0 & 0 \\ 
        0 & 0 & 0 & \mathcal{I}_y & 0 \\
        0 & 0 & 0 & 0 & \mathcal{I}_z
    \end{pmatrix}.
\end{equation}

Therefore, the determinant of $B$ in the intrinsic frame can be expressed in terms of $(\beta,\gamma)$ as
\begin{equation}
\begin{split}
    \sqrt{\vert \text{det}(B) \vert } &= 2 \beta^4  \vert\sin(3 \gamma)\vert \sqrt{G} \,,
\end{split}
\end{equation}
 with $G = G_\text{vib} B_x B_y B_z $ and $G_\text{vib}= (B_{\beta\beta}B_{\gamma\gamma}-B_{\beta\gamma}^2)$. 
After some algebraic manipulations, we express the vibrational part of the Hamiltonian as 
\begin{eqnarray}\label{H:general:vib}
    \hat{H}_\text{vib} &=&  -\frac{\hbar^2}{2\sqrt{G}} \left[ \frac{1}{\beta^4} \frac{\partial }{\partial \beta} \beta^4 \sqrt{G} \frac{B_{\gamma\gamma}}{G_\text{vib} } \frac{\partial }{\partial \beta} + \frac{1}{\beta^2 \sin 3\gamma}\frac{\partial }{\partial \gamma} \sin 3 \gamma \sqrt{G} \frac{B_{\beta\beta}}{G_\text{vib} } \frac{\partial }{\partial \gamma} \right. \nonumber \\
    && \left. - \frac{1}{\beta^4} \frac{\partial }{\partial \beta}  \beta^3 \sqrt{G} \frac{B_{\beta\gamma}}{G_\text{vib} } \frac{\partial }{\partial \gamma}  - \frac{1}{\beta \sin 3 \gamma}\frac{\partial }{\partial \gamma} \sin 3 \gamma \sqrt{G} \frac{B_ {\beta\gamma}}{G_\text{vib} } \frac{\partial }{\partial \beta} \right]  \, .
\end{eqnarray}

\noindent For the rotational part, we get
\begin{equation}\label{H:general:rot}
  \hat{H}_\text{rot} = \frac{1}{2} \left[\frac{\hat{L}_x^2}{\mathcal{I}_x}  +   \frac{\hat{L}_y^2}{\mathcal{I}_y}  +   \frac{\hat{L}_z^2}{\mathcal{I}_z} \right] .
\end{equation}

The Hamiltonian is fully determined once the collective potential, the mass parameters, and the moments of inertia are specified. These quantities can therefore be chosen in a simple form, allowing for an analytical solution of the problem, as done in Sec. \ref{sec:analytical}, where constant mass parameters and moments of inertia are assumed together with simple potentials. Alternatively, one may consider more elaborate descriptions in which the mass bitensor components depend explicitly on the deformation parameters $\beta$ and $\gamma$ and are obtained, for example, from mean-field calculations. In that case, the problem must be treated numerically as an eigenvalue problem.

In order to find the eigenvalues associated with this Hamiltonian, several techniques have been used: either by applying finite-difference methods \cite{Kumar_1967,hoo72} or by expanding the Hamiltonian over a basis \cite{Libert_1999}. We have chosen the second approach and describe it in the following section. To date, no comparison of the numerical accuracy of the two methods has been carried out, and further work in this direction is therefore required.

\section{Basis wave functions}\label{basis:wf}

We expand the Hamiltonian given in Eq.~\eqref{H:general} over a suitable orthonormal basis. 
For this purpose, we follow the procedure presented in Refs. \cite{Libert_Quentin_1982,pro99b,Libert_1999}. We assume that the basis functions are expressed in the laboratory frame as

\begin{equation}\label{wf:guess}
    \varphi^{I(lab)}_{Mmn} \left( \beta,\gamma,\Omega \right) = \sum_{L=-I}^{I} \varphi^{IM}_{Lmn} \left( \beta,\gamma,\Omega \right) = \sum_{L=-I}^{I} \varphi^{I\mathrm{(in)}}_{Lmn}\left( \beta,\gamma \right) \sqrt{\frac{2I+1}{8\pi^2}} \mathcal{D}_{ML}^{I} \left(\Omega \right)\,,
\end{equation}

\noindent where $I$ denotes the total angular momentum, while $M$ and $L$ denote its projections onto the $z$-axis in the laboratory and intrinsic frames, respectively. We further assume that $\varphi^{I\mathrm{(in)}}_{Lmn}$ is separable in the $(\beta,\gamma)$ coordinates as

\begin{equation}\label{guess:psi1}
    \varphi^{I\mathrm{(in)}}_{Lmn} \left( \beta,\gamma \right) = \text{e}^{-\mu^2 \beta^2/2} \beta^n \left\{ \begin{array}{c}
         \cos m \gamma\\
         \sin m \gamma
    \end{array} \right\}\,.
\end{equation} 
The indices run over $m=0,1,\ldots \,,m_\mathrm{max}$ and $n=m,m+2,\ldots \,,m_\mathrm{max}$. Eq.~\eqref{guess:psi1} can be expressed in a simpler way in terms of the $a_0,a_2$ variables. In this case, $m_\mathrm{max}$ represents simply the maximum degree of the resulting polynomial.

The above functions actually constitute a complete, non-orthogonal set of square-integrable functions. Therefore, in order to obtain an orthonormal basis, we first have to apply the appropriate symmetry constraints, as described in the next section.

\subsection{Symmetries \label{sec:sym}}

All the 24 transformations of the $\mathcal{O}$ group can be obtained as a combination of three generators  $R_1,R_2,R_3$  \cite{Bohr_1952,Kumar_1967,Corrigan_1976}. Moreover, from these generators, we can build three projectors (up to a constant factor) as
\begin{equation}\label{eq:projectors}
    \begin{array}{l}
        P_1 = \hat{\mathbb{I}} + R_1,  \\
        P_2 = \hat{\mathbb{I}} + R_2 + R_2^2 + R_2^3, \\
        P_3 = \hat{\mathbb{I}} + R_3 + R_3^2\,,
    \end{array}
\end{equation}

\noindent where $\hat{\mathbb{I}}$ is the identity operator. Notice that invariance under the product $P_3 P_2P_1$ automatically ensures that the functions in Eq.~\eqref{wf:guess} have the correct symmetries, \emph{i.e.}, that they are invariant under the 24 transformations of the group.

To start, we consider the triplet of Euler angles  $\Omega_k=\{\phi_k,\theta_k,\psi_k\}$ characterising the position of the intrinsic frame after one of the 24 transformations of the octahedral group. Using Eq.~\eqref{wigner:transform}, it is possible to relate this transformation to $\Omega$ used to move from the laboratory to the intrinsic frame as

\begin{equation}\label{eq:Rwf}
\begin{split}
   \varphi^{I(lab)}_{Mmn} \left( \beta_k,\gamma_k,\Omega_k \right) = & \sum_{L=-I}^{I}  \varphi^{I\mathrm{(in)}}_{Lmn}\left( \beta_k,\gamma_k \right) \sqrt{\frac{2I+1}{8\pi^2}} \mathcal{D}_{ML}^{I} \left(\Omega_k\right) \\
    = & \sum_{L=-I}^{I}  \left[ \sum_{K=-I}^{I} \varphi^{I\mathrm{(in)}}_{Kmn}\left( \beta_k,\gamma_k \right)\mathcal{D}_{LK}^{I} \left( \theta_k\right) \right] \sqrt{\frac{2I+1}{8\pi^2}} \mathcal{D}_{ML}^{I} \left(\Omega \right)\,,
\end{split}
\end{equation}

\noindent where the angles $\theta_k =\{\theta_k^1,\theta_k^2,\theta_k^3\}$ corresponding to each rotation are given in Tab.~\ref{tab:rotations}. From the above equation, the function $\varphi^{I\mathrm{(in)}}_{Lmn}$ must obviously satisfy the following condition to be invariant under the transformations of $\mathcal{O}$

\begin{equation}
  \varphi^{I\mathrm{(in)}}_{Lmn}\left( \beta, \gamma \right) = \sum_{K=-I}^{I} \varphi^{I\mathrm{(in)}}_{Kmn}\left( \beta_k,\gamma_k \right) \mathcal{D}_{LK}^{I}\left( \theta_k\right) \, .
     \label{eq:inv}
\end{equation} 

\noindent With this constraint in mind, we apply the various rotations $R_{k=1,2,3}$ to the functions given above. Details are provided in Appendix \ref{appendix:wf:rot}. After some tedious but straightforward algebraic manipulations \cite{Kumar_1967}, invariance with respect to $P_1, P_2,P_3$ leads to the expression

\begin{equation}\label{eq:basis:final}
    \varphi^{IM}_{Lmn} \left( \beta,\gamma,\Omega \right) =  \text{e}^{-\mu^2\beta^2/2} \beta^n \sum_{K\in \Delta_I} f^I_{LmK}\left(\gamma\right) \Phi^I_{MK} \left( \Omega \right),
\end{equation}

\noindent where we defined the normalized function $
    \Phi^{I}_{MK} (\Omega)= \sqrt{\frac{2I+1}{16\pi^2(1+\delta_{K0})}} \left[ \mathcal{D}^I_{MK} (\Omega)+ (-1)^{I} \mathcal{D}^I_{M-K} (\Omega) \right] $ and $K\in \Delta_I$ means that $K = 2(I \bmod 2 ), 2(I \bmod 2 )+2, \ldots , 2 \left\lfloor \frac{I}{2} \right\rfloor$ \footnote{The symbol $\left\lfloor \cdot \right\rfloor$ stands for the floor function defined for example in Ref.~\cite{iverson1962programming}.}. The $\gamma$ dependent part of the function is now defined as $f^I_{LmK}\left(\gamma\right) = c^I_{LmK} p_{Km} \left(\gamma\right) $ with 

\begin{equation} \label{cILmK:def}
     c^I_{LmK} =  \delta_{LK} +2  (-1)^{K/2} \left[ (1+\delta_{L0}) (1+\delta_{K0}) \right]^{-1/2} d^I_{KL} \left(\frac{\pi}{2}\right)  \times \left\{ \begin{array}{cl}
         2 \cos \left(\frac{2\pi}{3} m\right) & (K+L)/2 \, \text{even}\\
         2 \sin \left(\frac{2\pi}{3} m\right) & K/2 \, \text{even,}\,  L/2 \, \text{ odd}\\
         - 2 \sin \left(\frac{2\pi}{3} m\right) & K/2 \, \text{odd,}   L/2 \, \text{even}
    \end{array} \right. 
\end{equation}
and
 \begin{equation} \label{pKm:def}
   p_{Km} \left( \gamma \right) = \left\{ \begin{array}{ll}
         \cos (m \gamma) & K/2 \, \text{ even}\,,\\
         \sin (m \gamma) & K/2 \, \text{ odd}\,.
    \end{array} \right.
\end{equation}

\noindent Notice that $d^I_{KL}$ is the so-called small Wigner function, whose explicit expression can be found, for example, in Ref.~\cite{Varshalovich_1988}.
The functions given in Eq.~\eqref{eq:basis:final} have been constructed to be invariant under the transformations of the group $\mathcal{O}$. However, they are not yet orthogonal and include linearly dependent functions. We therefore apply a truncation scheme to obtain a suitable basis.

\subsection{Removing linearly dependent functions }

The symmetrization procedure has led to a set of functions, among which some appear to be redundant. Therefore, we have to go further and identify the subsets of linearly independent functions. In this process, special attention must be paid to the boundary conditions imposed by the physical problem (see Sec. \ref{sec:divergence}). In Ref. \cite{Libert_1999}, this is done via a Gram-Schmidt procedure, but this needs to be handled carefully in order to ensure numerical stability. Here, we prefer to use an analytical approach based on group theory that does not introduce any additional numerical uncertainty into the solution of the problem.

\subsubsection{Number of linearly independent functions}

The functions generated from distinct combinations of quantum numbers $(I,M,m,n)$ are, by construction, linearly independent. Consequently, the problem reduces to counting, within each subset associated with a fixed quadruplet $(I,M,m,n)$, the number of effectively independent functions. In practice, this number depends only on the values of $I$ and $m$, since $M=-I, \dots , I$ and the $\beta$-dependent sector, and thus $n$, are invariant under rotations. Furthermore, since the coefficients $c^I_{LmK}$ are cyclic, \emph{i.e.}, $c^I_{Lm+3K} = c^I_{LmK}$, we limit ourselves to $m=0,1,2,3$.

We adopt a group-theoretical approach: each set of functions can be viewed as forming a reducible representation $\mathcal{R}$ of $\mathcal{O}$, and the number of linearly independent functions must correspond exactly to the multiplicity of the trivial (identity) representation in the decomposition of $\mathcal{R}$ into irreducible components. The reason is simple: each occurrence of the trivial representation identifies one invariant subspace, \emph{i.e.}, one function that is linearly independent and fully symmetric under $\mathcal{O}$. Therefore, we decompose the representation $\mathcal{R}$ as a tensor product of two representations: $\mathcal{R}_\gamma$, which acts on the space of functions depending on $\gamma$, and $\mathcal{R}_\Omega$, which acts on the space of functions depending on the Euler angles $\Omega$.
The character of $\mathcal{R}$ can be expressed as
\begin{equation}
    \chi^{\mathcal{R}}(g) = \chi^{\mathcal{R}_\gamma}(g) \times \chi^{\mathcal{R}_\Omega}(g)\,,
\end{equation}

\noindent with $g$ an element of $\mathcal{O}$. 

We first focus on the two-dimensional representation corresponding to the functions depending on $\gamma$. The action of an element $g$ on the space spanned by $\cos(m\gamma)$ and $\sin(m\gamma)$ can be expressed as a linear transformation
\begin{equation}
    g \cdot 
\begin{pmatrix}
\cos(m\gamma) \\
\sin(m\gamma)
\end{pmatrix}
=\begin{pmatrix}
\cos(m\gamma_g)\\
\sin(m\gamma_g)
\end{pmatrix}
=
A_g
\begin{pmatrix}
\cos(m\gamma) \\
\sin(m\gamma)
\end{pmatrix} \,
\end{equation}

\noindent where $\gamma_g$ denotes the value of $\gamma$ after the transformation by the group element $g$. In this basis, the representation matrix of $g$ is a standard $2$-dimensional rotation matrix, and its character is given by $\chi^{\mathcal{R}_\gamma}(g) = \text{Tr}(A_g)$. Although the individual values of $\gamma_g$ may differ within a given class, all transformations belonging to the same class yield equivalent representation matrices and therefore the same character value. This allows us to systematically construct the character table shown in Tab.~\ref{tab:caracter_Rgamma}. See also Appendix \ref{app:rotations} for more details.

\begin{table}[h]
\begin{center}
    \caption{Character table of the $\mathcal{R}_\gamma$ representation for each conjugacy class of the group $\mathcal{O}$.}
    \label{tab:caracter_Rgamma}
\renewcommand*{\arraystretch}{1.5}
\setlength{\tabcolsep}{6pt}
    \begin{tabular}{lccccc}
    \hline
        Class & $\mathscr{C}_1$ & $\mathscr{C}_2$ & $\mathscr{C}_3$ & $\mathscr{C}_4$ & $\mathscr{C}_5$\\
         \hline \hline 
        $\gamma_g$ & $  \gamma$ & $ -\gamma,-\gamma\pm\frac{2\pi}{3}$ & $ \gamma\pm\frac{2\pi}{3}$ & $ \gamma$ & $ -\gamma,-\gamma\pm\frac{2\pi}{3}$ \\
         \hline \hline 
         $m$ $= 0$ & $1$ & $1$ & $1$ & $1$ & $1$ \\ 
         $m$ mod $3 = 1$ or $2$ & $2$ & $0$ & $-1$ & $2$ & $0$ \\
         $m$ mod $3 = 0$ ($m\ne0$) & $2$ & $0$ & $2$ & $2$ & $0$ \\ 
         \hline
    \end{tabular}
    \end{center}
\end{table}

Let us now turn to the $\mathcal{R}_\Omega$ representation. The set of functions defined by the Wigner functions $\mathcal{D}^{I}_{M L}$ naturally forms an irreducible representation of the group $SO(3)$, which is completely characterized by the angular momentum quantum number $I$. The corresponding character is given by \cite{hamermesh2012group}

\begin{equation}
\chi^{\mathcal{R}_\Omega}(g) =  \frac{\sin\bigl[(2I+1)\frac{\omega_g}{2}\bigr]}{\sin\bigl(\frac{\omega_g}{2}\bigr)},
\end{equation}
\noindent where $\omega_g$ denotes the rotation angle associated with the group element $g$. The character table is thus given in Tab.\ref{tab:character_Romega}.

\begin{table}[h]
\caption{Character table of the $\mathcal{R}_\Omega$ representation for each conjugacy class of the group $\mathcal{O}$.}
\label{tab:character_Romega}
\renewcommand{\arraystretch}{1.3}
\setlength{\tabcolsep}{5pt}
\centering
\begin{tabular}{c|ccccc}
\hline
$\mathscr{C}_i$ & $\mathscr{C}_1$ & $\mathscr{C}_2$ & $\mathscr{C}_3$ & $\mathscr{C}_4$ & $\mathscr{C}_5$\\
\hline \hline
$\omega_g$ & $0$ & $\pm \frac{\pi}{2}$ & $\pm \frac{2\pi}{3}$ & $\pi$ & $\pi$ \\ 
\hline
$\chi^{\mathcal{R}_\Omega}(\mathscr{C}_i)$\rule{0pt}{4.5ex}
& $\displaystyle (2I+1)$ 
& $\displaystyle \frac{\sin \bigl[(2I+1)\frac{\pi}{4}\bigr]}{\sin\bigl(\frac{\pi}{4}\bigr)}$ 
& $\displaystyle \frac{\sin \bigl[(2I+1)\frac{\pi}{3}\bigr]}{\sin\bigl(\frac{\pi}{3}\bigr)}$ 
& $\displaystyle (-1)^I$ 
& $\displaystyle (-1)^I$ \\ 
\hline
\end{tabular}
\end{table}

\noindent By definition, the multiplicity of the trivial representation is given by the following formula \cite{cornwell1997group}
\begin{equation}
    N = \frac{1}{|G|}\sum_\mathscr{C} |\mathscr{C}| \chi^{\mathcal{R}}(\mathscr{C}) \,,
\end{equation} 
where the sum runs over all classes $\mathscr{C}$, $|\mathscr{C}|$ denotes the size of the class, $|G|$ the order of the group, and $\chi^{\mathcal{R}}(\mathscr{C})$ the character of the representation $\mathcal{R}$ evaluated within the class $\mathscr{C}$. By combining the previous results, we obtain a general expression for the multiplicity
\begin{equation}\label{eq:multiplicity}
\begin{split}
    N(I,m) =& \frac{(2I+1)}{24} \chi^{\mathcal{R}_\gamma}(\mathscr{C}_1) + \frac{\sin \left( (2I+1) \pi/4 \right)}{2\sqrt{2}}  \chi^{\mathcal{R}_\gamma}(\mathscr{C}_2) \\
    & +  \frac{2\sin \left( (2I+1) \pi/3 \right)}{3\sqrt{3}}  \chi^{\mathcal{R}_\gamma}(\mathscr{C}_3) + \frac{(-1)^I}{8} \chi^{\mathcal{R}_\gamma}(\mathscr{C}_4) + \frac{(-1)^I}{4} \chi^{\mathcal{R}_\gamma}(\mathscr{C}_5) \,.
\end{split}
\end{equation}
The result can be made more explicit by inserting the possible values of $m$ into Eq.~\eqref{eq:multiplicity}. For $m=0$, we obtain

\begin{equation}
        N(I,m=0) = \left\lfloor\frac{I}{12}\right\rfloor + \left\{ \begin{array}{cc}
        1 & \text{, } I \text{ mod } 12 = 0,4,6,8,9,10 \, , \\
        0 & \text{, } I \text{ mod } 12 = 1,2,3,5,7,11 \, .  \\
    \end{array} \right.
\end{equation}
For $m\ne0$, the result depends on $I$, as summarised below
\begin{table}[h]
    \centering
     \renewcommand{\arraystretch}{1.6} 
     \begin{tabular}{c|c|c} 
     \hline 
     & $m$ $\bmod$ 3 = 0 & $m$ $\bmod$ 3 = 1,2 \\ 
     \hline \hline 
     $I$ $\bmod$ 3 = 0 & $\left\lfloor \frac{\text{I}}{6} \right\rfloor$ + 1 & $\left\lfloor \frac{\text{I}}{6} \right\rfloor$ \\ 
     $I$ $\bmod$ 3 = 1 & $\left\lfloor \frac{\text{I}}{6} \right\rfloor + \delta_{\text{I} \text{ even}}$ & $\left\lfloor \frac{\text{I}}{6} \right\rfloor + \delta_{\text{I} \text{ even}}$ \\ 
     $I$ $\bmod$ 3 = 2 & $\left\lfloor \frac{\text{I}}{6} \right\rfloor$ & $\left\lfloor \frac{\text{I}}{6} \right\rfloor + 1$ \\
     \hline 
     \end{tabular}
\end{table}

By applying these rules, we obtain $N(I,m)$ independent functions by limiting $L$ to values up to $L_\text{max}(I,m)$ within the interval $\Delta_I$.

\subsubsection{Divergence at $\gamma=n\pi/3$ \label{sec:divergence}}

Examination of Eqs. \eqref{eq:Ixyz}-\eqref{H:general:rot} shows that the rotational term diverges along the axes $\gamma=n\pi/3$, where the moments of inertia vanish. To ensure the regularity of the solution, the condition $f^I_{LmK}(\gamma=0)=0$ must be imposed.\footnote{Since the wave function is invariant under transformations between the different sextants, this condition is automatically fulfilled along the other axes.} Since this condition is clearly not fulfilled for $K>0$ and $I \geq 4$ with $K/2$ even, we construct a linear combination of basis functions that exhibits the correct behaviour. Using the periodicity condition $\cos((m-6)\gamma)=\cos(m\gamma)$ and the one of $c^I_{LmK}$ coefficients, we define regular functions \cite{pro99b}.
A thorough examination of all possible cases, particularly those with $m = 0 \pmod{3}$ and $m < 6$, leads to simple combinations that satisfy the boundary conditions. The results can be summarised as follows for $I \ge 4$

\begin{equation}
\begin{array}{ll}
& m<6 \\
\Tilde{f}_{02K}^I = f_{02K}^I + \tfrac{1}{2} f_{00K}^I  &\quad \text{for L=0} \\
\Tilde{f}_{L3K}^I = f_{L3K}^I + 2 \delta_{L0} f_{01K}^I &  \quad \text{for $L=(0),2,6,10,14,\dots,L_\text{max}(I,3)$} \\
\Tilde{f}_{L4K}^I = f_{L4K}^I - (-1)^{L/2} f_{L2K}^I & \quad \text{for $L=(0),2,4,\dots,L_\text{max}(I,2)$} \\
\Tilde{f}_{L5K}^I = f_{L5K}^I - (-1)^{L/2} f_{L1K}^I & \quad \text{for $L=(0),2,4,\dots,L_\text{max}(I,1)$} \\
&\\
& m \ge 6 \\
m \bmod 3 \ne 0 & \\
\Tilde{f}_{LmK}^I = f_{LmK}^I - f_{Lm-6K}^I & \quad \text{for $L=(0),2,4,\dots,L_\text{max}(I,m)$} \\
m \bmod 3 = 0 & \\
\Tilde{f}_{LmK}^I = f_{LmK}^I - \delta_{(L \bmod 4), 0} f_{Lm-6K}^I & \quad \text{for $L=(0),2,4,\dots,L_\text{max}(I,m)$} .
\end{array}
\end{equation}

\noindent The lowest value of $L$ is equal to 0 (2) for $I$ even (odd). The highest value $L_\text{max}$ is determined by previously discussed number of linear independent functions defined in Eq.\eqref{eq:multiplicity}.
This construction preserves invariance under the action of the octahedral group.

\subsection{Orthogonalisation }

The basis functions must be orthonormalised using the integration measure $\mathrm{d} \tau_\mathrm{coll} = \sqrt{G} \mathrm{d} \tau_0 \mathrm{d}\Omega = \sqrt{G} \beta^4 \vert \sin(3\gamma) \vert \mathrm{d}\beta \mathrm{d}\gamma \mathrm{d}\Omega$. However, the factor $\sqrt{G}$ has a non-trivial dependence on $\beta$ and $\gamma$. We drop this factor for the orthonormalisation procedure and multiply the final expression by $G^{-1/4}$  to obtain the correct normalisation. Moreover, since the dependence on $\beta$ and $\gamma$ is factorised, the procedure can now be safely separated into two independent steps: one for the angular part and one for the radial part.

\subsubsection{Orthogonalisation of the angular part}

The first step consists in orthonormalising\footnote{The functions $\Phi^I_{MK}$ are actually already orthonormal with respect to the angular integration measure $\text{d}\Omega= d\phi \sin \theta d\theta d\psi$, as $\int_\Omega \Phi^{I*}_{MK} \Phi^{I'}_{M'K'} \text{d}\Omega = \delta_{II'}\delta_{MM'}\delta_{KK'}$.} the functions $t^{IM}_{Lm}\left( \gamma,\Omega \right) \equiv \sum_{K\in \Delta_I} \Tilde{f}^I_{LmK}(\gamma)\Phi^I_{MK}(\Omega)$. The simplest way to deal with the problem is to consider a matrix $T^{IM}$ whose matrix elements are the functions $t^{IM}_{Lm}$ for fixed $(I, M)$. Then, the scalar product of two functions corresponds to the product of two matrices, $(T^{IM})^T T^{IM}$.

The result is a square matrix that is symmetric and positive definite, so that we can apply the Cholesky decomposition~\cite{cho24}, $(T^{IM})^T T^{IM} = C^{IM}(C^{IM})^T$, where $C$ is a lower triangular matrix.  We can now write the angular part of the basis as

\begin{equation}
    \sum_{K \in \Delta_I} \sum_{L',m'} \left(C^{-1}\right)^{IM}_{Lm,L'm'} \tilde{f}^{I}_{L'm'K} (\gamma) \Phi^I_{MK} (\Omega)\,,
\end{equation}

\noindent which is orthogonal by construction. 

\subsubsection{Orthogonalisation of the radial part}

The initial \emph{ansatz} for the $\beta$-dependent part has been taken as $\beta^n \text{e}^{-\mu^2 \beta^2 /2 }$. This choice was mainly dictated by the fact that the nuclear wave function should be integrable over the deformation space.

Since the angular part is already orthonormalised, for the radial part we have to consider only pairs of monomials with indices $n,n'$. Notice that these indices are not free, but for a fixed value of $m$, they obey the relations $n=m+2k$ and $n'=m+2k'$, with $k,k'=0,1,\dots$. More explicitly, the quantity we need to orthonormalise reads

\begin{eqnarray}
    \int_0^{+\infty} \beta^{m+2k}\beta^{m+2k'}\beta^4e^{-\mu^2\beta^2} d\beta\,.
\end{eqnarray}

\noindent Making a simple change of variable, $y=(\mu\beta)^2$, we see that the functions of $y$ that are orthogonal, for a fixed value of $m$, with respect to the measure $y^{m+\frac{3}{2}}e^{-y}dy$, are the generalised Laguerre polynomials $L^{m+\frac{3}{2}}_{(n-m)/2}(y)$ \cite{abr68}.
The function thus takes the form
\begin{equation}\label{basis:beta}
    B_{nm}(\beta) = \mu^{5/2} N_{nm} (\mu\beta)^m L^{(m+\frac{3}{2})}_{(n-m)/2}(\mu^2\beta^2) \text{e}^{-\mu^2\beta^2 /2 },
\end{equation}
with 
\begin{equation}
    N_{nm} = \left[ \frac{\Gamma((m+n+5)/2)}{2((n-m)/2)!} \right]^{-1/2} \, ,
\end{equation}

\noindent where $\Gamma$ is the gamma function \cite{abr68}. It is interesting to note that Eq.\eqref{basis:beta} is also the solution of the Bohr Hamiltonian for the case of a harmonic potential \cite{Fortunato_2005}.
Such a choice still leaves a free parameter $\mu$, which needs to be adjusted in order to obtain the best results when using a finite basis. See Sec.\ref{sec:analytical} for a more detailed discussion.
Other forms for the $\beta$ sector may be selected \cite{Libert_1999}, but we have found that Eq.~\eqref{basis:beta} provides good accuracy when using a limited number of basis states.

\subsection{Final expression}

\begin{figure}
    \centering
        \centering
          \includegraphics[width=0.6\textwidth]{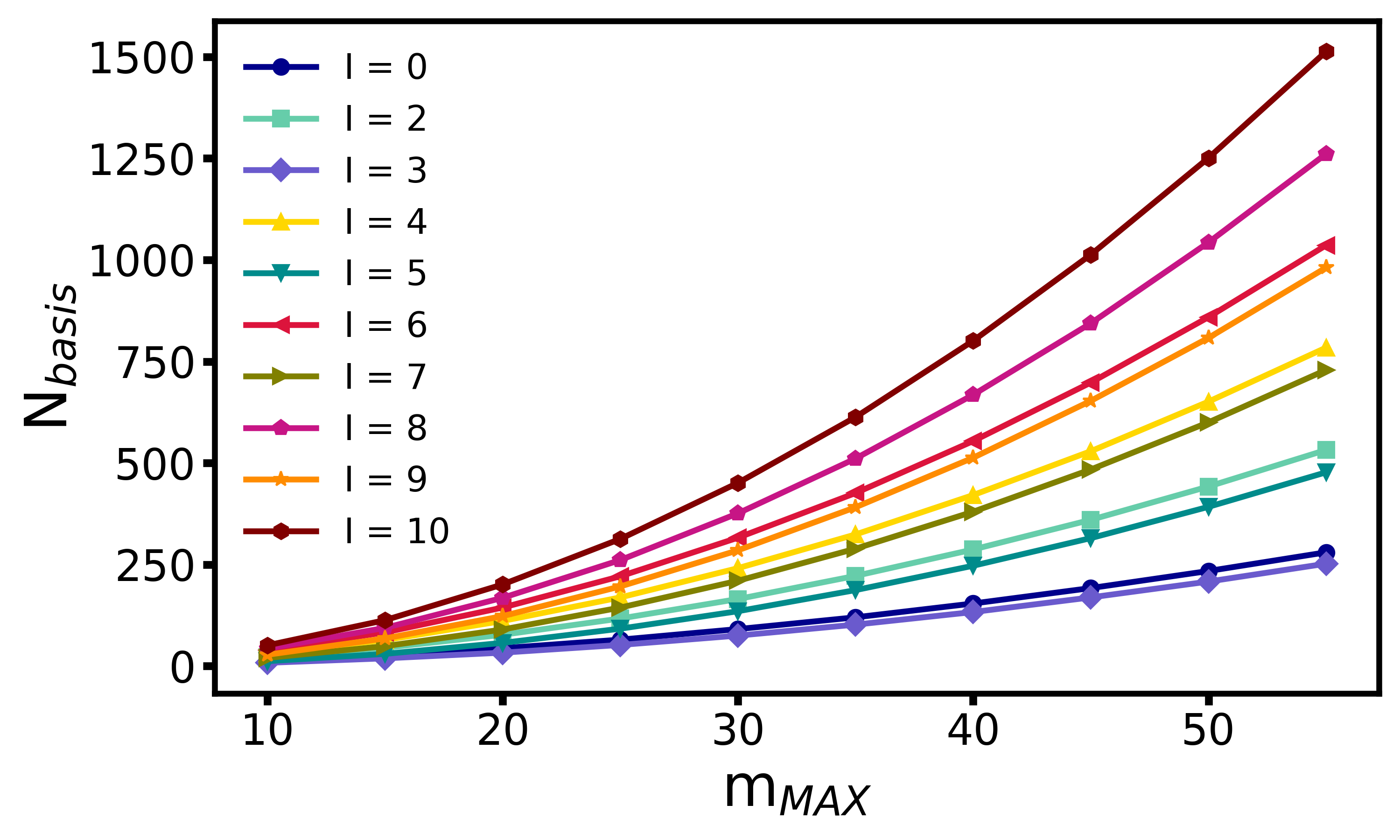}
        \caption{Evolution of the number of basis states included in our calculations as a function of the maximum value of the $m_{max}$ quantum number.}
    \label{fig:basis}
\end{figure}

Grouping the different functions discussed above, we can finally write the intrinsic basis wave functions in a compact form as

\begin{eqnarray}\label{eq:basis:final:TOT}
   \varphi^{IM}_{i}\left( \beta, \gamma,\Omega \right) &= & \sum_{K,L',m'} G^{-1/4} B_{nm}(\beta) \left(C^{-1}\right)^{IM}_{Lm,L'm'} \tilde{f}^{I}_{L'm'K} (\gamma) \Phi^I_{MK} (\Omega) \nonumber \\
  & = & \sum_{K\in \Delta_I} G^{-1/4} F^{I}_{K i}(\beta,\gamma) \Phi^I_{MK} (\Omega) = \sum_{K\in \Delta_I} \Tilde{F}^{I}_{K i}(\beta,\gamma) \Phi^I_{MK} (\Omega)\,.
\end{eqnarray}
where $i$ designates the triplet of quantum numbers $(L,m,n)$.
We observe that the only independent quantum number for a fixed angular momentum $I$ is $m$.
In Fig.~\ref{fig:basis}, we thus show the evolution of the number of basis states, $N_{basis}$, for a given $I$ as a function of the maximum value of $m$, $m_{max}$.
At first sight, we notice that even values of angular momentum require more basis states than odd ones. This fact originates from the selection rules discussed above, which effectively suppress the number of odd-$I$ basis states. The growth in the number of basis states is essentially quadratic, but with different slopes depending on the value of $I$.

The angular momentum $I=1$ is excluded. This follows from the selection rules imposed by the symmetry $R_1$ (see Eq.~\eqref{eq:inv1}), which indicate that, for $L=0$, $I$ must be even. Since $L=0$ is the only allowed projection when $I=1$, this angular momentum is therefore forbidden. Parity is conserved for quadrupole degrees of freedom; consequently, all basis functions have positive parity.


\section{Analytical potentials}\label{sec:analytical}

In order to validate our approach, we tested our numerical code against some simple potentials whose solutions are known analytically \cite{Fortunato_2005}. In order to fully preserve the symmetries of the problem, we considered only $\gamma$-unstable potentials \cite{wil56} and assumed that the mass parameters are constants, $B_{\beta\beta}=B_{\gamma\gamma}=B_x=B_y=B_z=B_m$, and that $B_{\beta\gamma}=0$. Since the potential has the property $V(\beta,\gamma) = V(\beta)$, we can separate the variables appearing in Eq.~\eqref{H:general} as~\cite{wil56}

\begin{eqnarray}
\left\{\frac{\hbar^2}{2B_m} \left( -\frac{\partial^2}{\partial \beta^2}  + \frac{(\tau+1)(\tau+2)}{\beta^2} \right) +V(\beta)\right\}(\beta^2f(\beta))&=&E(\beta^2f(\beta)) \, , \\
\left\{-\frac{1}{ \sin 3 \gamma} \frac{\partial}{\partial \gamma} \sin 3 \gamma\frac{\partial}{\partial \gamma}+\frac{1}{4} \sum_k\frac{\hat{L}_k^2}{ \hbar^2 \sin^2\left( \gamma-\frac{2\pi}{3}k\right)}\right\} \phi(\gamma,\Omega)&=&\tau(\tau+3) \phi(\gamma,\Omega) \, ,
\end{eqnarray}

\noindent where $\tau=0,1,2,\dots$.
Since the potential is only a function of $\beta$, the squared angular momentum and its third component are constants of motion, and thus $L$ and $M$ are good quantum numbers. This leads to some accidental degeneracies, as discussed in Ref.~\cite{Fortunato_2005}.
We first consider the infinite square well potential

\begin{eqnarray}
    V(\beta)=\left\{ \begin{array}{cc}
    0 & \text{for }\beta\le 1 \, ,\\
    +\infty & \text{for }\beta > 1 \, .
    \end{array}\right.
\end{eqnarray}

\noindent The eigenvalues are given by \cite{iachello2000dynamic}
\begin{equation}\label{eq:e+:exact}
    E_{\xi,\tau}=\frac{\hbar^2}{2B_m}k_{\xi\tau}^2 \, ,
\end{equation}
\noindent where $k_{\xi\tau}$ is the $\xi$-th zero of the Bessel function $J_{\tau+3/2}$ \cite{abr68}.

\begin{figure}
    \centering
        \centering
          \includegraphics[width=0.48\textwidth]{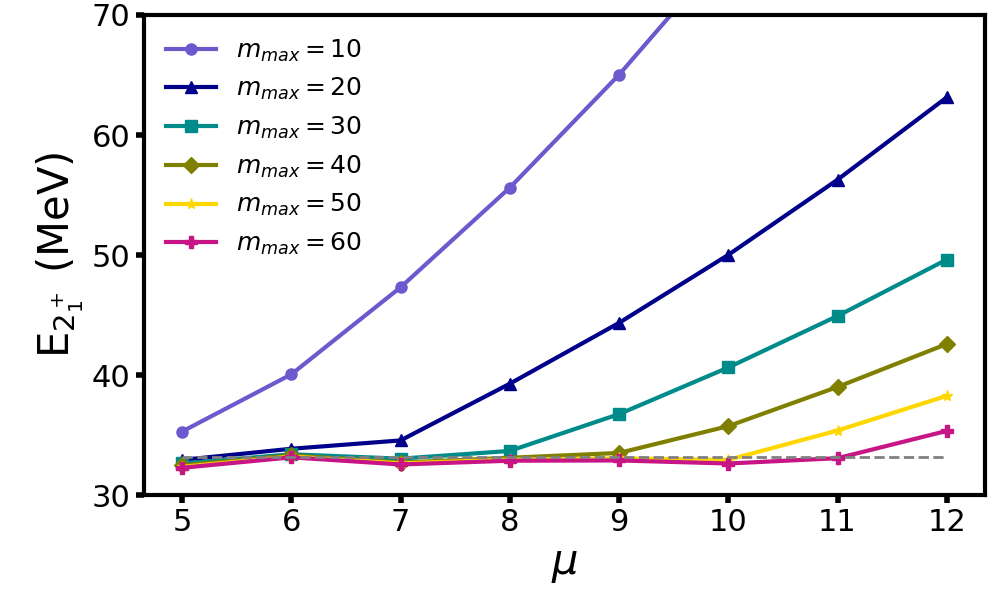}
        \includegraphics[width=0.48\textwidth]{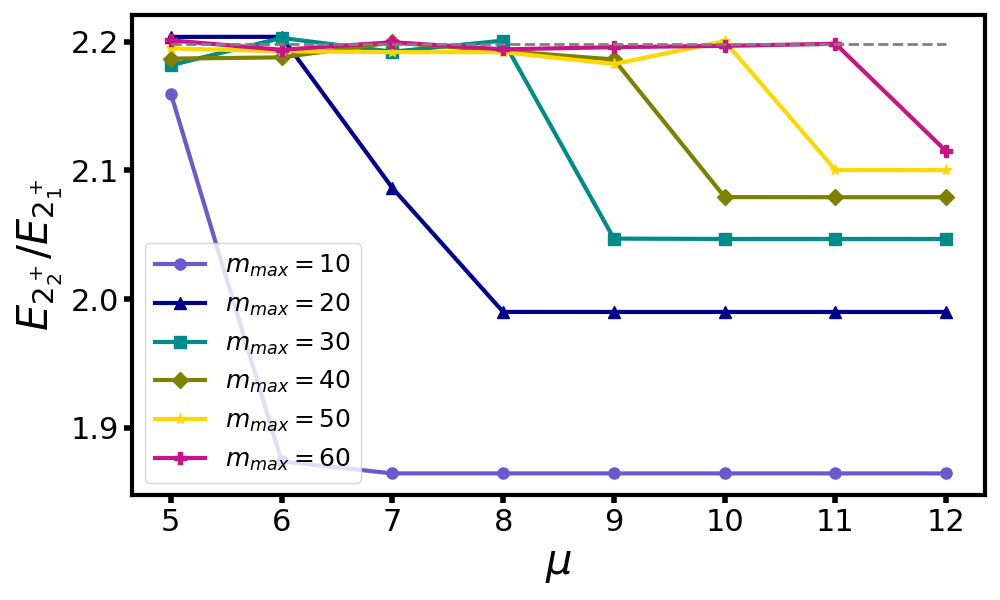}
        \caption{Convergence of the energy of the first $E_{2^+_1}$ (left panel) and the ratio of energy of the second to first $2^+$ (right panel) as a function of $\mu$ for different choices of $m_{max}$. The dashed lines correspond to the exact value of the energy as given by Eq.~\eqref{eq:e+:exact}. See text for details.}
    \label{fig:convergenceE2}
\end{figure}

Following the above discussion, we have developed a numerical code to solve Eq.~\eqref{H:general} and determine the size of the basis required to achieve good convergence of the results. For a given angular momentum $I$, the size of the available basis is governed by the quantum number $m$, as illustrated in Eq.~\eqref{eq:basis:final}. The constant $\mu$ is a free parameter related to the decay rate of the tails of the basis wave functions. In Ref.~\cite{nikvsic2009beyond}, the authors claim that the results are roughly independent of $\mu$ over the range $8\le \mu \le 15$. In Fig.~\ref{fig:convergenceE2}, we illustrate the convergence of the energy of the first $2^+_1$ state for a fixed value of $m_{max}$ as a function of $\mu$ (left panel), and the convergence of the ratio of the energy of the second to the first $2^+$ state (right panel). As a reference, we report the exact value of the energy of the $2^+_1$ state, as obtained from Eq.~\eqref{eq:e+:exact}, as a dashed line. Following Ref.~\cite{iachello2000dynamic}, the analytical value of $(E_{2^+_2}-E_{0^+_1})/(E_{2^+_1}-E_{0^+_1})$ is $2.198$.
We observe that the larger the number of basis states included ($m_{max}=60$), the more independent the results become of $\mu$. However, it will not always be possible to reach such large values of $m_{max}$, since, for large values of $m$, the basis wave functions become more delocalised towards large values of $\beta$. This is clearly not a problem for analytical potentials, but in the case of microscopic ones, we have to take into account the maximum extent of the numerical grid used in our calculations. We observe that, for smaller values of $\mu$, convergence is reached faster for small values of $m_{max}$. Since the parameter $\mu$ is directly related to the number of Gauss-Laguerre integration points in $\beta$, it is preferable to avoid choosing it too small; otherwise, we lose accuracy when calculating our matrix elements numerically.

We conclude that $\mu=9$ and $m_{max}=40$ seem to provide the best compromise in terms of basis size and accuracy of the results. For comparison, the authors of Ref.~\cite{nikvsic2009beyond} selected $\mu=9$, while in Ref.~\cite{mui21} $\mu=10.5$ was used.

\begin{figure}
    \centering
        \centering
           \includegraphics[width=0.49\textwidth]{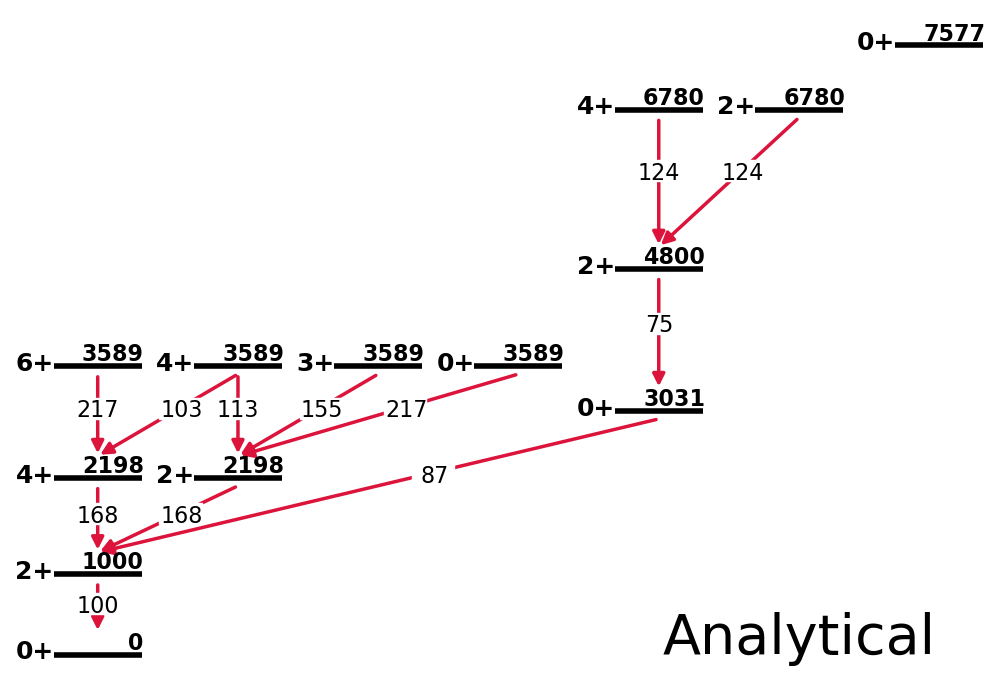}
          \includegraphics[width=0.49\textwidth]{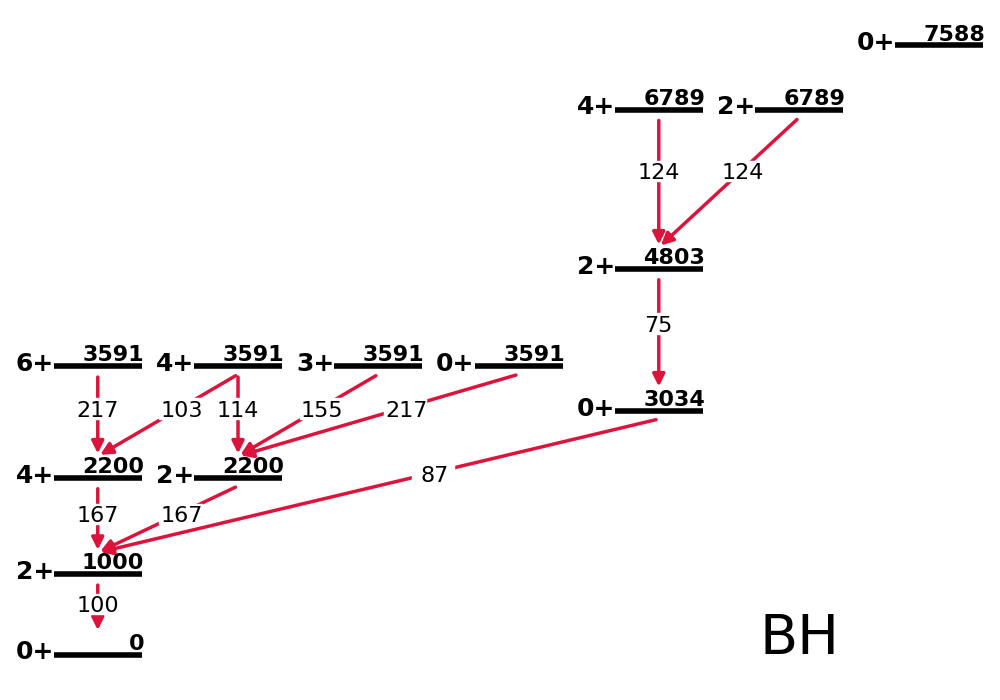}
        \caption{ Analytical (left) and numerical (right) low energy spectrum for the infinite square well obtained with $\mu=9, m_{max}=40$. See text for details.}
    \label{fig:spectrum}
\end{figure}

As a final check, in Fig.~\ref{fig:spectrum}, we compare our numerical calculations, obtained with $\mu=9$ and $m_{max}=40$, with the analytical spectrum derived in Ref.~\cite{iachello2000dynamic}\footnote{We noticed that the value of electromagnetic transition between $2_4^+\rightarrow2_3^+$ reported in Ref.~\cite{iachello2000dynamic} is most likely incorrect since it  does not agree with the value reported in Refs.~\cite{bonatsos2004sequence,cap02,Gar08} for the same potential. We checked numerically that the value reported in the present article is thus in agreement with the scientific literature.}. The energy levels are rescaled to the first $2^+$ state, and the electromagnetic transitions are also rescaled so that $\text{B}(\text{E2},2_1^+\rightarrow0_1^+)=100$.
See Ref.~\cite{Fortunato_2005} for more details concerning the scaling procedure, as well as the exact definition of the quadrupole operators.

We observe that the agreement between our numerical and analytical spectra is very good, even for higher-energy states, with accuracy up to the third significant digit. On the contrary, the electromagnetic transitions are much more sensitive to the size of the basis. For the transition from $2_2^+\rightarrow2_1^+$, we find a discrepancy of approximately $0.6\%$. A better agreement can be achieved simply by increasing the value of $m_{max}$. Therefore, the basis developed here appears to be very accurate and suitable for the description of low-lying states.

\begin{figure}
    \centering
        \centering
         \includegraphics[width=0.7\textwidth]{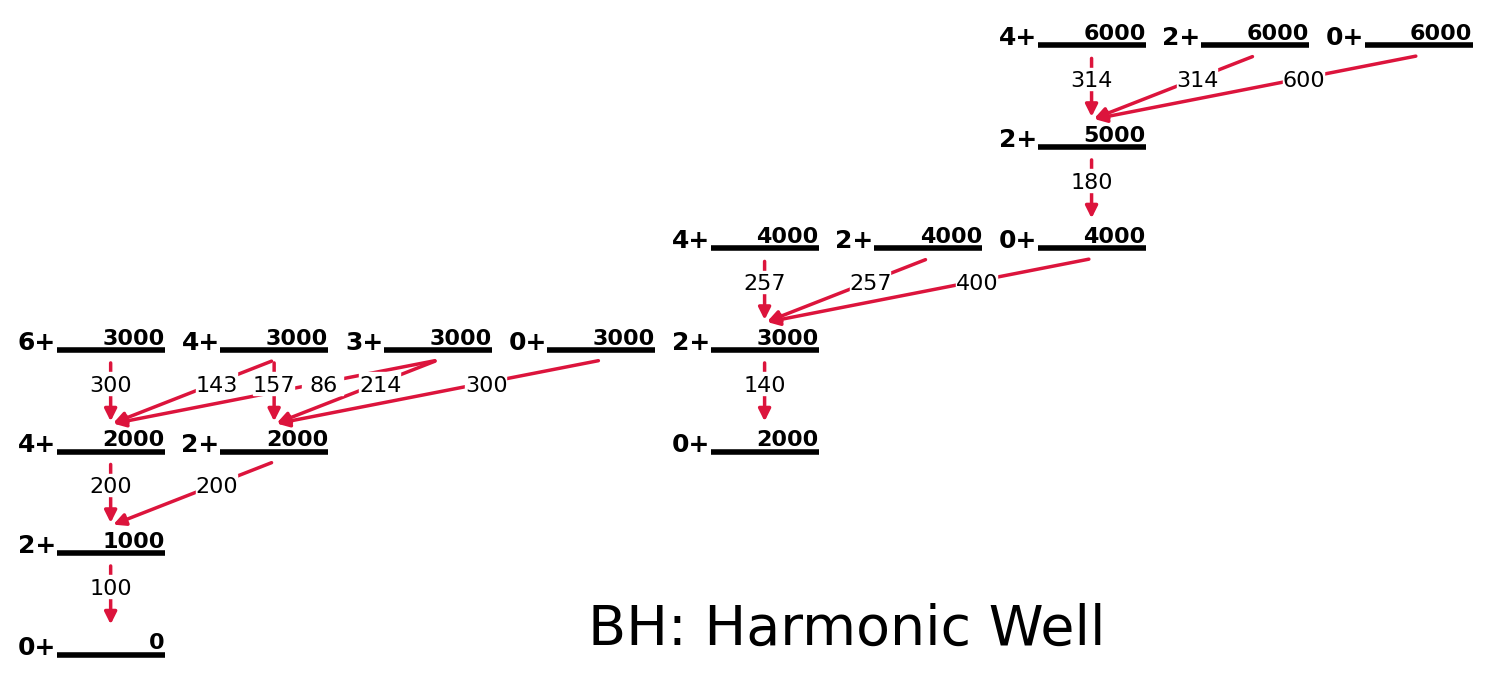}
        \caption{ Numerical results for the  low energy spectrum for the harmonic potential obtained with  $\mu=9, m_{max}=40$. See text for details.}
    \label{fig:spectrum:HO}
\end{figure}

Using the same configuration as discussed above, we now test our numerical code with a harmonic potential defined as
\begin{eqnarray}
    V(\beta)=\frac{1}{2}C\beta^2\,.
\end{eqnarray}

with $\omega=\sqrt{C/B_m}$.
The eigenvalues are given by \cite{Fortunato_2005}
\begin{equation}\label{eq:Harmo:exact}
    E_{n_\beta,\tau}=\hbar\omega\left[(2n_\beta+\tau) +\frac{5}{2}\right]\, ,
\end{equation}

where $n_\beta$ is the radial quantum number. The analytical expressions for the electromagnetic transitions are given in Ref.~\cite{bonatsos2004sequence}. Following the previous conventions, we scale the energies and the electromagnetic transitions so that $E_{2_1^+}=1000$ and $\text{B}(\text{E2},2_1^+\rightarrow0_1^+)=100$. The degeneracy of the spectrum is exactly the same as that illustrated in Fig.~\ref{fig:spectrum}. We obtain remarkable accuracy in terms of relative energies, since all our states are exactly equally spaced, and the energies agree with the analytical values up to the fourth significant digit. This agreement is even better than that observed using the infinite-well potential.
The transitions are also very well reproduced. In Fig.~\ref{fig:spectrum:HO}, we report the numerical results obtained with our numerical code. The figure can be directly compared with Fig. 3a of Ref.~\cite{bonatsos2004sequence}. We observe that the transitions agree essentially exactly, up to rounding errors. The remarkable accuracy of the method allows us to describe perfectly the degeneracies of the various multiplets.

These examples clearly validate the robustness of our numerical approach to solving the generalised BH.

\section{Microscopic calculations}\label{sec:micro}

\subsection{Ground state}

Instead of using simple forms for the potential and the mass parameters, we can evaluate them using a fully microscopic calculation. More precisely, this can be done via the full adiabatic time-dependent Hartree-Fock-Bogoliubov model (ATDHFB) developed in Ref.~\cite{bar78} and recently applied to systematic calculations in Ref.~\cite{sun25}. Following Ref.~\cite{bar11}, in order to reduce the computational cost, it is possible to simplify the calculations using the so-called cranking approximation. In this case, the time-odd fields of the residual interaction are neglected. Following Ref.~\cite{Libert_1999}, the non-diagonal terms of the  Hartree-Fock-Bogoliubov (HFB) energy matrix can also be neglected, and the derivatives of the particle-hole potential can be approximated by terms proportional to the quadrupole operator, leading to the perturbative cranking approximation. In this case, the mass parameters $B_{ij}$, defined in Eq.\eqref{eq:appendix:B:B0B1B2BxByBz}, with indices $i,j$ taking the values 0 and 2, are expressed as \cite{Sad13}   \footnote{Notice that there is a misprint in Ref.~\cite{bar11}.}

\begin{equation}
 B_{ij}= 
\hbar^2
\left[ \left(\mathbf{M}^{(1)}\right)^{-1}
\mathbf{M}^{(3)}
\left(\mathbf{M}^{(1)}\right)^{-1}\right]_{ij} \, ,
\end{equation}

\noindent where $\mathbf{M}^{(k)}$ are the energy-weighted moment matrices whose matrix elements read

\begin{eqnarray}\label{eq:moments:masses}
    M_{ij}^{(k)}=\sum_{\alpha\beta} \frac{\langle 0 | \hat{Q}_{2i}|\alpha \beta\rangle \langle \alpha \beta| \hat{Q}^\dagger_{2j}| 0\rangle }{(E_\alpha+E_\beta)^k}\,.
\end{eqnarray}

\noindent $|\alpha\beta\rangle$ are two-quasiparticle wave functions, and $E_{\alpha(\beta)}$ are the quasiparticle energies. The proton and neutron contributions are simply added~\cite{Sad13,sch16}.

As extensively discussed in the literature \cite{bar11,Sad13,sat13}, this approximation tends to underestimate the values of the masses by roughly $30\%$ and, as a consequence, it is customary to rescale them by a factor of $1.3$.
The mass parameters are then expressed in $(\beta,\gamma)$ coordinates using the transformation given in Eq.~\eqref{transformation:B}. Similarly, the three moments of inertia appearing in the collective Hamiltonian are calculated using the Inglis-Belyaev formulas \cite{ing56,bel61,li12}

\begin{eqnarray}
    \mathcal{I}_k&=& \sum_{\alpha \beta}\frac{| \langle \alpha \beta| \hat{J}_k^\dagger| 0\rangle |^2}{E_\alpha+E_\beta}\,,
\end{eqnarray}

\begin{figure}
    \centering
        \centering
          \includegraphics[width=0.6\textwidth]{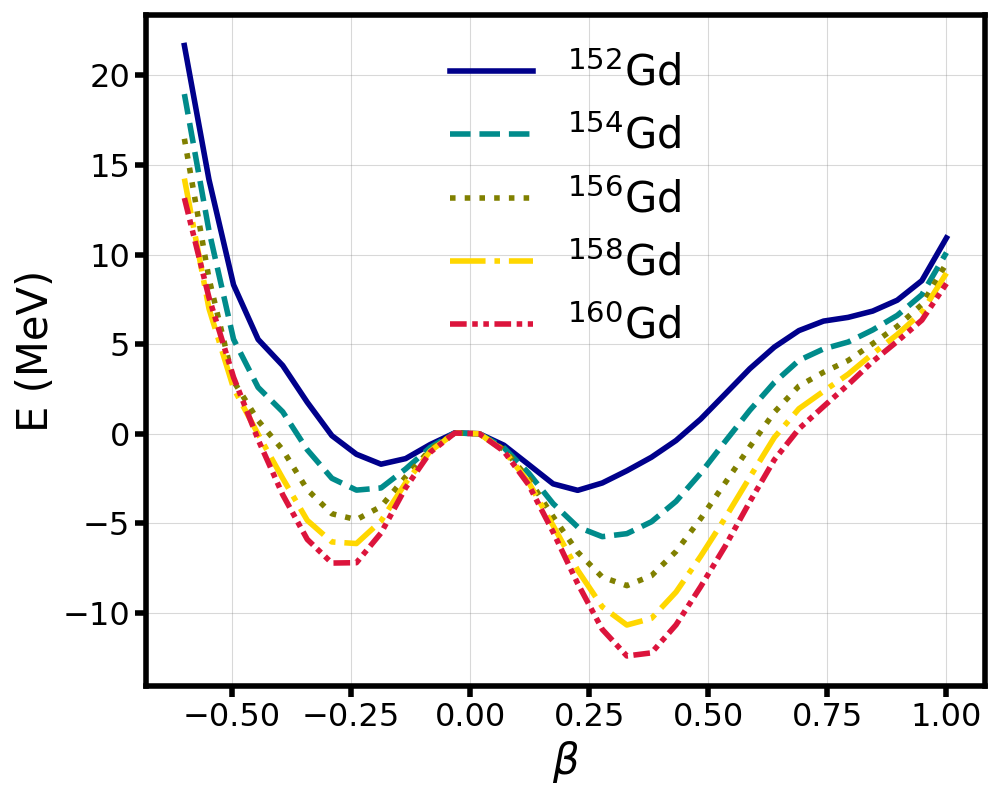}
        \caption{Axial potential energy surfaces for selected Gd isotopes calculated using the D3G3M interaction. The spherical point is fixed arbitrarily to zero to facilitate the comparison.  Negative values of $\beta$ correspond to the $(\beta>0,\gamma = \pi)$ axis on the $(\beta,\gamma)$ plane.}
    \label{fig:PES:ax:Gdchain}
\end{figure}

\noindent where $\hat{J}_k$ is the $k$-component of the total angular momentum \cite{Varshalovich_1988} and the sum is
over the whole quasi-particle space.

The HFB equations are solved on a cylindrical harmonic oscillator basis \cite{fro21} using a Gogny interaction in a fully self-consistent way, \emph{i.e.}, the same interaction is used for both the particle-hole and particle-particle channels. In order to treat triaxial configurations on an equal footing, we decided to use the same oscillator length along the axial and perpendicular directions. By selecting 14 major shells, $N_{shell}$, we have checked that our results are well converged. We refer to Appendix \ref{app:num:det} for a more detailed discussion.
As discussed in Ref.~\cite{chappert2008towards}, the Coulomb exchange term is neglected in both the particle-hole and particle-particle channels, and the contribution of the zero-range spin-orbit term in the pairing channel is also suppressed. For the present article, we have selected four representative Gogny interactions: D1S \cite{decharge1980hartree}, D1N \cite{chappert2008towards}, D1M \cite{goriely2009first}, and D3G3M \cite{batail2025three}.
These interactions have been adjusted using different fitting protocols, and we refer to Refs.~\cite{sellahewa2014isovector,davesne2025isovector} for a detailed comparison of their properties in infinite nuclear matter.

Of particular interest are D1M and D3G3M, since they have been adjusted to reproduce nuclear masses with high accuracy and also include the coupling to collective excitations via the BH method \cite{girod1979zero} in their fitting procedures.

We focus our attention on the isotopic chain of Gd isotopes in order to compare our results with existing literature. In Fig.~\ref{fig:PES:ax:Gdchain}, we show the evolution of the potential energy surface (PES), restricted to the axial case, as a function of the deformation parameter $\beta$, obtained using constrained HFB calculations with the D3G3M interaction. The PES is rescaled such that the energy at the spherical point is zero. We notice that all Gd isotopes considered here present a well-pronounced prolate minimum and a secondary oblate one. The position of the main minimum along the $\beta$-axis increases progressively with $N$, shifting from approximately $\beta = 0.2$ for $^{152}$Gd to $\beta = 0.35$ for $^{160}$Gd. This evolution is accompanied by an increase in the depth of the minima. Thus, $^{152}\text{Gd}$ exhibits a relatively flat PES, suggesting a strong coupling between the two configurations, while the PES of $^{160}\text{Gd}$ rather points to a well-deformed prolate nucleus in its ground state. These results are fairly consistent across all the Gogny interactions considered here and are also in good agreement with previous findings obtained with zero-range Skyrme interactions \cite{baran1996gd} and relativistic mean-field models \cite{nikvsic2009beyond}.

As discussed in Ref.~\cite{mui21}, these nuclei are ideal candidates for testing the BH method since they have open shells in the medium-mass region.
\begin{figure}
    \centering
        \centering
          \includegraphics[width=0.4\textwidth]{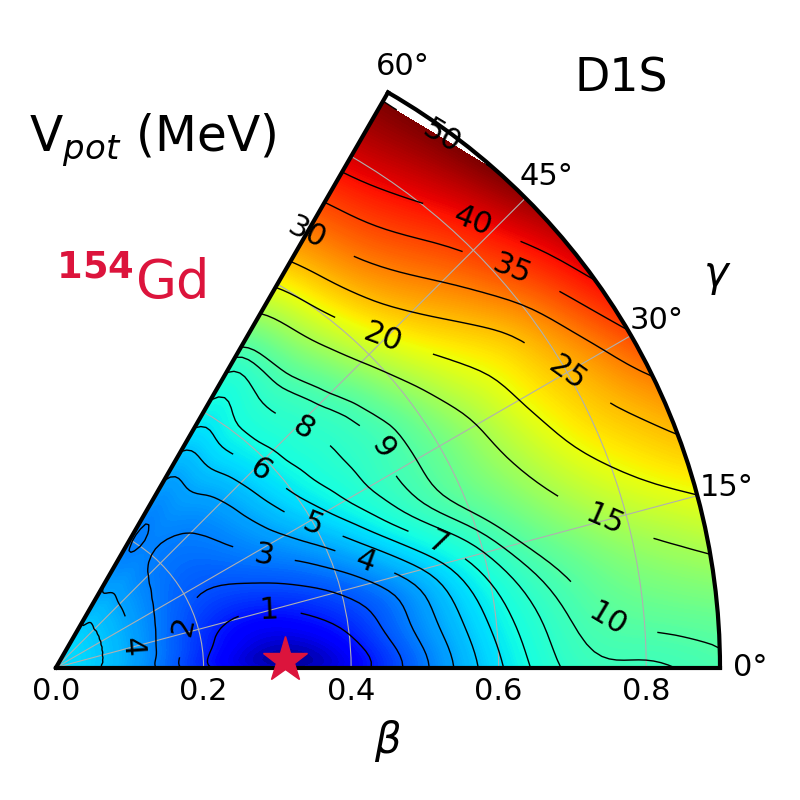}
         \includegraphics[width=0.4\textwidth]{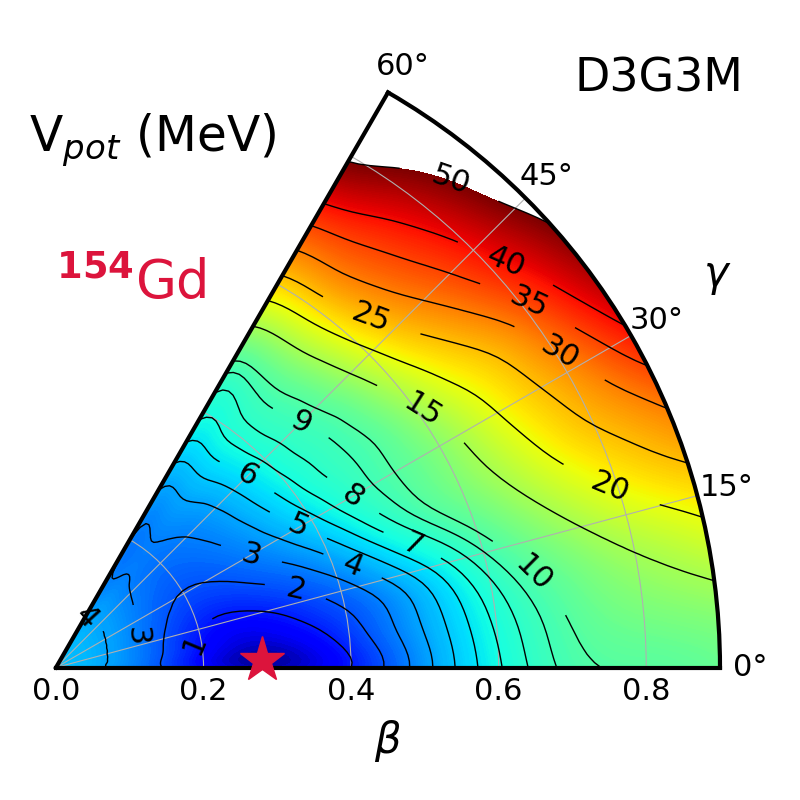}
        \caption{PES for $^{154}$Gd in the $(\beta,\gamma)$ sextant for the Gogny D1S (left panel) and the Gogny D3G3M (right panel) interactions. The zero of the energy is fixed at the absolute minimum.}
    \label{fig:PES:Gd154}
\end{figure}

\begin{figure}
    \centering
    \centering
         \includegraphics[width=0.32\textwidth]{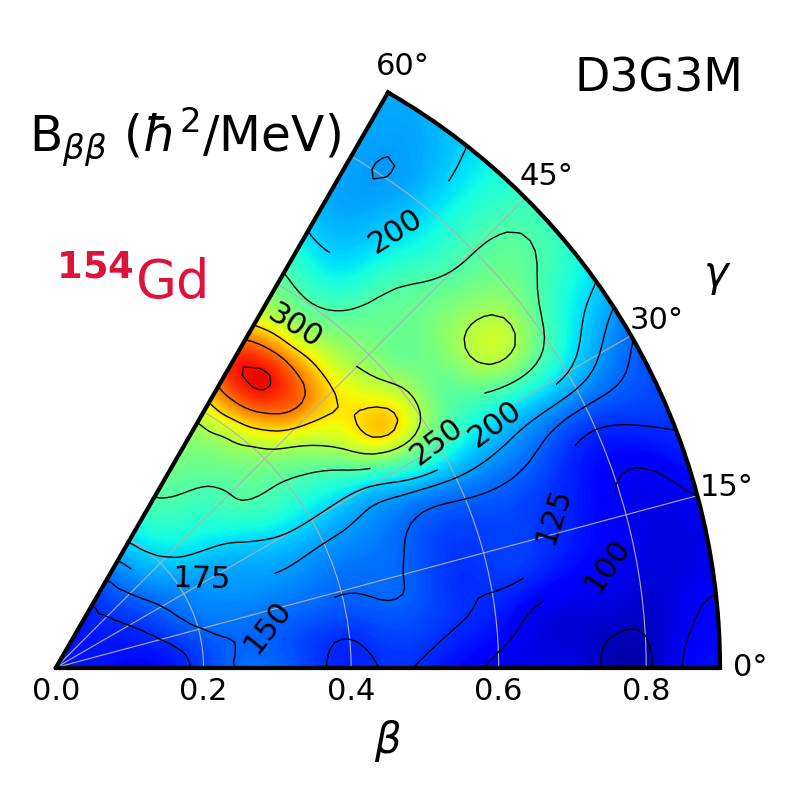}
         \includegraphics[width=0.32\textwidth]{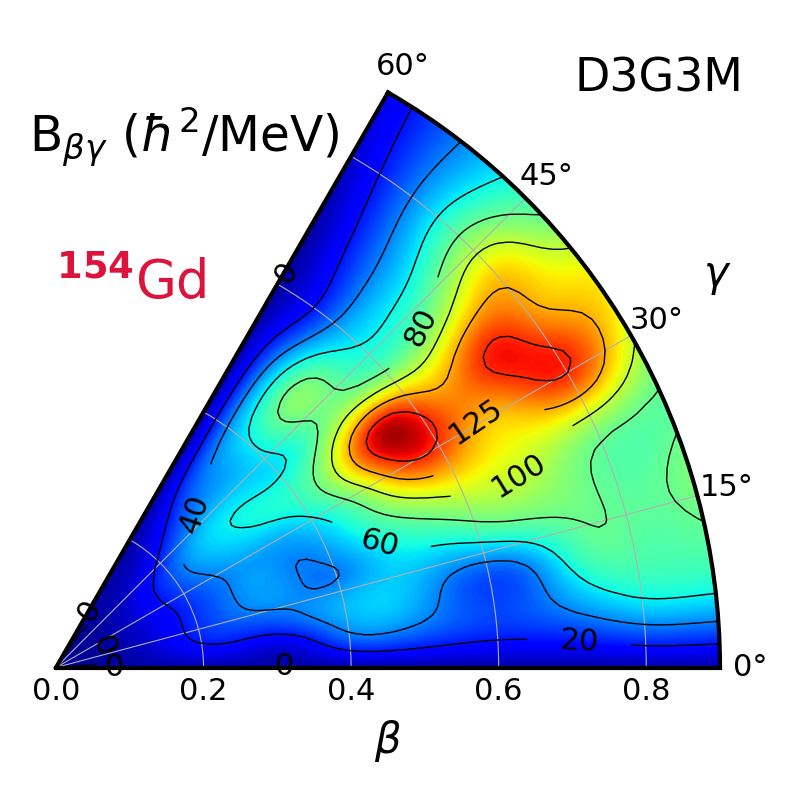}
         \includegraphics[width=0.32\textwidth]{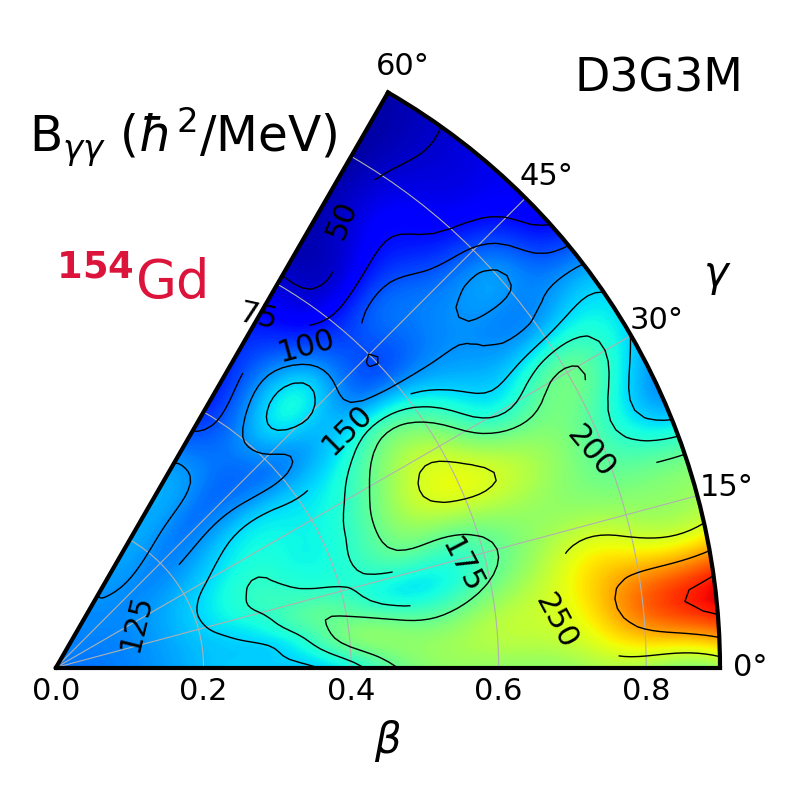}\\
         \includegraphics[width=0.32\textwidth]{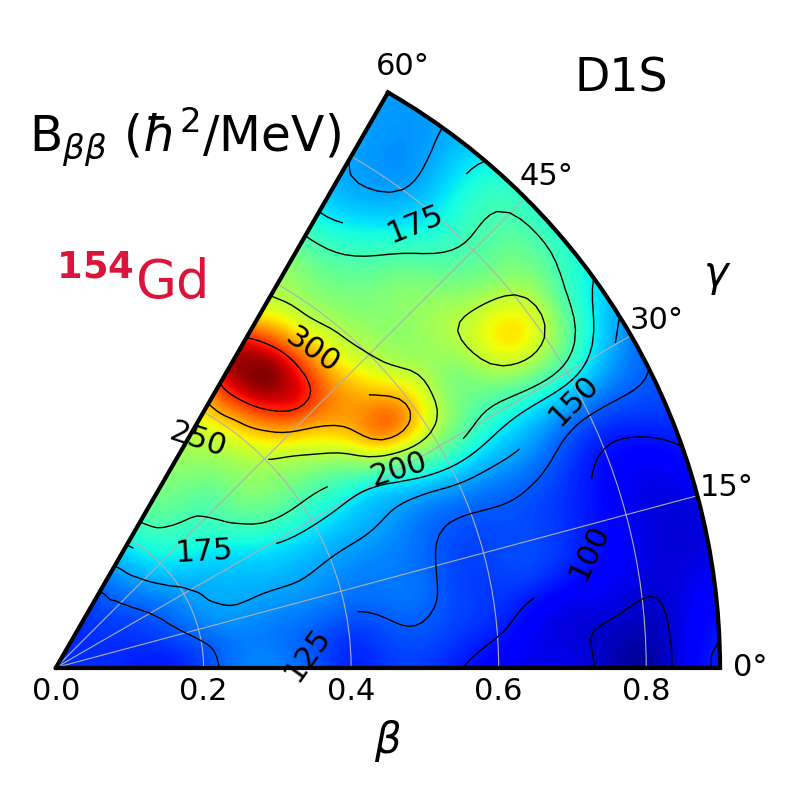}
         \includegraphics[width=0.32\textwidth]{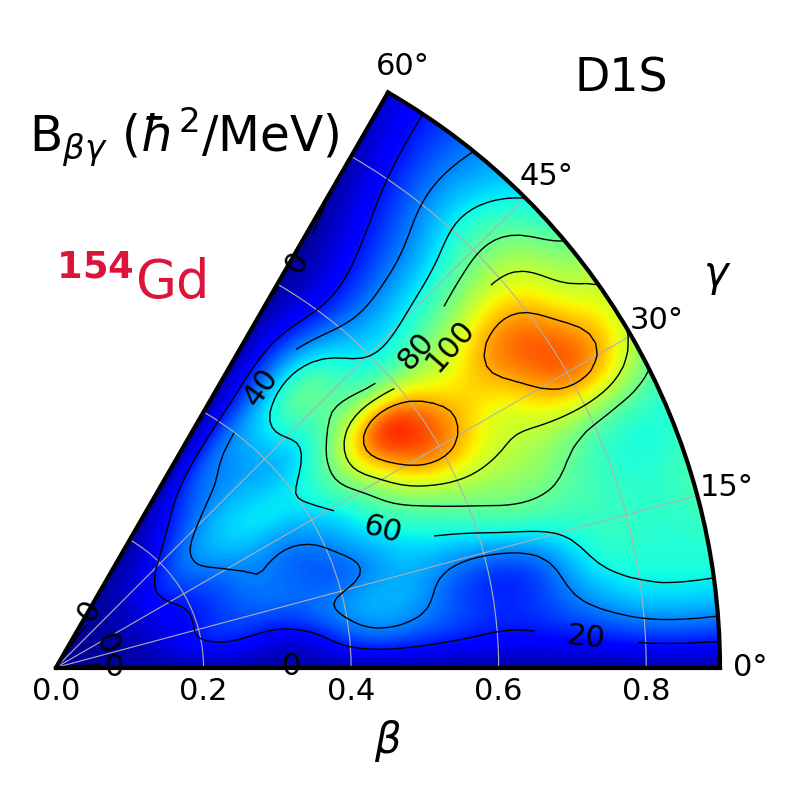}
         \includegraphics[width=0.32\textwidth]{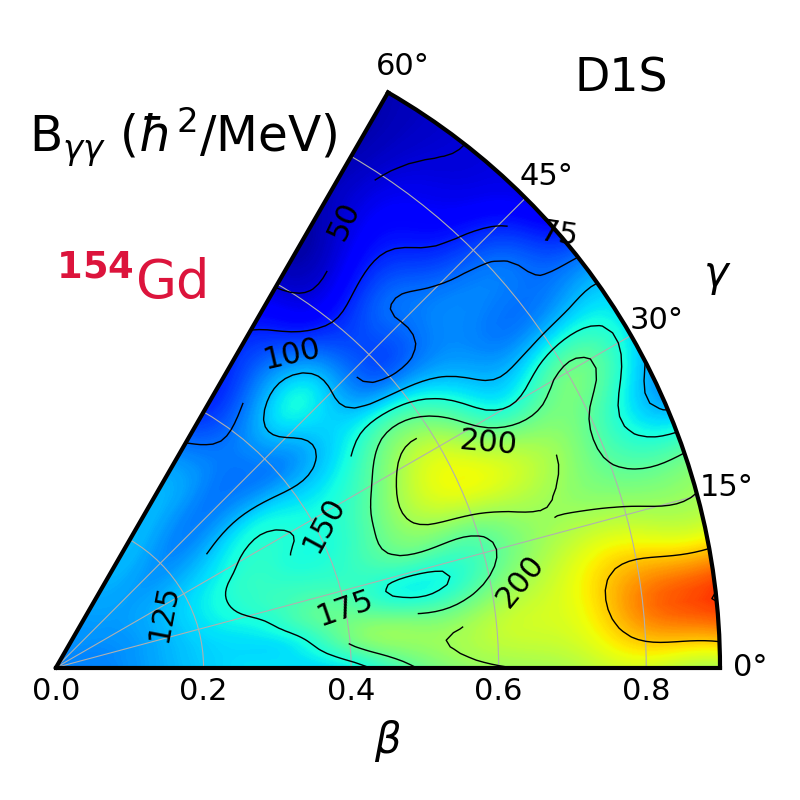}
        \caption{Mass parameters $B_{\beta\beta}$, $B_{\beta\gamma}$ and $B_{\gamma\gamma}$  in $^{154}$Gd calculated using the D3G3M (top panels) and D1S (bottom panels) interaction. See text for details.}
    \label{fig:Bbb:gd154}
\end{figure}

In Fig.~\ref{fig:PES:Gd154}, we compare the full PES in the $\beta,\gamma$ sextant for two Gogny interactions in $^{154}$Gd, namely D1S (left panel) and D3G3M (right panel). We notice that the PESs are very similar to each other, both presenting a strong prolate and axial minimum at $\beta\approx0.3$. We also notice that D3G3M tends to be slightly softer in the $\gamma$ direction compared to D1S.
As discussed, for example, in Ref.~\cite{llewellyn2020establishing}, two nuclei having very similar PESs can show very different spectra, since the mass parameters can vary significantly from one nucleus to the other. For this reason, we present in Fig.~\ref{fig:Bbb:gd154} the three mass parameters in $^{154}$Gd obtained with the two Gogny interactions. Although the interaction does not play a direct role in the evaluation of the mass parameters themselves, we clearly observe differences in their structure. In particular, we notice that the mass parameters in both cases are not constant but exhibit a very rich structure, with well-localised peaks.

The presence of these peaks can be easily understood by inspecting Eq.~\eqref{eq:moments:masses}: when the pairing gap \cite{ring2004nuclear} is very small, the quasiparticle energy is also small. Since this quantity appears in the denominator of our expression, we therefore obtain a very large value for the mass. See Ref.~\cite{yul99} for a more detailed discussion.

\begin{figure}
    \centering
          \includegraphics[width=0.4\textwidth]{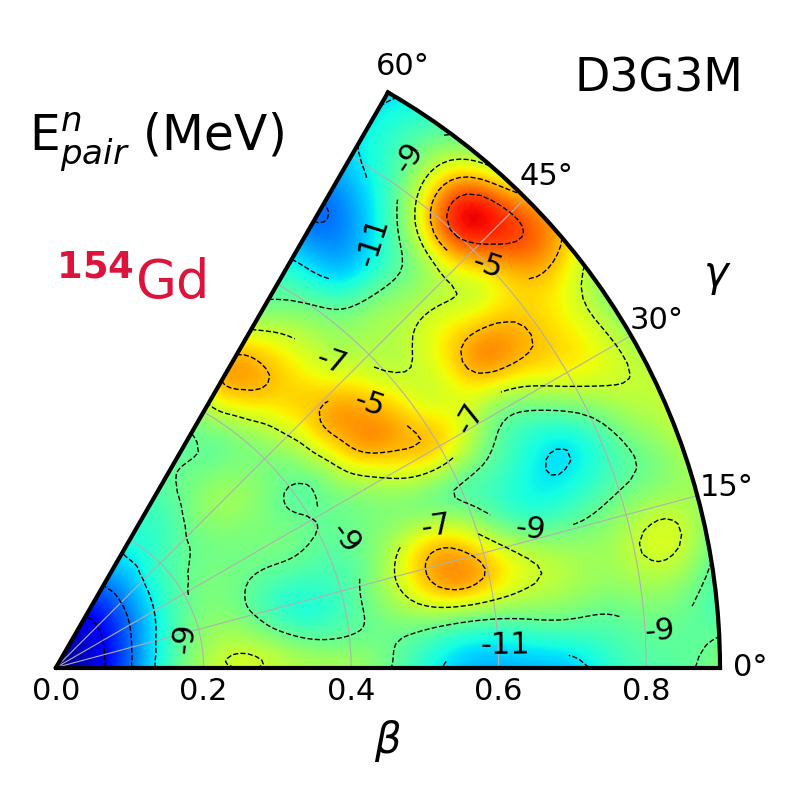}
             \includegraphics[width=0.4\textwidth]{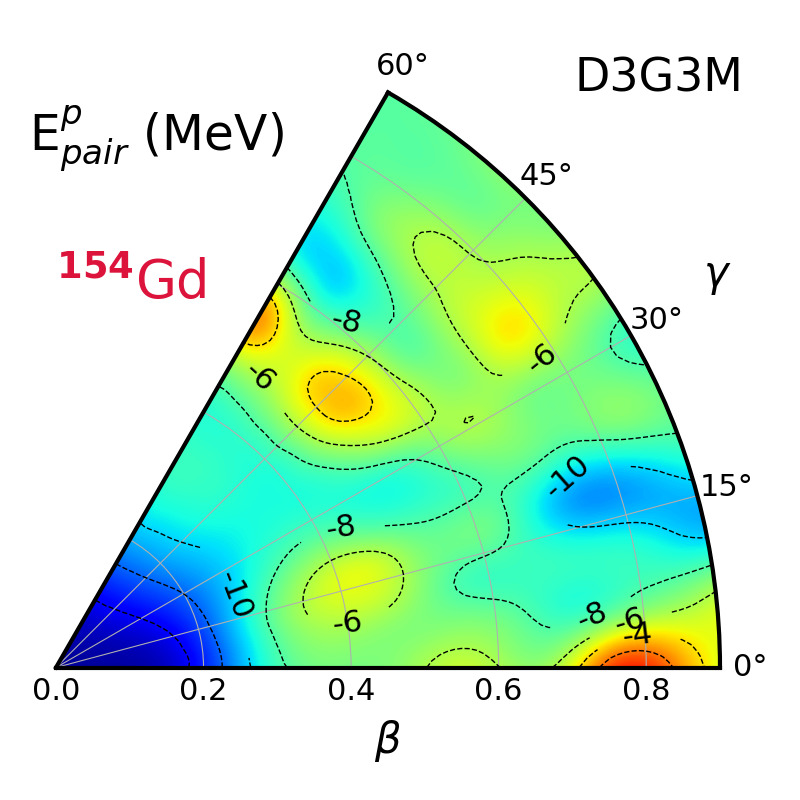}
          \caption{Neutron (left) and proton (right) pairing energies (expressed in MeV) in ($\beta,\gamma$) coordinates for $^{154}$Gd calculated with D3G3M interaction.}
    \label{fig:pairing:energy}
\end{figure}

To get a better insight, we show in Fig.~\ref{fig:pairing:energy} the neutron (left panel) and proton (right panel) pairing energies, expressed as $E_{pair}^{q=n,p}=-Tr(\Delta^q \kappa^q)/2$, where $\Delta^q$ is the pairing field and $\kappa^q$ is the abnormal density \cite{ring2004nuclear}. We observe that the peak in the $B_{\beta\beta}$ mass at $\beta\approx0.4$ and $\gamma=60^{\circ}$ arises from the strong reduction of the neutron pairing energy at the same location, while the suppression around $\beta\approx0.5$ and $\gamma=30^{\circ}$ is reflected in a peak in the $B_{\beta\gamma}$ mass parameter. Finally, the strong suppression of the neutron pairing energy around $\beta\approx0.8$ leads to the peak in the $B_{\gamma\gamma}$ mass.

\subsection{Excited states}

\begin{figure}
    \centering
          \includegraphics[width=0.7\textwidth]{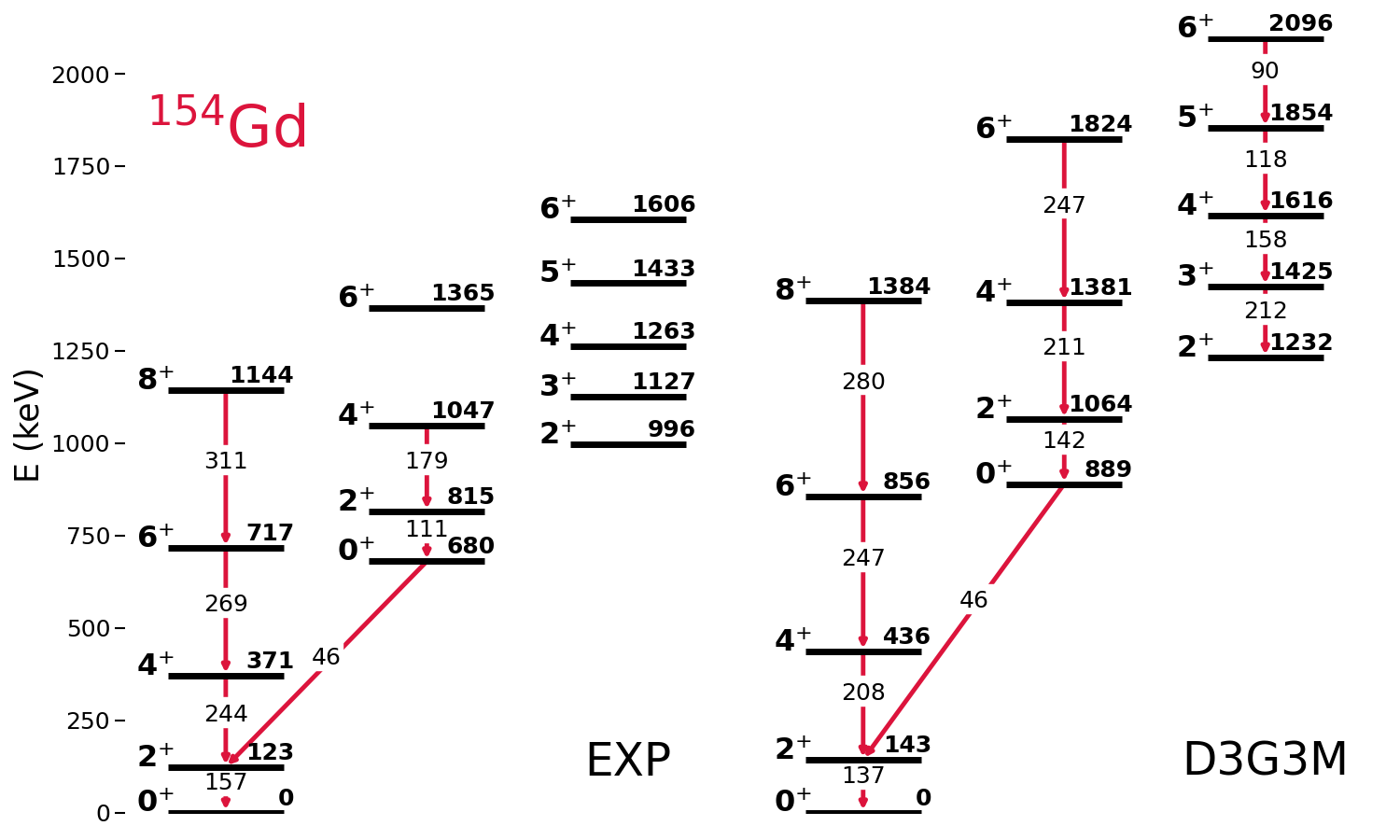}
          \caption{Experimental (left panel) and calculated (right panel) energy spectra of $^{154}$Gd obtained within the Bohr Hamiltonian using the D3G3M Gogny interaction. Excitation energies are given in keV, while electromagnetic transition strengths are expressed in Weisskopf units. See text for details.
}
    \label{fig:Spectrum_Gd154_EXP_D3G3M_14_100}
\end{figure}

In Fig.~\ref{fig:Spectrum_Gd154_EXP_D3G3M_14_100}, we show the excited-state spectrum of $^{154}$Gd obtained using the Bohr Hamiltonian code with the D3G3M Gogny interaction, compared with experimental data taken from the ENSDF database \cite{ENSDF}. All calculations are performed using a scaling parameter (1.3) and including the zero-point energy contribution, as discussed in Secs.~\ref{sec:scaling} and \ref{sec:ZPE}.

We performed a convergence study to determine the optimal values of $m_{\max}$ and $\mu$, which we fixed at $m_{\max} = 40$ and $\mu = 11$. We observe that the BH is capable of reproducing the low-lying energy states, as well as the electromagnetic transitions between them, but the agreement quickly deteriorates as the energy increases. This result is well known in the literature and most likely arises from the perturbative cranking approximation adopted here. As shown in Ref.~\cite{was24}, the use of the local quasiparticle random-phase approximation allows the spectrum to be compressed and results closer to the experimental data. Although very promising, there are still no systematic calculations using such a technique, and further studies in this direction are required.
The current implementation of the BH  model is capable of reproducing the excited states of the Gd isotopes with a difference of $20\text{ keV}$ for the $2^+_1$ levels and less than $100\text{ keV}$ for the $4^+_1$ states. This is quite interesting since the Gogny interaction has not been explicitly adjusted to these data, and thus this result may be considered a prediction of the model. The agreement deteriorates at higher energies, where the energies are systematically overestimated. Despite using the scaling factor for the vibrational masses and moments of inertia, we observe that the density of states is lower in the BH case than in the experimental data.

Excited states are conventionally grouped into bands, of which only the first three are represented here. The ground-state band, built upon the lowest $0_1^+$ state, includes the first excited states of the nucleus and exhibits a pronounced rotational structure. The $\beta$ band is interpreted as an axial shape vibration, with its band-head corresponding to the second $0^+_2$ state. Finally, the $\gamma$ band, whose band-head is the first $2^+$ state outside the ground-state band, is associated with triaxial vibrations. The band structure is reconstructed based on electromagnetic transitions. We assume that states belong to the same band if they exhibit strong intra-band transitions. For this nucleus, the band structure is correctly reproduced.

\begin{figure}
    \centering
          \includegraphics[width=0.7\textwidth]{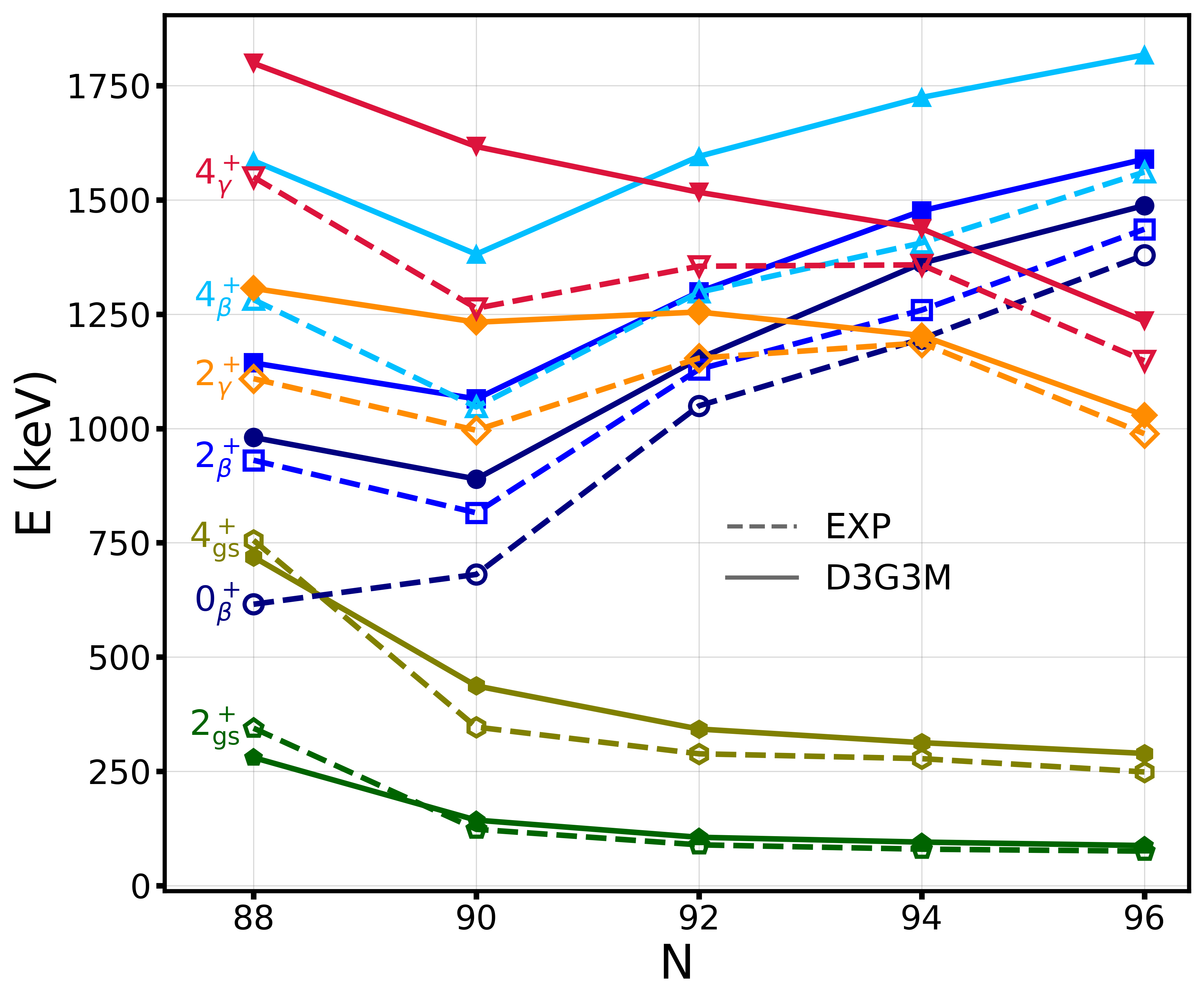}
          \caption{Evolution of energies of the ground-state-band (gs),  $\beta$-band and $\gamma$-band first levels in keV. The experimental results are given by dashed lines and the Bohr Hamiltonian using the D3G3M Gogny interaction by solid lines.}
    \label{fig:E_band_D3G3M}
\end{figure}

In Fig.~\ref{fig:E_band_D3G3M}, we show the evolution of the first few states (expressed in keV) of the ground-state (gs), $\beta$, and $\gamma$ bands in various Gd isotopes as a function of their neutron number (N), as obtained using the Bohr Hamiltonian with D3G3M, together with the corresponding experimental values.
We notice that, with increasing neutron number, the second $0^+$ state increases in energy, while the $2^+$ band-head of the $\gamma$ band decreases in energy, leading to a crossing and then an inversion in $^{160}$Gd. The BH results obtained with D3G3M reproduce this feature very nicely, although they overestimate the energies of the states in the upper part of the spectrum, as already pointed out above.
The worst agreement between theory and experiment is observed in $^{152}\text{Gd}$, where the BH predicts the $\beta$-band-head state to lie at a higher energy than the $4_1^+$ state of the ground-state band, and thus very close to the $\gamma$-band head. Such a feature is not observed experimentally.

Systematic calculations in Ref.~\cite{delaroche2010structure} have shown that the BH performs best for well-deformed nuclei. As observed in Fig.~\ref{fig:PES:ax:Gdchain}, $^{152}$Gd does not exhibit the same strongly deformed character as the other Gd isotopes.

\subsection{Sensitivity analysis}

Within the scientific literature, different authors adopt different strategies for performing BH calculations. Several methodological choices may vary, including the interaction employed, the value of the scaling parameter, and the treatment of the zero-point energy, which may or may not be included. In this section, we therefore provide a comprehensive assessment of these variations, with the aim of qualitatively estimating the uncertainty associated with these methodological choices and evaluating their impact on the final results.

\subsubsection{NN interaction}\label{Sec:NN}

\begin{figure}
    \centering
         \includegraphics[width=0.8\textwidth]{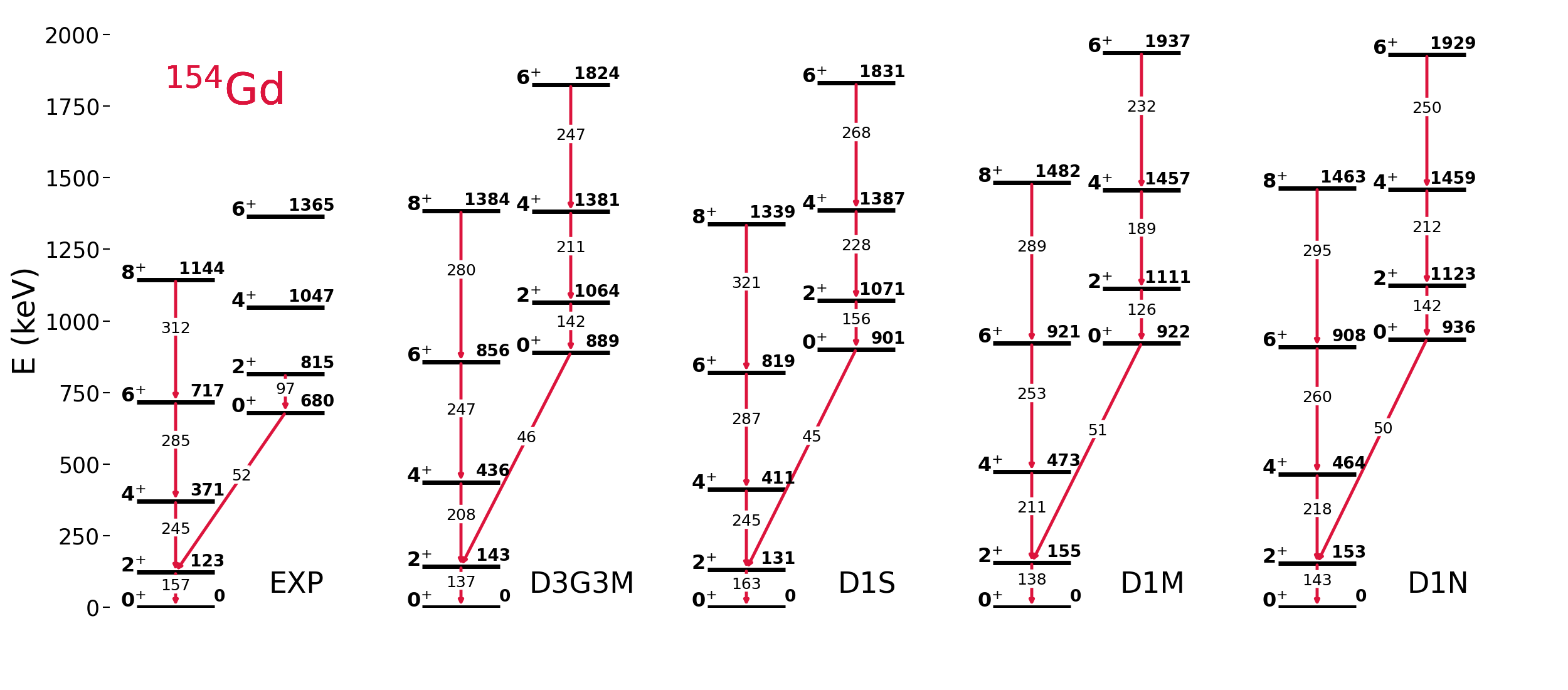}
        \caption{Experimental positive parity excited states, expressed in keV, in $^{154}$Gd (left panel) compared with theoretical ones obtained  using the BH with D3G3M, D1S, D1M and D1N interactions. The electromagnetic transitions between states are indicated by arrows and expressed in Weisskopf units. See text for details.}
    \label{fig:spectrum:gd154}
\end{figure}

In Fig.~\ref{fig:spectrum:gd154}, we compare the theoretical calculations of the first two bands of the nuclear spectrum in $^{154}$Gd, obtained using the microscopic input generated with four different Gogny interactions, with the experimental data. The spectra generated with the four interactions are in fair agreement with each other, although we observe that D1S and D3G3M tend to reproduce the experimental energies more accurately. Qualitatively, the various interactions provide similar spectra, although it is interesting to note that D1S predicts a clear inversion in the positions of the $6_1^+$ and second $0_2^+$ states, while the other interactions tend to predict them at almost the same energy. Experimentally, the situation is reversed, but the two levels are quite close in energy.

The electromagnetic transitions between states are calculated using Eq.~\eqref{q:transition}. Notice that, in Ref.~\cite{delaroche2010structure}, the authors used a different procedure based on the solution of the cranked HFB equations to calculate the moments of inertia in order to include the so-called Thouless-Valatin dynamical rearrangement contribution \cite{thou62}. This explains the discrepancy with previously published results \footnote{The original values of D1S are accessible at \cite{del26}.}.

\begin{figure}
    \centering
          \includegraphics[width=0.46\textwidth]{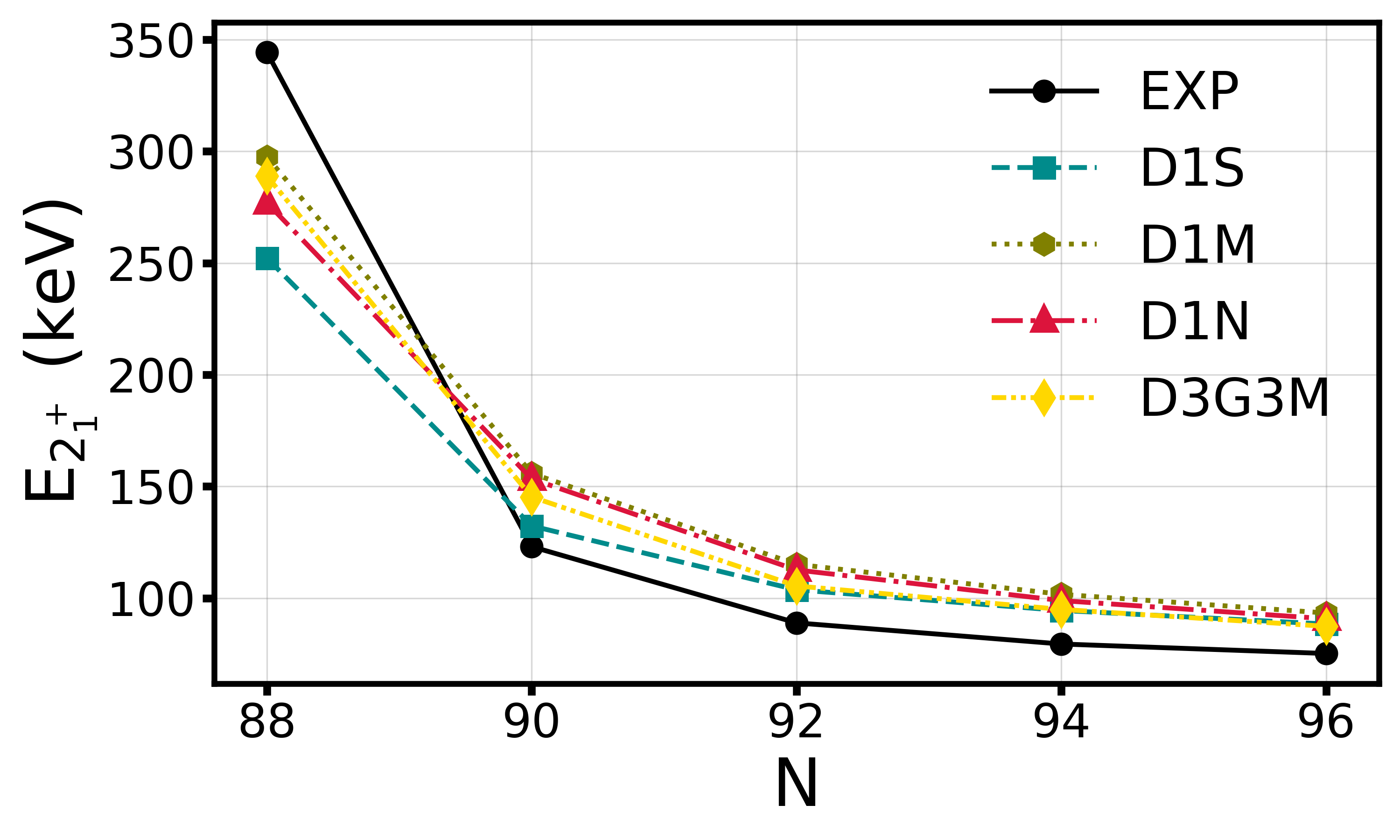}
 \includegraphics[width=0.46\textwidth]{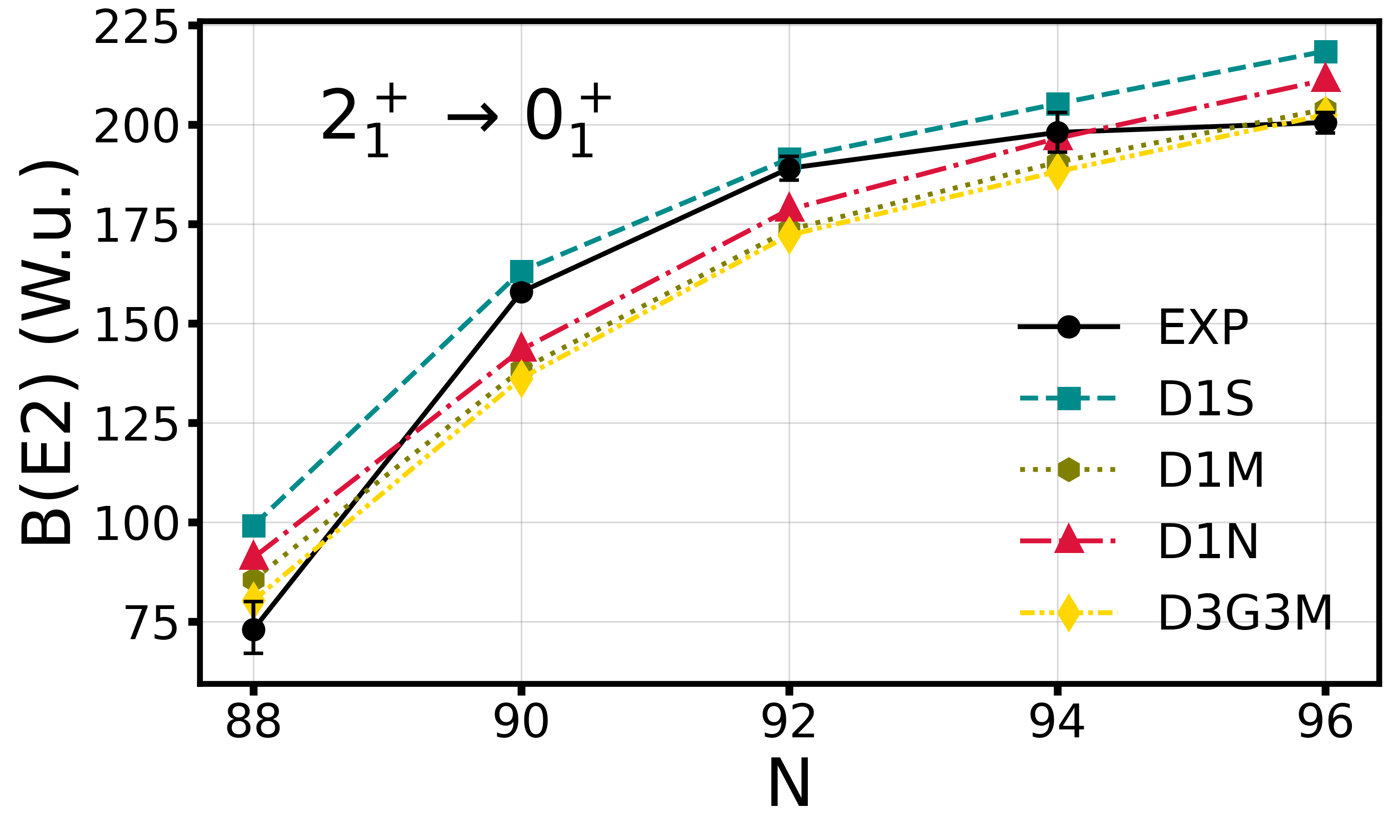}
        \caption{Energy of the first $2^+$ states (left panel) and electromagnetic transitions $2^+ \to 0^+$ in Weisskopf units (right panel) for the selected Gd isotopes. The black dots represent the experimental value, while the dashed lines are the results obtained with the four different Gogny interactions. See text for details. }
    \label{fig:2+:chain}
\end{figure}

\begin{figure}
    \centering
          \includegraphics[width=0.46\textwidth]{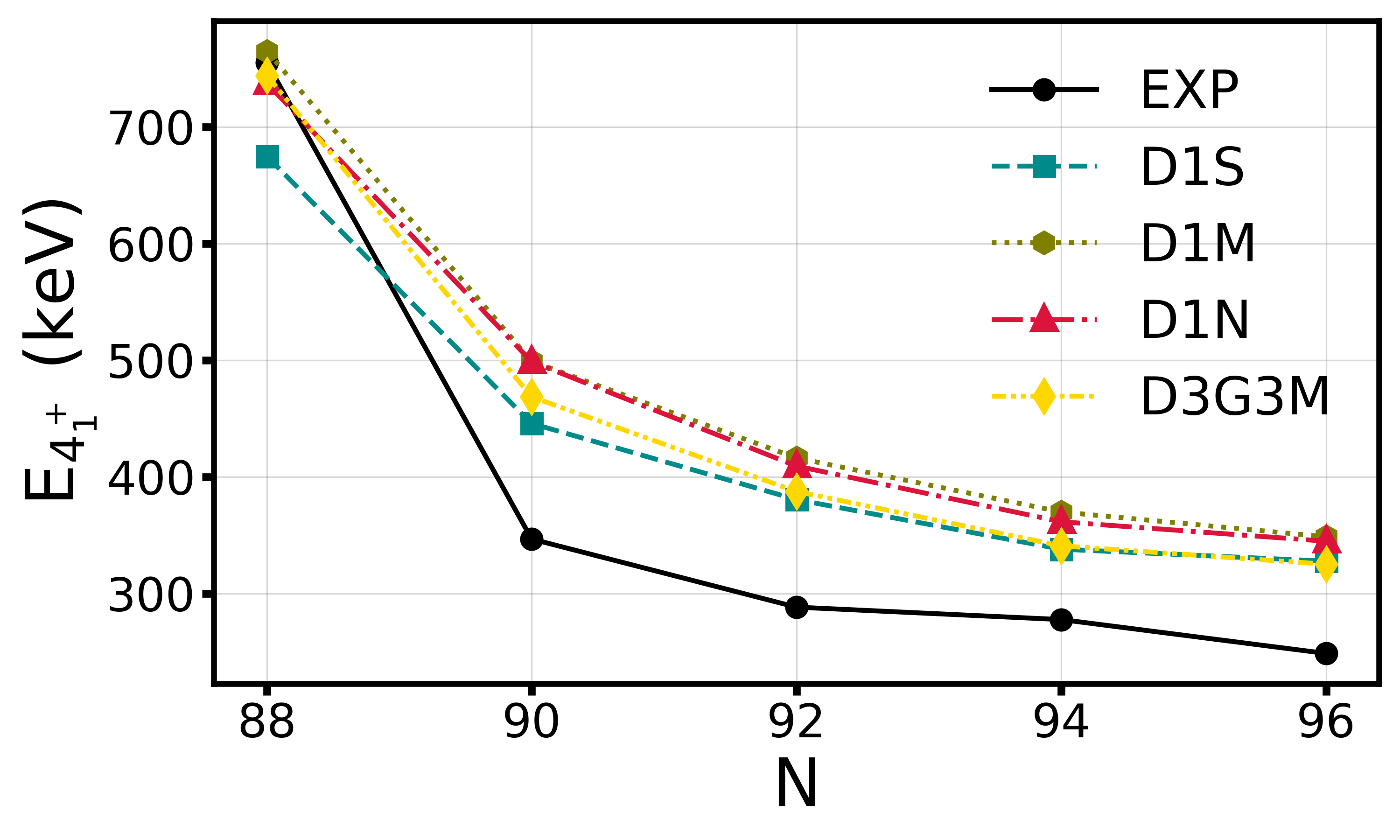}
            \includegraphics[width=0.46\textwidth]{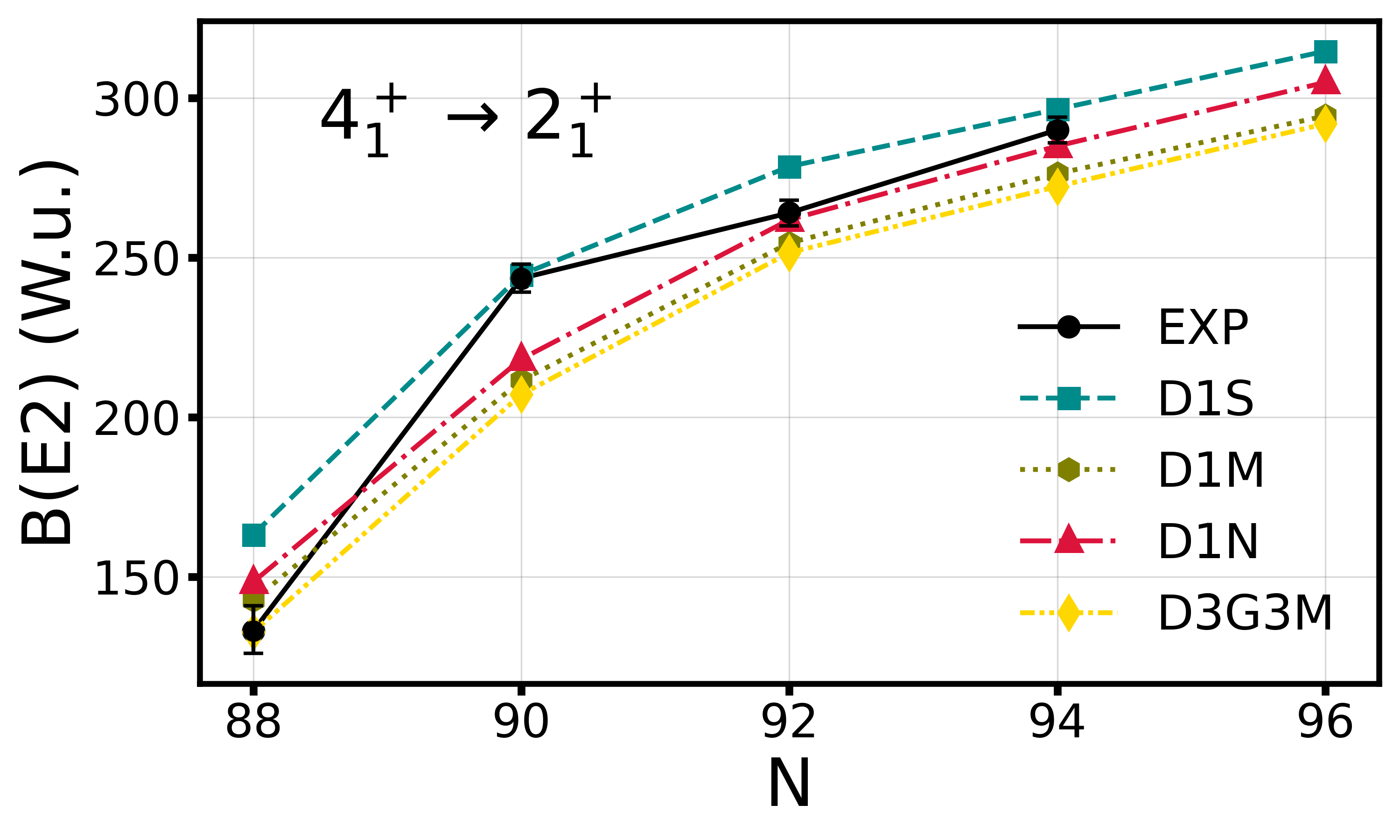}
          \caption{Energy of the first $4^+$ states (left panel) and electromagnetic transitions $4^+ \to 2^+$ in Weisskopf units (right panel) for the selected Gd isotopes. The black dots represent the experimental value, while the dashed lines are the results obtained with the four different Gogny interactions. See text for details.}
    \label{fig:4+:chain}
\end{figure}

In order to better assess the role of the NN interaction, in Fig.~\ref{fig:2+:chain}, we present the evolution of the lowest $2^+$ state (left panel) and its transition to the ground state, expressed in Weisskopf units (W.u.) (right panel), for the selected Gd isotopes using the four Gogny interactions. We observe that, in general, the energies are not very sensitive to the effective interaction, although D1S and D3G3M show better agreement with the experimental data. On the contrary, when comparing the electromagnetic transitions, we observe a larger variance. D1S is capable of reproducing the experimental data well for $N=90$ and $92$, while for $N=88$ and $96$ the best results are obtained with D3G3M, and for $N=94$ with D1N.
In Fig.~\ref{fig:4+:chain}, we repeat the same analysis, but for the first $4^+_1$ state (left panel) and its transition to the first $2^+_1$ state (right panel). Also in this case, D1S and D3G3M give results close to the experimental values for $N=92-96$, with D1S being closer for $N=90$ and D3G3M for $N=88$. By inspecting the electromagnetic transitions, we see that D1S provides the results closest to experiment, except at $N=88$, where D1M and D3G3M lie on top of the experimental value. Also in this case, we observe a larger variance in the results for the electromagnetic transitions than for the energies, with D1S behaving slightly differently from the three other Gogny interactions.

To characterise the nature of each state and classify it into the appropriate band, several quantities derived from the corresponding wave functions can be employed. One example is the mixing of different intrinsic configurations, which can be assessed by evaluating the distribution of the projection $K$ of the total angular momentum $I$ along the $z$-axis in the intrinsic frame.
It is defined as
\begin{eqnarray}\label{eq:NK}
    N_K=\int_0^{2\pi}\int_0^\infty |\psi^I_{K \alpha}|^2 d\tau_0\,.
\end{eqnarray}
\noindent where the wave-function $\psi^I_{K \alpha}$ is obtained from the full diagonalisation of the BH, as given in Eq.~\eqref{eq:diago:psi} and  $d\tau_0$ is defined in Eq.\eqref{eq:dtau0}.

\noindent A sharp $K$-distribution characterises a purely rotational state, whereas a broader distribution is expected for transitional or vibrational states. A distribution dominated by the $K=0$ component is associated with axial deformation and corresponds to the ground-state and $\beta$ bands. Conversely, a distribution dominated by the $K=2$ component indicates triaxial deformation, which is typical of $\gamma$-band excitations. Mixed distributions, or those exhibiting higher-order $K$ components, generally point to a more complex nuclear structure.

\begin{figure}
    \centering
          \includegraphics[width=0.99\textwidth]{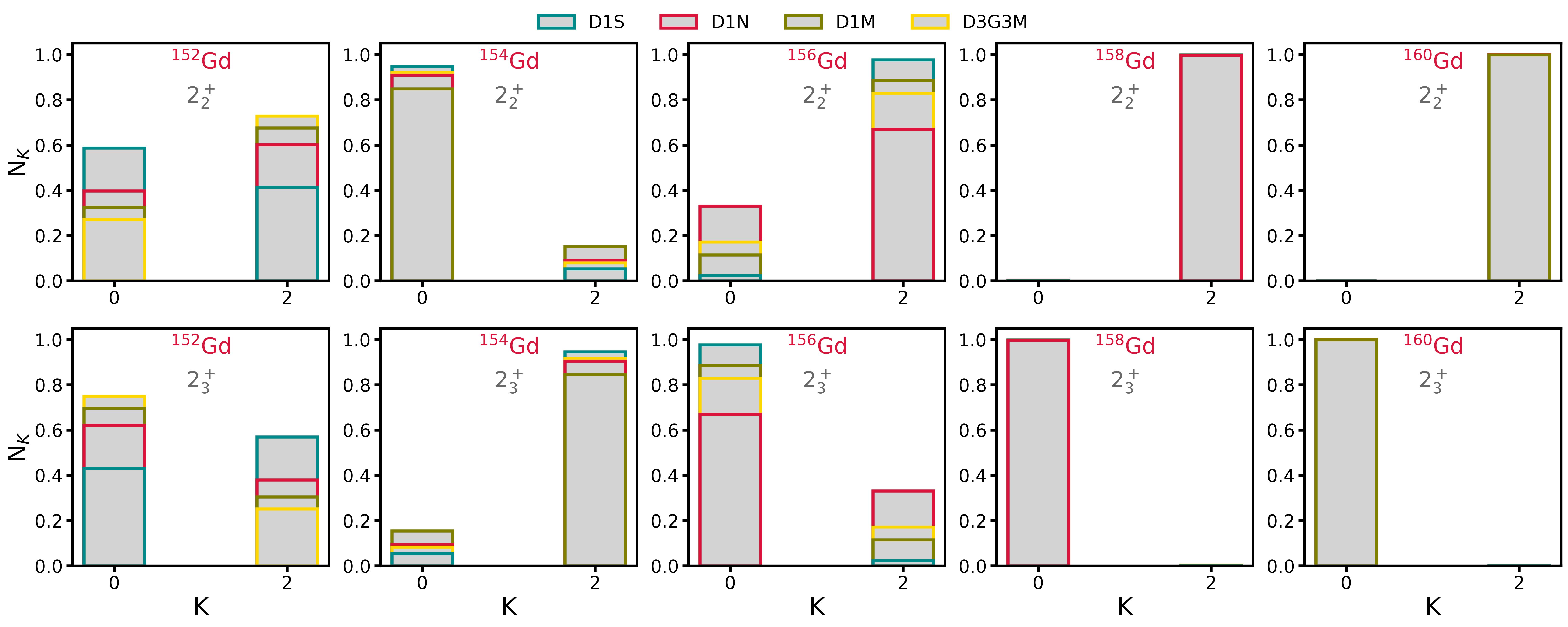}
          \caption{Distribution of $N_K$, as given in Eq.~\eqref{eq:NK}, for the $2^+_2$ and $2_3^+$ states. See text for details.}
    \label{fig:Nk}
\end{figure}

In Fig.~\ref{fig:Nk}, we present the intrinsic angular momentum distributions for the second and third (in energy) $2^+$ levels across the gadolinium isotopic chain for the four Gogny interactions. For $^{158}$Gd and $^{160}$Gd, the distributions are sharp, with the dominant components being very close to $1$, confirming the rotational character of these well-deformed nuclei. Since these distributions are highly peaked, band assignment becomes straightforward. As the neutron number decreases, the distributions become less peaked. For $^{154}\text{Gd}$, the $K=0$ dominance in the $2^+_2$ states identifies the second band as a $\beta$ band, while the $K=2$ character of the $2^+_3$ state establishes it as part of the $\gamma$ band. For $^{156}\text{Gd}$, a slight mixing is observed between the two structures, which can be attributed to the close energetic proximity of the bands \cite{nikvsic2009beyond}. Finally, in $^{152}$Gd, the strong mixing between the different $K$ components points to the transitional and vibrational nature of this nucleus. Consequently, band assignments in this case are only approximate. We therefore associate the $2^+_3$ levels with a quasi-$\beta$ structure and the $2^+_2$ levels with a quasi-$\gamma$ band. Comparing the distributions obtained with the different interactions reveals more pronounced discrepancies. In $^{156}$Gd, the D1S interaction yields a strongly localised distribution, whereas the other three interactions predict a more moderate localisation of the states. These differences become particularly significant in the case of $^{152}$Gd. Indeed, with D1S, we assign the $2^+_2$ state to the $\beta$ band and the $2^+_3$ state to the $\gamma$ band, whereas with the other three interactions the opposite assignment is obtained.

\begin{figure}[h]
    \centering
        \includegraphics[width=0.28\textwidth]{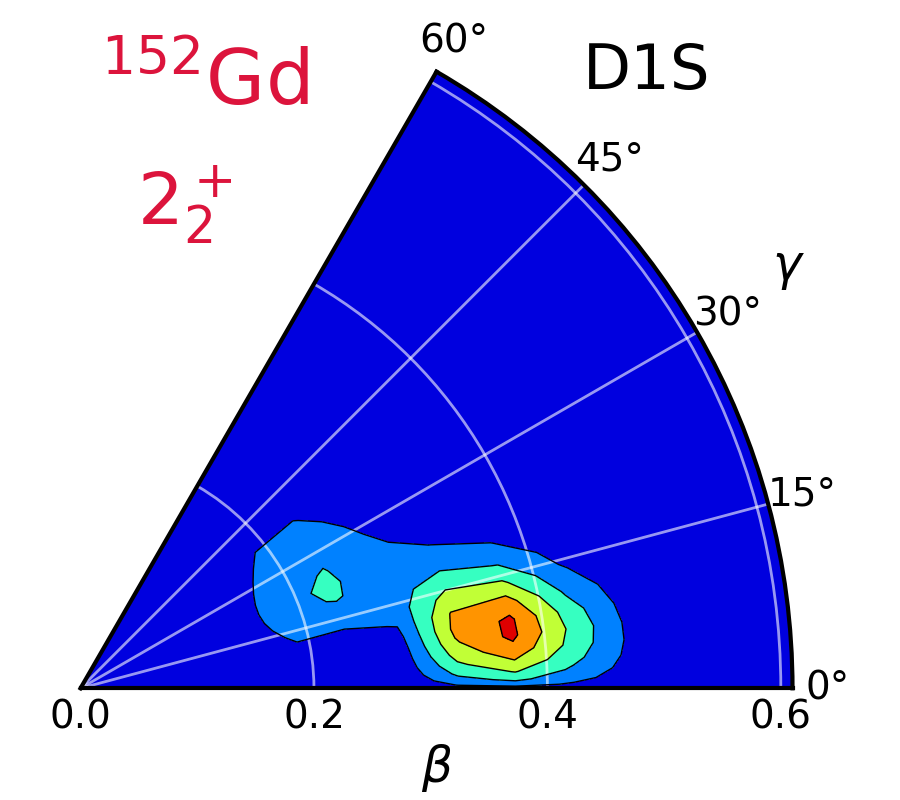}
        \includegraphics[width=0.30\textwidth]{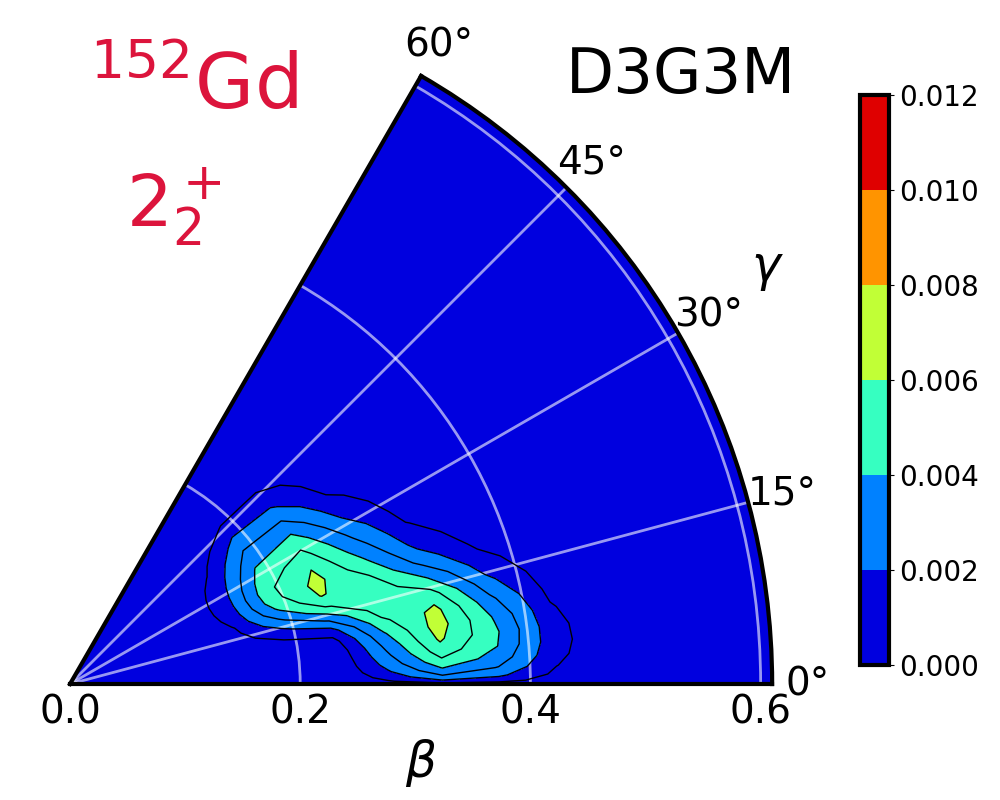}
        \\
        \includegraphics[width=0.28\textwidth]{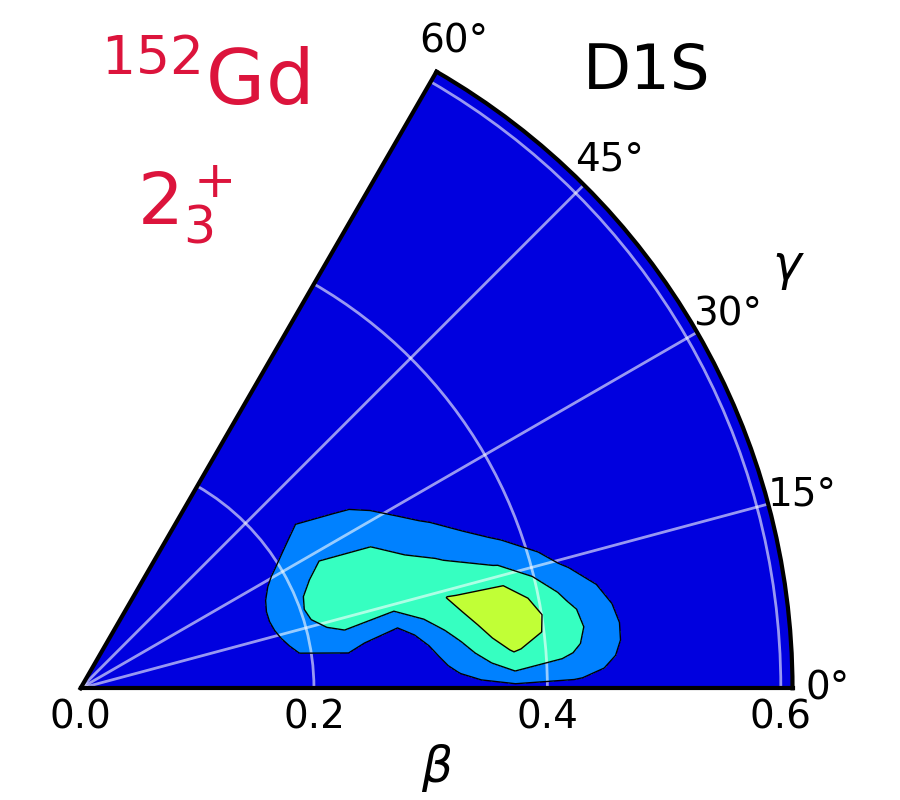}
        \includegraphics[width=0.30\textwidth]{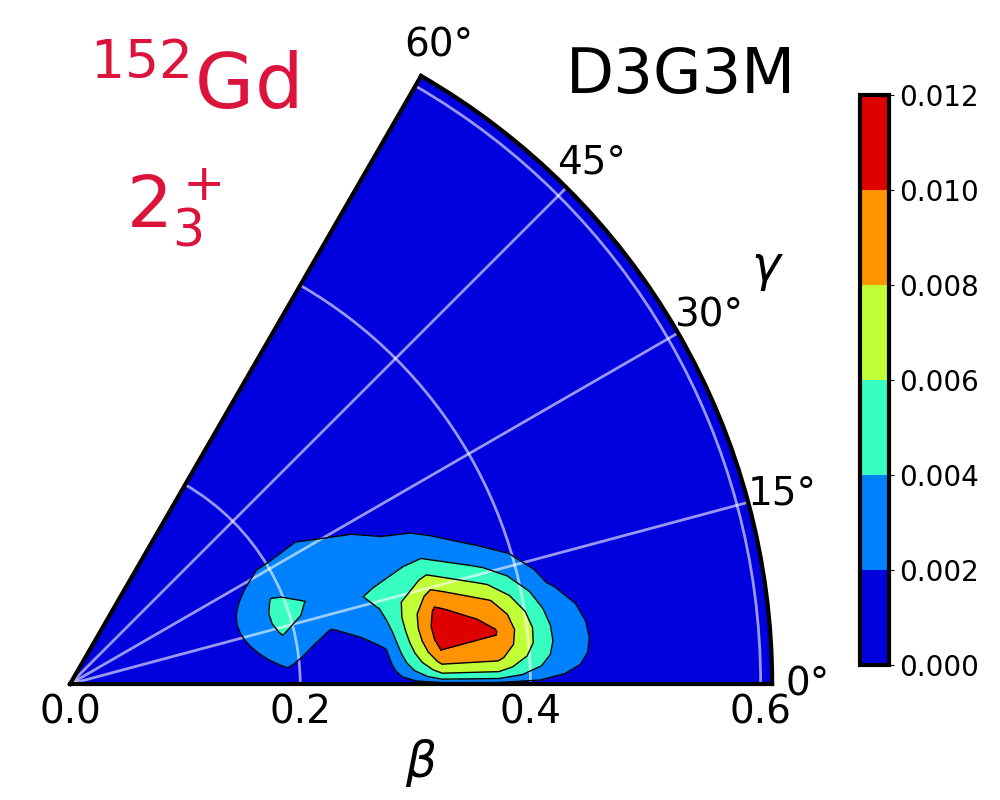}
        \caption{Norm of the wave functions of the $2^+_2$ and $2^+_3$ states in $^{152}$Gd for the Gogny D1S (left panel) and D3G3M (right panel) interactions. See text for details.}
    \label{fig:wf}
\end{figure}

In order to clarify this difference, in Fig. \ref{fig:wf} we present the norm of the wave functions $\sqrt{G} \beta^4 \sum_K \vert \psi^I_{K \alpha} \vert^2$, for the $2^+_2$ and $2^+_3$ states in $^{152}$Gd, obtained with the Gogny D1S (left column) and D3G3M (right column) interactions. We observe that, for both states, the wave functions are broadly distributed in the $\beta$ and $\gamma$ directions, confirming the ambiguity in the band assignment. However, we notice a strong similarity between the wave function of the $2^+_2$ state obtained with D1S and that of the $2^+_3$ state obtained with D3G3M, as well as between the $2^+_3$ state obtained with D1S and the $2^+_2$ state obtained with D3G3M. Since the wave functions are used to calculate quantities such as electromagnetic transition probabilities, this comparison illustrates how even small differences in the wave functions can affect the final observables.

Having access to the wave functions, we can further characterise the properties of the excited states by defining the average values of $\beta$ and $\gamma$ as
\begin{eqnarray}
    \langle \beta \rangle_{I\alpha} &=& \sqrt{ \langle \beta^2 \rangle_{I\alpha}}  \label{eq:inv:beta} \\
    \langle \gamma \rangle_{I\alpha} &=& \frac{1}{3} \arccos \left( \frac{\langle \beta^3\cos 3\gamma \rangle_{I\alpha}}{\sqrt{ \langle \beta^2 \rangle_{I\alpha}\langle \beta^4 \rangle_{I\alpha}}} \right)
    \label{eq:inv:gamma}
\end{eqnarray}
where the notation $\langle x \rangle_{I\alpha} $ means

\begin{equation}\label{eq:invariants}
    \langle x \rangle_{I\alpha}= \sum_{K \in \Delta_I} \iint x \, |  \psi^{I}_{ K \alpha}\left( \beta, \gamma \right)|^2  d\tau_0 \, .
\end{equation}

The results obtained for the first states of the $\beta$ and $\gamma$ bands are reported in Tab. \ref{tab:betagamma_excited}. We observe that the average value of $\beta$ is similar for all interactions and all states, and gradually increases as we move from $^{152}$Gd to $^{160}$Gd. Although the mean-field ground state is purely axial, we observe that the average value of $\gamma$ is different from zero.
In the case of states belonging to the $\gamma$ band, we observe values of $\langle \gamma \rangle_{I\alpha}$ larger than those of states in the $\beta$ band. In general, the various Gogny interactions provide similar results, although D3G3M and D1M tend to show a slight preference for triaxial configurations.

As already discussed in Figs.~\ref{fig:Nk} - \ref{fig:wf}, the different structure of the wave functions obtained with D1S and the other interactions for the second and third $2^+$ states in $^{152}$Gd leads to a different band assignment. This is also confirmed by the values of $\langle \gamma \rangle_{2^+_3}=17.73$ and $\langle \gamma \rangle_{2^+_2}=16.35$, indicating a larger degree of triaxiality for the $2^+_3$ state. This behaviour is opposite to that obtained with the other interactions.

\begin{table*}[t]
\centering
\renewcommand{\arraystretch}{1.25}
\setlength{\tabcolsep}{7pt}
\caption{Average values of $\beta$ and $\gamma$ (in degrees) for the $0_2^+$, $2_2^+$, $2_3^+$, $4_2^+$ and $4_3^+$ states obtained with the selected Gogny interactions.}
\label{tab:betagamma_excited}
\begin{tabular}{c c cc cc cc cc cc}

&
&
\multicolumn{2}{c}{$0^{+}_{2}$} &
\multicolumn{2}{c}{$2^{+}_{2}$} &
\multicolumn{2}{c}{$2^{+}_{3}$} &
\multicolumn{2}{c}{$4^{+}_{2}$} &
\multicolumn{2}{c}{$4^{+}_{3}$}
\\
\cline{3-12}
&
&
$\langle \beta \rangle_{I\alpha}$ & $\langle \gamma \rangle_{I\alpha}$ &
$\langle \beta \rangle_{I\alpha}$ & $\langle \gamma \rangle_{I\alpha}$ &
$\langle \beta \rangle_{I\alpha}$ & $\langle \gamma \rangle_{I\alpha}$ &
$\langle \beta \rangle_{I\alpha}$ & $\langle \gamma \rangle_{I\alpha}$ &
$\langle \beta \rangle_{I\alpha}$ & $\langle \gamma \rangle_{I\alpha}$
\\
\hline

\multirow{4}{*}{$^{152}$\!Gd}
& D3G3M & 0.287 & 13.83 & 0.260 & 19.73 & 0.291 & 15.74 & 0.299 & 14.54 & 0.282 & 19.33 \\
& D1S   & 0.311 & 13.44 & 0.296 & 16.35 & 0.296 & 17.73 & 0.323 & 13.78 & 0.301 & 19.26 \\
& D1M   & 0.293 & 13.77 & 0.264 & 19.70 & 0.296 & 15.66 & 0.301 & 14.98 & 0.290 & 18.74 \\
& D1N   & 0.297 & 13.57 & 0.273 & 18.59 & 0.295 & 16.19 & 0.309 & 14.18 & 0.289 & 19.43 \\
\hline

\multirow{4}{*}{$^{154}$\!Gd}
& D3G3M & 0.313 & 12.22 & 0.318 & 12.23 & 0.293 & 19.42 & 0.324 & 12.15 & 0.310 & 17.91 \\
& D1S   & 0.329 & 12.42 & 0.334 & 12.37 & 0.316 & 18.48 & 0.339 & 12.41 & 0.334 & 16.91 \\
& D1M   & 0.312 & 12.75 & 0.313 & 13.51 & 0.296 & 18.86 & 0.320 & 13.45 & 0.316 & 17.13 \\
& D1N   & 0.319 & 12.38 & 0.323 & 12.64 & 0.298 & 19.19 & 0.329 & 12.50 & 0.316 & 17.79 \\
\hline

\multirow{4}{*}{$^{156}$\!Gd}
& D3G3M & 0.327 & 12.08 & 0.322 & 16.42 & 0.332 & 12.58 & 0.330 & 15.21 & 0.339 & 12.95 \\
& D1S   & 0.346 & 11.97 & 0.342 & 16.53 & 0.350 & 11.76 & 0.348 & 15.80 & 0.358 & 11.75 \\
& D1M   & 0.327 & 12.33 & 0.321 & 17.03 & 0.334 & 12.37 & 0.330 & 15.75 & 0.342 & 12.62 \\
& D1N   & 0.334 & 12.16 & 0.328 & 15.83 & 0.337 & 13.53 & 0.336 & 14.64 & 0.345 & 13.72 \\
\hline

\multirow{4}{*}{$^{158}$\!Gd}
& D3G3M & 0.341 & 11.74 & 0.342 & 15.48 & 0.344 & 11.54 & 0.346 & 15.18 & 0.349 & 11.32 \\
& D1S   & 0.359 & 11.58 & 0.361 & 14.85 & 0.361 & 11.41 & 0.364 & 14.61 & 0.365 & 11.22 \\
& D1M   & 0.341 & 11.89 & 0.342 & 15.68 & 0.345 & 11.60 & 0.347 & 15.30 & 0.351 & 11.32 \\
& D1N   & 0.348 & 11.79 & 0.347 & 15.70 & 0.352 & 11.53 & 0.352 & 15.30 & 0.357 & 11.26 \\
\hline

\multirow{4}{*}{$^{160}$\!Gd}
& D3G3M & 0.348 & 11.22 & 0.356 & 13.16 & 0.349 & 11.10 & 0.359 & 13.00 & 0.353 & 10.92 \\
& D1S   & 0.364 & 11.06 & 0.369 & 12.91 & 0.366 & 10.96 & 0.371 & 12.79 & 0.369 & 10.81 \\
& D1M   & 0.348 & 11.37 & 0.355 & 13.59 & 0.351 & 11.18 & 0.358 & 13.40 & 0.355 & 10.95 \\
& D1N   & 0.357 & 11.20 & 0.363 & 13.37 & 0.359 & 11.03 & 0.365 & 13.18 & 0.363 & 10.82 \\
\hline

\end{tabular}

\end{table*}

\subsubsection{Pairing correlations}

In order to better clarify how pairing correlations affect our results, we decided to perform a series of calculations in $^{154}$Gd, varying only the pairing strength in D3G3M. Although this breaks the consistency of the interaction between the particle-hole and particle-particle channels, it provides a simple way to qualitatively assess the impact of small changes in the pairing term on the nuclear spectrum.

\begin{figure}
    \centering
        \centering
          \includegraphics[width=0.7\textwidth]{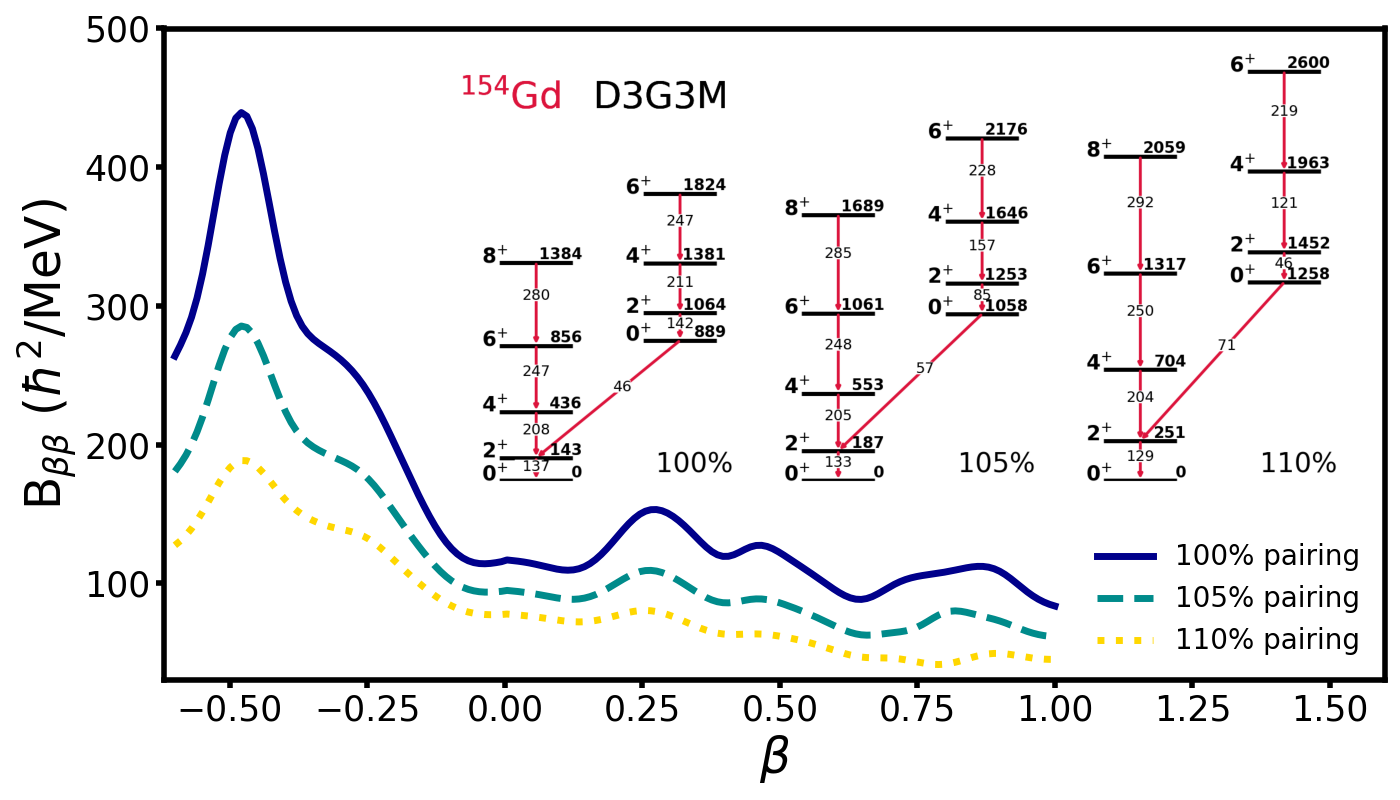}
        \caption{Axial mass parameter $B_{\beta\beta}$ as a function of the nuclear deformation $\beta$ in $^{154}$Gd calculated using D3G3M and three choices of the pairing strength. In the inset we show the low-energy spectra corresponding to the solutions of the BH in the three cases.}
    \label{fig:massparam:pairing}
\end{figure}

In Fig.~\ref{fig:massparam:pairing}, we show the mass parameter $B_{\beta\beta}$ in the axial case, obtained with the original pairing interaction (solid line), and with the pairing strength increased by 5\% (dashed line) and 10\% (dotted line), as a function of the $\beta$ deformation. We see that the peak at $\beta\approx-0.5$ disappears as the pairing strength increases, demonstrating that even a small variation in the pairing strength can lead to significant changes in the structure of the mass parameters.

In the inset of this figure, we also show the spectra for the three cases. We observe that a 10\% increase in the pairing strength leads to a 70\% increase in the energy of the first $2_1^+$ state. More generally, the energies of all levels increase, but not uniformly, resulting in a dilution of the spectrum.

Similar conclusions have also been reached in Ref.~\cite{suz24} for the case of a BH based on relativistic mean-field calculations combined with a separable pairing interaction. An increase in the pairing strength typically leads to an increase in the excitation energies of the levels. The situation is less clear for electromagnetic transitions, since we observe a decrease in the transition probability for $2_1^+\rightarrow 0_1^+$, whereas the opposite trend is observed for the $0_2^+\rightarrow 2_1^+$ transition. Therefore, it is not possible to draw a firm conclusion. It is also interesting to point out that the exact form of the adopted pairing interaction does not seem to matter significantly, provided that its strength is properly adjusted, as discussed in Ref.~\cite{pro04}.

The importance of pairing correlations in BH calculations has been extensively discussed in the literature, with several studies suggesting that the pairing strength should be included directly as a collective variable. Several authors \cite{goz85,pil93,xia20,nom21,xia2024} have investigated this approach and reported improvements in the results. However, a full treatment of the collective space, including both quadrupole and pairing degrees of freedom, has not yet been achieved \cite{pro09}. Nevertheless, including a single collective pairing variable fully coupled to the quadrupole shape, as done in Ref.~\cite{xia2024}, produces a remarkable improvement in both the level density and the electromagnetic transition probabilities compared with available experimental data. This therefore represents a promising avenue for future developments of the model.
\subsubsection{Scaling parameter}\label{sec:scaling}

\begin{figure}
    \centering
        \centering
          \includegraphics[width=0.8\textwidth]{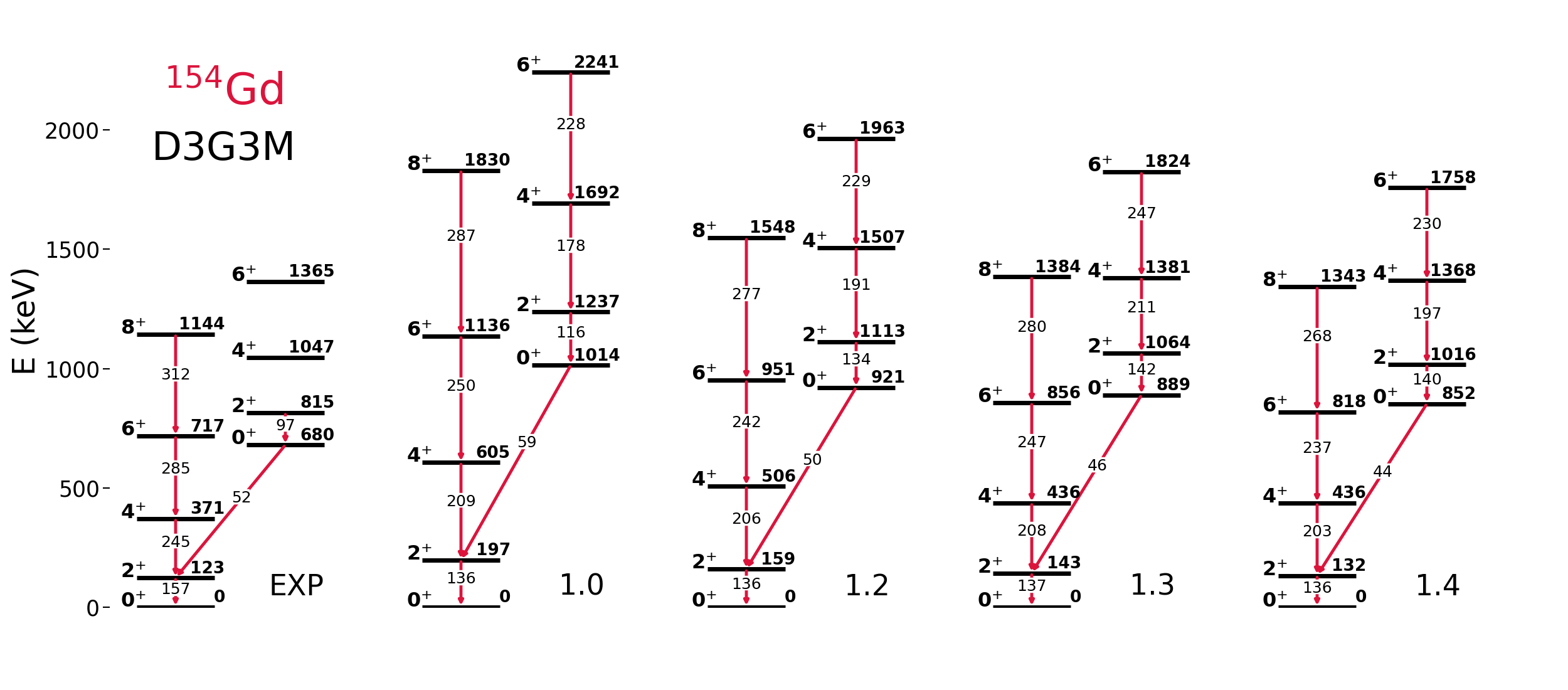}
        \caption{Experimental positive parity excited states, expressed in keV, in $^{154}$Gd (left panel) compared with theoretical ones obtained using the BH with D3G3M and various scaling parameters as reported in the figure. The electromagnetic transitions between states are indicated by arrows and expressed in Weisskopf units. See text for details.}
    \label{fig:scaling:Gd154_Spectrum}
\end{figure}

In order to compensate for the perturbative cranking approximation used to calculate vibrational masses, it is customary in the literature to introduce a scaling parameter to increase these quantities. In this article, we adopted a value of 1.3 to scale the masses and moments of inertia entering Eq.~\eqref{H:general}, although the exact value is somewhat arbitrary. In order to assess the sensitivity of the results to this parameter, in Fig.~\ref{fig:scaling:Gd154_Spectrum} we show the evolution of the spectrum of $^{154}$Gd obtained with D3G3M, using different scaling factors ranging from 1 (\emph{i.e.}, no scaling) to 1.4, thus covering the range of values adopted in the scientific literature. We observe that the energy of the lowest $2_1^+$ state decreases from 197 keV to 143 keV when the scaling factor is increased from 1 to 1.3, indicating a non-negligible impact. For smaller variations of the scaling factor, from 1.2 to 1.4, the energy changes by approximately 10 keV.
This shows that a significant difference is observed only when no scaling is applied, whereas the precise value of the scaling factor, provided it remains within this reasonable range, does not significantly alter the results.

The electromagnetic transition probabilities are only marginally affected by the scaling factor. Therefore, this parameter cannot be used to correct the systematic overestimation observed in the systematic calculations of Ref.~\cite{delaroche2010structure}.

\subsubsection{Zero point energy}\label{sec:ZPE}

\begin{figure}
    \centering
        \centering
          \includegraphics[width=0.4\textwidth]{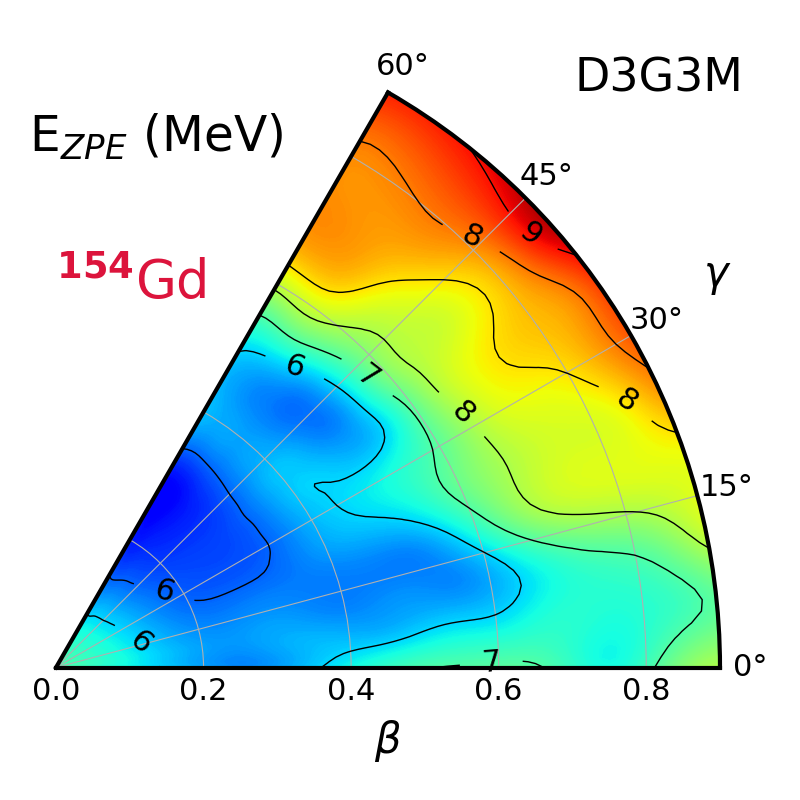}
        \caption{Total ZPE for $^{154}$Gd in the $(\beta,\gamma)$ sextant for the Gogny D3G3M interaction, calculated using the expressions of Ref.~\cite{girod1979zero}. See text for details.}
    \label{fig:ZPE:Gd154}
\end{figure}

Within BH calculations, it is customary to add an additional potential to the PES, known as the zero-point energy (ZPE) correction \cite{girod1979zero}. The latter accounts for the quantum zero-point fluctuations associated with the collective degrees of freedom around a given nuclear shape. It includes contributions from both rotational and vibrational degrees of freedom and is presented in Fig.~\ref{fig:ZPE:Gd154} for $^{154}$Gd, calculated using the Gogny D3G3M interaction.

We notice that the variation of the ZPE across the $\beta,\gamma$ plane is much smaller than the variations in the potential landscape. Whether or not such a term should be included in the collective Hamiltonian remains a matter of debate, since, within the adopted approximation, a rotational contribution is also obtained at the spherical point \cite{rei75}.

\begin{figure}
    \centering
        \centering
          \includegraphics[width=0.7\textwidth]{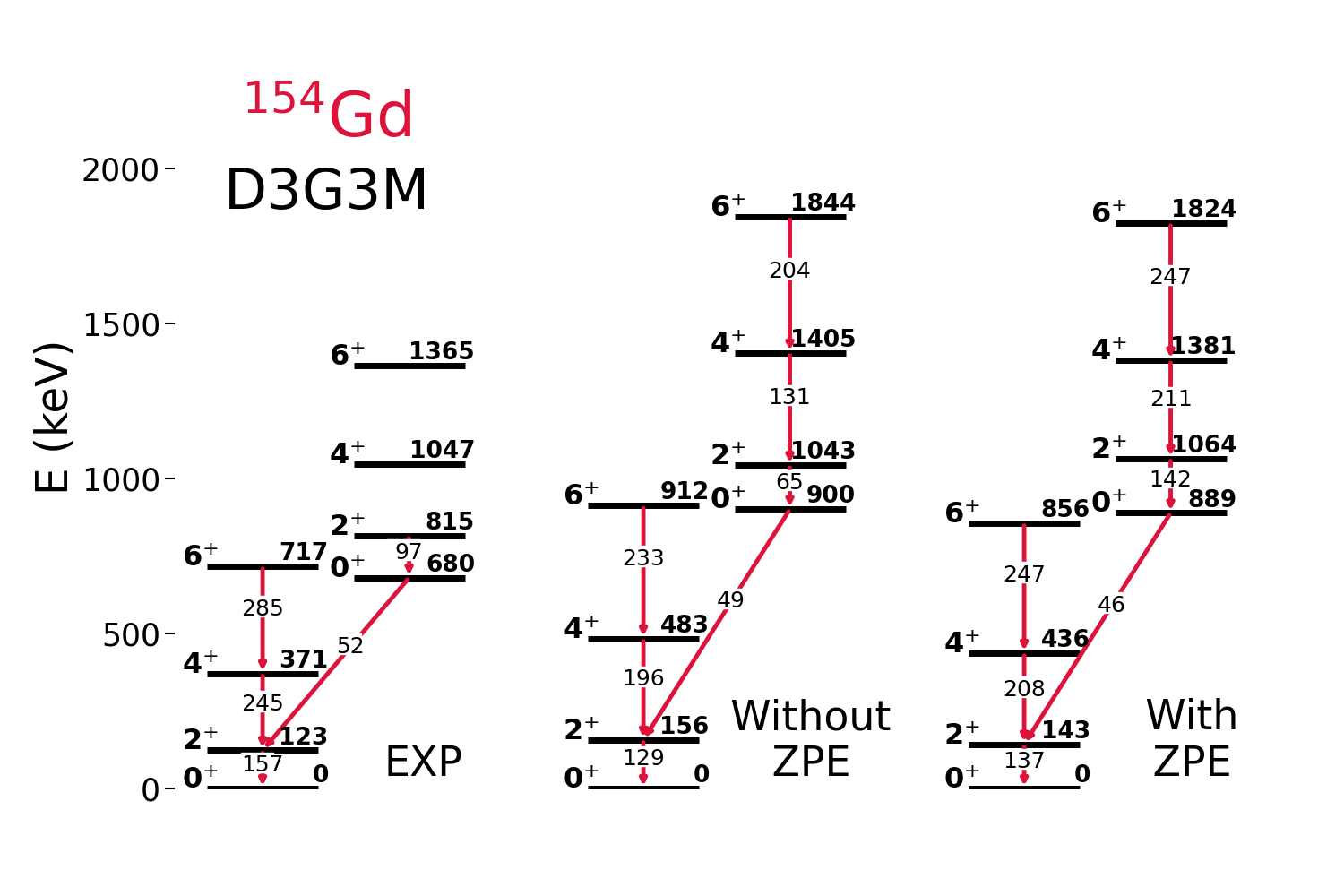}
        \caption{Same as Fig.~\ref{fig:scaling:Gd154_Spectrum}, but for a fixed scaling parameter of 1.3 and including or not the ZPE contribution to the collective Hamiltonian. See text for details.}
    \label{fig:ZPE:Gd154_Spectrum}
\end{figure}

In Fig.~\ref{fig:ZPE:Gd154_Spectrum}, we show the spectrum of $^{154}$Gd calculated with and without the ZPE included in the collective Hamiltonian. We see that its effect on the energy of the lowest $2_1^+$ state is approximately 13 keV, and is therefore comparable to the uncertainty arising from the choice of the scaling parameter. The impact on the electromagnetic transition probabilities is also quite marginal, with changes of the order of a few Weisskopf units. The ZPE may play a more important role in specific cases, such as shape-coexisting \cite{gir09} or super-deformed nuclei, as discussed in Ref.~\cite{lib16}. However, a more detailed analysis is still required to assess the relevance of this term.

\section{Conclusions}\label{sec:conclusion}
We have presented in detail the formalism required to derive and solve the generalised quadrupole Bohr Hamiltonian in connection with a microscopic potential obtained from an effective Gogny interaction. We have also carefully assessed the performance of our numerical method for a simple potential for which exact analytical results are available in the literature. In particular, we have shown that the basis size must be chosen consistently with the value of the decay parameter $\mu$ in order to optimise the numerical results.

We have subsequently performed a systematic series of mean-field calculations using the recently adjusted Gogny D3G3M interaction to generate all the microscopic input quantities entering the generalised Bohr Hamiltonian. Through systematic comparisons with experimental data, we have found that the results obtained with this interaction are in good overall agreement with experiment. In particular, the calculations reproduce rather well the low-energy part of the spectra of the studied Gd isotopes, as well as the corresponding electromagnetic transition strengths. At higher excitation energies, some discrepancies remain, although the calculations correctly reproduce the overall trends and the ordering of the various $\beta$- and $\gamma$-band structures.

To assess the robustness of these results, we have also performed a series of sensitivity analyses in which several ingredients of the Bohr Hamiltonian calculation were modified. We find that the strength of the pairing interaction has a significant impact on the calculated excitation energies, while its effect on electromagnetic transition strengths is comparatively small. Similarly, variations in the scaling parameter and the inclusion or omission of the zero-point energy correction affect the calculated energies, but have only a marginal influence on the transition strengths.

A different behaviour is observed when the underlying nuclear interaction used to generate the microscopic inputs---such as the potential energy surfaces, collective masses, and moments of inertia---is changed. In this case, the four Gogny interactions considered in this work lead to very similar predictions for the excitation energies, whereas substantially larger differences are found for the electromagnetic transition strengths. In particular, the results obtained with D1S exhibit some noticeable deviations from those obtained with the other three interactions. One illustrative example is the different assignment of the $2^+$ states belonging to the $\beta$- and $\gamma$-bands in $^{152}$Gd.

Overall, our results indicate that repeating the systematic calculations of Ref.~\cite{delaroche2010structure} with a different effective interaction, such as D3G3M, can lead to noticeable modifications in the predicted systematics of electromagnetic transition strengths, without significantly degrading the already good agreement obtained for excitation energies across the nuclear chart.

\section*{Acknowledgments}

We are indebted to J. Libert for useful discussions on the properties of the Bohr Hamiltonian model.
We are also grateful to P. Tamagno for providing valuable support in dealing with HFB calculations.
We also thank S. P\'eru, S. Hilaire, D. Reigner, N. Dubray and W. Ryssens for providing useful information concerning the microscopic calculations of the mass parameters.


\appendix

\section{Octahedral group $\mathcal{O}$}\label{app:rotations}

\subsection{Group description}\label{app:rotations_O}
The octahedral group $\mathcal{O}$ describes the symmetries of an octahedron \cite{hamermesh2012group}. Since an octahedron is the dual of a cube, it is equivalent to discuss the symmetry properties of a cube, which simplifies the visualisation. By symmetries of a cube, we mean all transformations that map the cube onto itself. Thus, the full octahedral group $\mathcal{O}_h$ consists of 48 transformations, including 24 rotations that preserve orientation and constitute the rotation subgroup $\mathcal{O}$. The remaining transformations are obtained by combining these rotations with inversion. In the following, we restrict ourselves to the rotation subgroup $\mathcal{O}$.

\begin{figure}
    \centering
        \centering
        \includegraphics[width=0.99\textwidth]{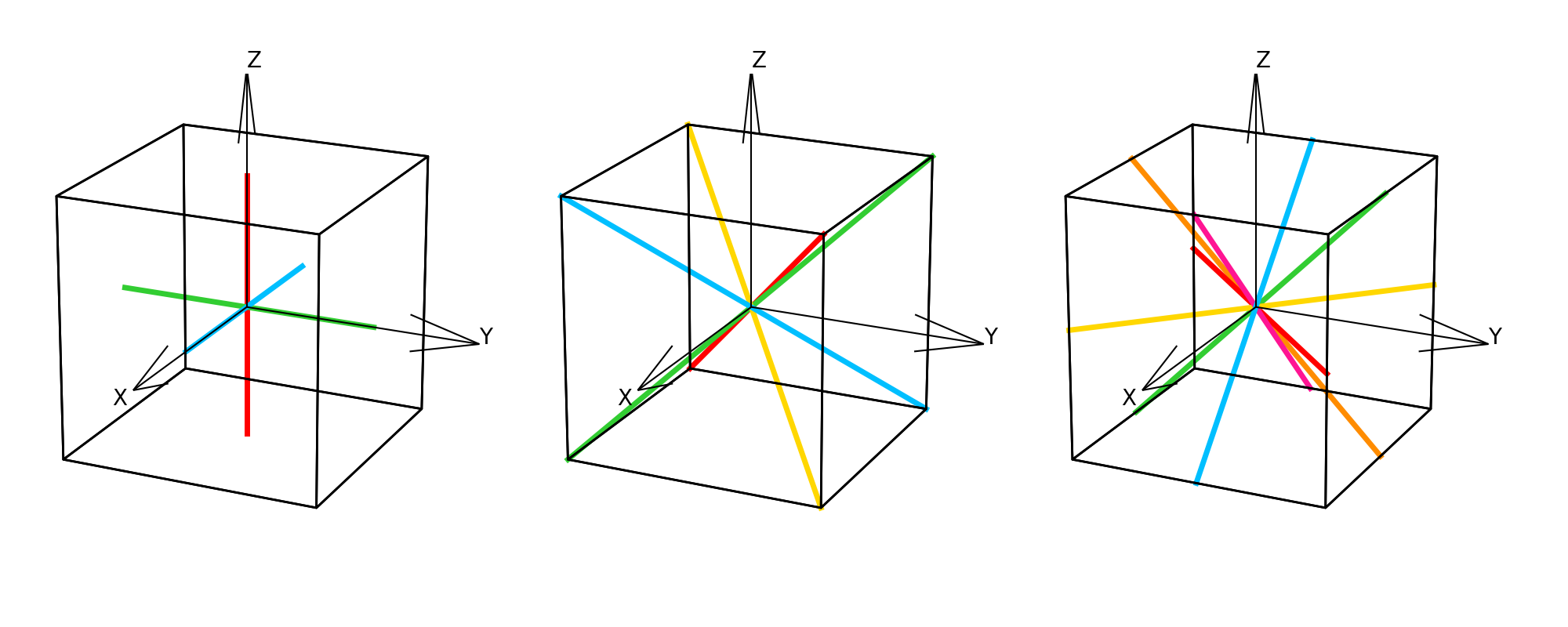}
        \caption{The rotation axes discussed in the text:  axis passing through the center of two opposite faces of the cube (left panel), 4 axes passing through two opposite vertices (middle panel) and  6 axes passing through the midpoints of two opposite edges of the cube (right panel).}
    \label{fig:cube}
\end{figure}

The 24 elements of the group are grouped into five conjugacy classes, each corresponding to a given type of symmetry, as follows.
\begin{itemize}
    \item[$\mathscr{C}_1$]: the identity ($E$),
    \item[$\mathscr{C}_2$]: 6 rotations by $\pm 90^{\circ}$ about the axes passing through centers of opposite faces ($6C_4$),
    \item[$\mathscr{C}_3$]: 8 rotations by $\pm 120^{\circ}$ about the axes passing through opposite vertices ($8C_3$),
    \item[$\mathscr{C}_4$]: 3 rotations by $180^{\circ}$ about the face-center axes ($3C_2$),
    \item[$\mathscr{C}_5$]: 6 rotations by $180^{\circ}$ about the axes passing through midpoints of opposite edges ($6C_2$).
\end{itemize}
The corresponding axes are illustrated in Fig.~\ref{fig:cube}.

The octahedral group $\mathcal{O}$ admits five irreducible representations, denoted $A_1$, $A_2$, $E$, $T_1$, and $T_2$, with respective dimensions $1$, $1$, $2$, $3$, and $3$. The character table of $\mathcal{O}$ summarizes, for each irreducible representation and each conjugacy class, the character of the group elements. Each character corresponds to the trace of the matrix associated with the group element in the given representation. This table is given in Tab. \ref{tab:char_O}.

\begin{table}[h]
\centering
\caption{Character table of the octahedral group $\mathcal{O}$.}
\label{tab:char_O}
\begin{tabular}{c||ccccc}
 & $\mathscr{C}_1$ & $\mathscr{C}_2$ & $\mathscr{C}_3$ & $\mathscr{C}_4$ & $\mathscr{C}_5$ \\
\hline \hline
$A_1$ & 1 & 1 & 1 & 1 & 1 \\
$A_2$ & 1 & -1 & 1 & 1 & -1 \\
$E$ & 2 & 0 & -1 & 2 & 0 \\
$T_1$ & 3 & 1 & 0 & -1 & -1 \\
$T_2$ & 3 & -1 & 0 & -1 & 1 \\
\end{tabular}
\end{table}

The group $\mathcal{O}$ has three generators, \emph{i.e.}, elements from which all group elements can be obtained through successive combinations. These three elements are also called Bohr rotations and are denoted by $R_1$, $R_2$, and $R_3$ \cite{Bohr_1952}. These three basic rotations act on the intrinsic axes $\xi_1, \xi_2, \xi_3$ \cite{pro09}, and are defined as follows
\begin{itemize}
    \item[$R_1$]: the transformation that reverses $\xi_2$ and $\xi_3$,
    \item[$R_2$]: a rotation by $\pi/2$ around the $\xi_3$ axis,
    \item[$R_3$]: a cyclic permutation of the axes, $\xi_i \rightarrow \xi_{i-1}$ (with $\xi_0=\xi_3$).
\end{itemize}
The expressions of all group elements in terms of these generators are summarised in Tab. \ref{tab:rotations} \cite{Corrigan_1976}.

Any rotation of the group can be expressed as a sequence of three elementary rotations defined with respect to the initial fixed reference frame, corresponding to the Euler-angle decomposition $R = R_z(\theta^1) R_y(\theta^2) R_z(\theta^3)$. These transformations can be described more compactly using a rotation operator $\hat{D}^1(\theta^1,\theta^2,\theta^3)$ \cite{Varshalovich_1988}. In this work, the adopted convention corresponds to the complex conjugate of this matrix, which leads to
\begin{equation}
    \mathcal{D}^{J*}_{KM}(\theta^1,\theta^2,\theta^3)
    = \left[\hat{\mathcal{D}}^{-1}(\theta^1,\theta^2,\theta^3)\right]^J_{MK}= \mathcal{D}^{J}_{MK}(\pi-\theta^3,\theta^2,-\pi-\theta^1)= \mathcal{D}^{J}_{MK}(-\theta^3,-\theta^2,-\theta^1).
\end{equation}
It is thus useful to define the angles $(\theta^{1'},\theta^{2'},\theta^{3'})=(\pi-\theta^3,\theta^2,-\pi-\theta^1)$ or $(-\theta^3,-\theta^2,-\theta^1)$. Table \ref{tab:rotations} gives, for each element of the group, the corresponding Euler angles and shows how the reference system $\{x,y,z\}$ is transformed. For the sake of consistency and clarity, the angles already given in the literature are presented according to our own conventions.

It is worth recalling that the orientation $\{\phi_k,\theta_k,\psi_k\}$ of the rotated frame for a given transformation $k$ is related to the orientation $\{\phi,\theta,\psi\}$ of the initial frame through the formula \cite{Varshalovich_1988}.

\begin{eqnarray}\label{wigner:transform}
    \mathcal{D}^\lambda_{\mu\nu}(\phi_k,\theta_k,\psi_k)=\sum_{m=-\lambda}^\lambda \mathcal{D}^\lambda_{\mu m}(\phi,\theta,\psi)\mathcal{D}^\lambda_{m\nu}(\theta_k^1,\theta_k^2,\theta_k^3).
\end{eqnarray}

\begin{table}[h]
    \centering
     \caption{Euler angle associated with the 24 rotations of a cube as well as the combinations of elementary transformations $R_i$ as discussed in the text.}
    \label{tab:rotations}
    \renewcommand*{\arraystretch}{1.25}
    \begin{tabular}{c|c|ccc|ccc|c}
    \hline
    \hline
        Type & Generator &\multicolumn{3}{c|}{Euler angle} & \multicolumn{3}{c|}{Coordinates} & $\gamma_k$\\
        & & $\theta_{k'}^1$& $\theta_{k'}^2$& $\theta_{k'}^3$ & $x_k$ & $y_k$ & $z_k$ &\\
        \hline \hline
        $\mathscr{C}_1$ & $R_1^2=R_2^4=R_3^3$ & $0$ & $0$ & $0$ & $x$ & $y$ & $z$ & $\gamma$\\
        \hline
        $\mathscr{C}_4$ & $R_1$ &  $\pi$ & $\pi$ & $0$ & $x$ & $-y$ & $-z$ & $\gamma$\\
        $\mathscr{C}_4$ & $R_1R_2^2$ & $0$ & $\pi$ & $0$ & $-x$ & $y$ & $-z$ & $\gamma$\\
        $\mathscr{C}_4$ & $R_2^2$ & $\pi$ & $0$ & $0$ & $-x$ & $-y$ & $z$ & $\gamma$\\
        \hline
       $\mathscr{C}_2$ & $R_2R_3^2$ & $\frac{\pi}{2}$ & $\frac{\pi}{2}$ & -$\frac{\pi}{2}$ & $x$ & $-z$ & $y$ & $-\gamma-\frac{2\pi}{3}$\\
       $\mathscr{C}_2$ & $R_1R_2R_3^2$ & $-\frac{\pi}{2}$ & $\frac{\pi}{2}$ & $\frac{\pi}{2}$ & $x$ & $z$ & $-y$ & $-\gamma-\frac{2\pi}{3}$\\
       $\mathscr{C}_2$ & $R_2^3R_3$ & $0$ & $\frac{\pi}{2}$ & $0$ & $-z$ & $y$ & $x$ & $-\gamma+\frac{2\pi}{3}$\\
       $\mathscr{C}_2$ & $R_1R_2R_3$ & $\pi$ & $\frac{\pi}{2}$ & $\pi$ & $z$ & $y$ & $-x$ & $-\gamma+\frac{2\pi}{3}$\\
       $\mathscr{C}_2$ & $R_2^3$& $0$ & $0$ & $-\frac{\pi}{2}$ & $-y$ & $x$ & $z$ & $-\gamma$\\
       $\mathscr{C}_2$ & $R_2$ & $0$ & $0$ & $\frac{\pi}{2}$ & $y$ & $-x$ & $z$ & $-\gamma$\\
       \hline
       $\mathscr{C}_5$ & $R_2^3R_3^2$ & $\frac{\pi}{2}$ & $\frac{\pi}{2}$ & $\frac{\pi}{2}$ & $-x$ & $z$ & $y$ & $-\gamma-\frac{2\pi}{3}$\\
       $\mathscr{C}_5$ & $R_1R_2^3R_3^2$ & $-\frac{\pi}{2}$ & $\frac{\pi}{2}$ & $-\frac{\pi}{2}$ & $-x$ & $-z$ & $-y$ & $-\gamma-\frac{2\pi}{3}$\\
       $\mathscr{C}_5$ & $R_1R_2$ & $-\frac{\pi}{4}$ & $\pi$ & $\frac{\pi}{4}$ & $y$ & $x$ & $-z$ & $-\gamma$\\
       $\mathscr{C}_5$ & $R_1R_2^3$ & $\frac{\pi}{4}$ & $\pi$ & $-\frac{\pi}{4}$ & $-y$ & $-x$ & $-z$ & $-\gamma$\\
       $\mathscr{C}_5$ & $R_1R_2^3R_3$ & $\pi$ & $\frac{\pi}{2}$ & $0$ & $-z$ & $-y$ & $-x$ & $-\gamma+\frac{2\pi}{3}$\\
       $\mathscr{C}_5$ & $R_2R_3$ & $0$ & $\frac{\pi}{2}$ & $\pi$ & $z$ & $-y$ & $x$ & $-\gamma+\frac{2\pi}{3}$\\
       \hline
       $\mathscr{C}_3$ & $R_2^2R_3$ & $0$ & $\frac{\pi}{2}$ & $-\frac{\pi}{2}$ & $-y$ & $-z$ & $x$ & $\gamma-\frac{2\pi}{3}$\\
       $\mathscr{C}_3$ & $R_1R_3^2$ & $-\frac{\pi}{2}$ & $\frac{\pi}{2}$ & $\pi$ & $z$ & $-x$ & $-y$ & $\gamma+\frac{2\pi}{3}$\\
       $\mathscr{C}_3$ & $R_1R_3$ & $\pi$ & $\frac{\pi}{2}$ & $-\frac{\pi}{2}$ & $y$ & $-z$ & $-x$ & $\gamma-\frac{2\pi}{3}$\\
       $\mathscr{C}_3$ & $R_1R_2^2R_3^2$ & $-\frac{\pi}{2}$ & $\frac{\pi}{2}$ & $0$ & $-z$ & $x$ & $-y$ & $\gamma+\frac{2\pi}{3}$\\
       $\mathscr{C}_3$ & $R_2^2R_3^2$ & $\frac{\pi}{2}$ & $\frac{\pi}{2}$ & $0$ & $-z$ & $-x$ & $y$ & $\gamma+\frac{2\pi}{3}$\\
       $\mathscr{C}_3$ & $R_1R_2^2R_3$ & $\pi$ & $\frac{\pi}{2}$ & $\frac{\pi}{2}$ & $-y$ & $z$ & $-x$ & $\gamma-\frac{2\pi}{3}$\\
       $\mathscr{C}_3$ & $R_3$ & $0$ & $\frac{\pi}{2}$ & $\frac{\pi}{2}$ & $y$ & $z$ & $x$ & $\gamma-\frac{2\pi}{3}$\\
       $\mathscr{C}_3$ & $R_3^2$ & $0$ & $\frac{\pi}{2}$ & $\pi$ & $z$ & $x$ & $y$ & $\gamma+\frac{2\pi}{3}$ \\
    \hline
    \hline
    \end{tabular}
\end{table}

\subsection{Transformation of the coordinates $\gamma$ under $\mathcal{O}$}

Following Ref.~\cite{Corrigan_1976}, we determine how the collective coordinate $\gamma$ transforms under a given rotation. The collective coordinates $\alpha_{2\mu}$ in the laboratory frame are unaffected by a rotation of the axes. Thus, we can write
\begin{equation}
    \beta \cos \gamma \mathcal{D}^{2}_{\mu 0 }(\Omega) + \beta \frac{\sin \gamma}{\sqrt{2}} \left[ \mathcal{D}^{2}_{\mu 2}(\Omega) + \mathcal{D}^{2}_{\mu -2}(\Omega) \right] = \beta_k \cos \gamma_k \mathcal{D}^{2}_{\mu 0}(\Omega_k) + \beta_k \frac{\sin \gamma_k}{\sqrt{2}}  \left[ \mathcal{D}^{2}_{\mu 2}(\Omega_k) + \mathcal{D}^{2}_{\mu -2}(\Omega_k) \right].
\end{equation}
where $\beta$ and $\gamma$ are the deformation parameters in the rotated frame characterised by the Euler angles $\Omega$, while $\beta_k$ and $\gamma_k$ denote the corresponding parameters in the frame characterised by the Euler angles $\Omega_k$. Using Eq.~\eqref{wigner:transform} and the fact that $\beta = \beta_k$,
\begin{equation}
    \cos \gamma \mathcal{D}^{2}_{\mu 0 }(\Omega) + \frac{\sin \gamma}{\sqrt{2}} \left[ \mathcal{D}^{2}_{\mu 2}(\Omega) + \mathcal{D}^{2}_{\mu -2}(\Omega) \right] =  \sum_m \mathcal{D}^{2}_{\mu m}( \Omega) \left[ \cos \gamma_k \mathcal{D}^{2}_{m 0}(\theta_k) + \frac{\sin \gamma_k}{\sqrt{2}} \left[ \mathcal{D}^{2}_{m 2}(\theta_k) + \mathcal{D}^{2}_{m -2}(\theta_k) \right] \right] \, .
\end{equation}
One obtains five equations for $\gamma$, corresponding to the different values of $\mu$.  Since the cases $\mu=\pm2$ are equivalent and the equations for $\mu=\pm1$ vanish, only two independent equations remain to be solved.  

\begin{equation}
\begin{split}
    \cos \gamma &= \cos \gamma_k \left[ \frac{3}{2} \cos^2(\theta_k^2)-\frac{1}{2} \right]  +  \frac{\sin \gamma_k}{\sqrt{2}} \left[ \frac{\sqrt{6}}{2}  \sin^2(\theta_k^2) \cos (2\theta_k^3) \right]\,,\\
    \frac{\sin \gamma}{\sqrt{2}} &= \cos \gamma_k \left[ \frac{\sqrt{6}}{4}  \sin^2 (\theta_k^2) \cos(2\theta_k^1) \right]  +  \frac{\sin \gamma_k}{\sqrt{2}} \left[\frac{(1+\cos^2(\theta_k^2))}{2} \cos(2\theta_k^3) \cos(2\theta_k^1)+ \cos(\theta_k^2) \sin(2\theta_k^3) \sin(2\theta_k^1) \right] \, .
\end{split}
\end{equation}

The resulting angles are given in Tab. \ref{tab:rotations} for each of the rotations.

\subsection{Transformation of the basis under $\mathcal{O}$}\label{appendix:wf:rot}

This appendix details the symmetrization procedure introduced in Sec. \ref{sec:sym}. 
For each of the three projectors defined in Eq.~\eqref{eq:projectors}, the corresponding generator is applied according to Eq.~\eqref{eq:Rwf}. 
The Euler angles associated with each rotation, the rotated deformation variables, and the corresponding Wigner functions are listed in Tab. \ref{tab:R1R2R3}.

\begin{table}[h]
    \centering
        \caption{From left to right: generators $R_k$ of the group $\mathcal{O}$; angles $(\theta^1_k, \theta^2_k, \theta^3_k)$ defined in Eq.~\eqref{wigner:transform}; deformation parameters $\beta_k$ and $\gamma_k$ after the rotation; Euler angles $\phi_k$, $\theta_k$, and $\psi_k$ defining the orientation of the intrinsic axes with respect to the laboratory axes; and Wigner functions $\mathcal{D}^I_{LK}(\theta_k^1,\theta_k^2,\theta_k^3)$ evaluated at the Euler angles $\theta_k^1$, $\theta_k^2$, and $\theta_k^3$ \cite{pro09}.} \label{tab:R1R2R3}
    \begingroup
    \renewcommand*{\arraystretch}{1.5}
    \begin{tabular}{c|cccc}
    \hline
    $R_k$ & $\theta_k^1,\theta_k^2,\theta_k^3$ & $\beta_k,\gamma_k$ & $\Omega_k=\phi_k,\theta_k,\psi_k$ & $\mathcal{D}^I_{LK}(\theta_k^1,\theta_k^2,\theta_k^3)$ \\
    \hline \hline
    $R_1$ & \begin{tabular}{cccc}
         $\pi$ & $\pi$ & 0    
    \end{tabular} & \begin{tabular}{cc}
         $\beta,$ & $\gamma$ 
    \end{tabular} & \begin{tabular}{ccc}
          $\phi \pm \pi$ & $\pi - \theta$ & $- \psi$  
    \end{tabular} & $(-1)^I \delta_{L\,-K}$ \\
    $R_2$  & \begin{tabular}{cccc}
         0 & 0 & $\pi/2$    
    \end{tabular} & \begin{tabular}{cc}
         $\beta,$ & $-\gamma$ 
    \end{tabular} & \begin{tabular}{ccc}
          $\phi$ & $\theta$ & $\psi+\pi/2$  
    \end{tabular}  & $i^{K} \delta_{LK}$  \\
    $R_3$ & \begin{tabular}{cccc}
         0 & $\pi/2$ & $\pi/2$
    \end{tabular} & \begin{tabular}{cc}
         $\beta,$ & $\gamma-2\pi/3$ 
    \end{tabular} & \begin{tabular}{ccc}
          $\phi_3$ & $\theta_3$ & $\psi_3$\footnotemark 
    \end{tabular} & $i^{K} d^I_{LK}(\frac{\pi}{2})$  \\
    \hline
    \end{tabular}
    \endgroup
\end{table}
\footnotetext{$\cos \theta_3 = - \sin \theta \cos \psi$, $\sin \theta_3 \sin \psi_3 = \cos \theta$ and $\cos (\psi_3 - \psi)= \cos \psi \sin \psi_3 $}

\subsubsection{Invariance against $R_1$}

The $R_1$ rotation simply reverses two axes, and by using the properties of Wigner matrices, we can write it as

\begin{equation}
     R_1\varphi^{IM}_{Lmn}\left( \beta,\gamma, \Omega \right)   = \sum_{K=-I}^{I} \varphi^{I\mathrm{(in)}}_{Lmn}\left( \beta,\gamma \right) \sqrt{\frac{2I+1}{8\pi^2}}  \mathcal{D}_{MK}^{I} \left(\Omega \right) \mathcal{D}_{KL}^{I} \left( \pi, \pi, 0 \right) \,,
\end{equation}

which leads to

\begin{equation}
    \varphi^{I\mathrm{(in)}}_{Lmn}\left( \beta,\gamma \right) = (-1)^{I} \varphi^{I\mathrm{(in)}}_{-Lmn}\left( \beta,\gamma \right) \,.
    \label{eq:inv1}
\end{equation}

This implies that, for odd values of $I$, the case $L=0$ is excluded. Eq.~\eqref{eq:inv1} shows that the wave functions with negative $L$ can be deduced from those with positive $L$. Thus, by applying the projector $P_1$, we can restrict the sum over $L$ and re-express the wave function as

\begin{equation}
     P_1 \varphi^{IM}_{Lmn}\left( \beta,\gamma, \Omega \right)  
    = \tilde{\varphi}^{I\mathrm{(in)}}_{Lmn}\left( \beta,\gamma \right) \Phi^I_{ML}(\Omega) \, .
\end{equation}

The relation between the two normalization is given by
\begin{equation}
        \tilde{\varphi}^{I\mathrm{(in)}}_{Lmn}\left( \beta,\gamma \right) = \sqrt{\frac{2}{(1+\delta_{L0})}} \varphi^{I\mathrm{(in)}}_{Lmn} \left( \beta,\gamma \right) \, .
        \label{appendix:wf:normalization}
\end{equation}

\subsubsection{Invariance against  $R_2$ }

 $R_2$ is a rotation by $\pi/2$ around the third axis of the intrinsic frame.
\begin{equation}
     R_2 \varphi^{IM}_{Lmn}\left( \beta,\gamma, \Omega \right)  
      = \sum_{K=-I}^{I} \varphi^{I\mathrm{(in)}}_{Kmn}\left( \beta,\gamma \right) \mathcal{D}_{ML}^{I} \left(\Omega \right) \mathcal{D}_{LK}^{I} \left( 0, 0 , \frac{\pi}{2} \right).  
\end{equation}

The function $\varphi^{I\mathrm{(in)}}_{Lmn}$ must therefore fulfil the following invariance condition. Applying the rotation $R_2$ a second time, we find that $L$ must be even.
\begin{equation}
     \varphi^{I\mathrm{(in)}}_{Lmn}\left( \beta,\gamma \right) = (-1)^{L/2} \varphi^{I\mathrm{(in)}}_{Lmn}\left( \beta,-\gamma \right).
\end{equation}

\noindent The action of the projector $P_2$ leads to
\begin{equation}
\begin{split}
    P_2 P_1 \varphi^{IM}_{Lmn}\left( \beta,\gamma, \Omega \right)  
    &= \sqrt{\frac{2}{(1+\delta_{L0})}}\mathrm{e}^{-\mu^2 \beta^2/2}\, \beta^n\, \Phi^I_{ML}\left( \Omega \right) p_{Lm}( \gamma ) \, .
\end{split}
\end{equation}

With these new conditions on the allowed values of $L$, we have, in particular, $L = 2(I \bmod 2), 2(I \bmod 2)+2, \ldots, 2 \left\lfloor \frac{I}{2} \right\rfloor \in \Delta_I$, and $p_{Lm}(\gamma)$ is expressed as
 \begin{equation}
   p_{Lm}( \gamma ) = \left\{ \begin{array}{cc}
         \cos m \gamma &\text {$L/2$ even}\,,\\
         \sin m \gamma &\text{$L/2$ odd}\,.
    \end{array} \right. 
\end{equation}
 
\subsubsection{Invariance against  $R_3$ }

The $R_3$ rotation is a cyclic permutation of the axes (see Appendix \ref{app:rotations_O}) and is the most complicated case to deal with. We follow closely the argument developed in Ref.~\cite{Kumar_1967} to prove the invariance. The action of $R_3$ on the wave function reads

\begin{equation}
     R_3 \varphi^{IM}_{Lmn}\left( \beta,\gamma, \Omega \right)  
      =  \sqrt{\frac{2I+1}{8\pi^2}} \sum_{K=-I}^{I} \varphi^{I\mathrm{(in)}}_{Kmn}\left( \beta,  \gamma-\frac{2\pi}{3} \right)\mathcal{D}_{LK}^{I} \left( 0, \frac{\pi}{2} , \frac{\pi}{2} \right)  \mathcal{D}_{ML}^{I} \left(\Omega \right) \, .
\end{equation}

In this case, the invariance condition leads to a more complicated relation than for the previous rotations. Using the following property of the small Wigner function $d^{I}_{L\,-K}\!\left(\frac{\pi}{2}\right) = (-1)^{I-K} \, d^{I}_{LK}\!\left(\frac{\pi}{2}\right)$ and the first invariance condition, we write the previous expression as a sum over non-negative even  values of $K$ as 
\begin{equation}
    \varphi^{I(in)}_{Lmn}\left( \beta,\gamma \right) 
    = \sum_{K \in \Delta_I} \frac{2}{(1+\delta_{K0})} \varphi^{I(in)}_{Kmn}\left( \beta,  \gamma-\frac{2\pi}{3} \right)  (-1)^\frac{K}{2} d_{LK}^{I}\left( \frac{\pi}{2} \right) \, .
\end{equation}
To recover the normalization corresponding to positive values of $L$, we do the following manipulation
{\small
\begin{equation}
\begin{split}
   \tilde{\varphi}^{I(in)}_{Lmn}\left( \beta,\gamma \right) =&\sqrt{\frac{2}{(1+\delta_{L0})}} \sum_{K \in \Delta_I}  \sqrt{\frac{2}{(1+\delta_{K0})}} \sqrt{\frac{2}{(1+\delta_{K0})}} \varphi^{I(in)}_{Kmn}\left( \beta,  \gamma-\frac{2\pi}{3} \right) (-1)^{\frac{K}{2}} d_{LK}^{I}\left( \frac{\pi}{2} \right)\\
    = &  \sum_{K \in \Delta_I} \frac{2}{\sqrt{(1+\delta_{L0})(1+\delta_{K0})}} \tilde{\varphi}^{I(in)}_{Kmn}\left( \beta,  \gamma-\frac{2\pi}{3} \right)  (-1)^{\frac{K}{2}} d_{LK}^{I}\left( \frac{\pi}{2} \right) \, .\\
\end{split}
\end{equation}}
So by identification we observe that 
\begin{equation}
   R_3 \varphi^{IM}_{Lmn}\left( \beta,\gamma, \Omega \right)   = \sum_{K \in \Delta_I} \frac{2 \, (-1)^{\frac{K}{2}} d_{LK}^{I}\left( \frac{\pi}{2} \right)}{\sqrt{(1+\delta_{L0})(1+\delta_{K0})}} \tilde{\varphi}^{I(in)}_{Kmn} \left( \beta,  \gamma-\frac{2\pi}{3} \right)  \Phi^I_{ML}\left( \Omega \right) \, .
\end{equation}
Using the summation properties of Wigner functions, the action of $R_3$ a second time results in
{
\begin{equation}
\begin{split}
    &R_3^2 \varphi^{I(lab)}_{Mmn}\left(\beta,\gamma, \Omega \right) \\
    &=  \sum_{L \in \Delta_I} \sum_{K \in \Delta_I}  \sum_{P \in \Delta_I} \frac{2 \, \mathcal{D}_{LK}^{I} \left( 0, \frac{\pi}{2} , \frac{\pi}{2} \right)}{\sqrt{(1+\delta_{L0})(1+\delta_{K0})}}   \frac{2 \, \mathcal{D}_{KP}^{I} \left( 0, \frac{\pi}{2} , \frac{\pi}{2} \right)}{\sqrt{(1+\delta_{K0})(1+\delta_{P0})}} \tilde{\varphi}^{I(in)}_{Pmn}\left( \beta,  \gamma+\frac{2\pi}{3}\right) \Phi^I_{ML}\left( \Omega \right) \\
    &= \sum_{L \in \Delta_I} \sum_{P \in \Delta_I} \frac{2 \, \mathcal{D}_{LP}^{I}  \left( \frac{\pi}{2} , \frac{\pi}{2} , \pi \right)}{\sqrt{(1+\delta_{L0})(1+\delta_{P0})}} \tilde{\varphi}^{I(in)}_{Pmn}\left( \beta,  \gamma+\frac{2\pi}{3} \right)  \Phi^I_{ML}\left( \Omega \right) \\
    &= \sum_{L \in \Delta_I} \sum_{K \in \Delta_I} \frac{2 \, (-1)^{\frac{L}{2}} d_{LK}^{I}\left( \frac{\pi}{2} \right) }{\sqrt{(1+\delta_{L0})(1+\delta_{K0})}} \tilde{\varphi}^{I(in)}_{Kmn} \left( \beta,  \gamma + \frac{2\pi}{3} \right) \Phi^I_{ML}\left( \Omega \right) \, .
 \end{split}
 \end{equation}}

Consequently, we successively apply the three projectors to the initial functions as
{\small
\begin{equation}
\begin{split}
    & P_3 P_2 P_1 \varphi^{I(lab)}_{Mmn}\left(\beta,\gamma, \Omega\right) \\
    &= \sum_{L \in \Delta_I}\sum_{K \in \Delta_I} \Phi_{ML}^{I} \left(\Omega\right) \Biggl[ \delta_{KL} \tilde{\varphi}^{I(in)}_{Kmn} \left(\beta,\gamma\right) \\
    & + \frac{2 \,  d_{LK}^{I}\left(\frac{\pi}{2}\right) }{\sqrt{(1+\delta_{L0})(1+\delta_{K0})}} \left( (-1)^{\frac{L}{2}} \tilde{\varphi}^{I(in)}_{Kmn} \left(\beta,\gamma+\frac{2\pi}{3}\right) + (-1)^{\frac{K}{2}} \tilde{\varphi}^{I(in)}_{Kmn} \left(\beta, \gamma-\frac{2\pi}{3}\right)\right)\Biggr] \\
    &= \sum_{L \in \Delta_I}\sum_{K \in \Delta_I} \Phi_{MK}^{I} \left(\Omega\right) \Biggl[ \delta_{KL} \tilde{\varphi}^{I(in)}_{Lmn} \left(\beta,\gamma\right) \\
    & + \frac{2  d_{KL}^{I}\left(\frac{\pi}{2}\right) }{\sqrt{(1+\delta_{L0})(1+\delta_{K0})}} \left( (-1)^{\frac{K}{2}} \tilde{\varphi}^{I(in)}_{Lmn} \left(\beta,\gamma+\frac{2\pi}{3}\right) + (-1)^{\frac{L}{2}} \tilde{\varphi}^{I(in)}_{Lmn} \left(\beta, \gamma-\frac{2\pi}{3}\right)\right)\Biggr] \\
    &= \sum_{L \in \Delta_I} \sum_{K \in \Delta_I} \Phi_{MK}^{I} \left(\Omega \right) \mathrm{e}^{-\mu^2 \beta^2/2}\, \beta^n\,  \left\{ \cos(m\gamma) \biggl[ \delta_{KL} +  \frac{2  \, d_{KL}^{I}\left( \frac{\pi}{2} \right)}{\sqrt{(1+\delta_{L0})(1+\delta_{K0})}} \right. \\
    & \left. \biggl( \cos\left(\frac{2\pi}{3} m\right) ( (-1)^{\frac{L}{2}}+(-1)^{\frac{K}{2}} ) + \sin\left(\frac{2\pi}{3} m\right) ((-1)^{\frac{K}{2}} - (-1)^{\frac{L}{2}}) \biggr) \biggr] + \sin\left(m\gamma\right) \Biggl[ \delta_{KL} + \right. \\
    & \left.  \frac{2 \, d_{KL}^{I}\left( \frac{\pi}{2} \right) }{\sqrt{(1+\delta_{L0})(1+\delta_{K0})}}   \biggl( \cos\left(\frac{2\pi}{3} m\right) ((-1)^{\frac{L}{2}}+(-1)^{\frac{K}{2}}) + \sin\left(\frac{2\pi}{3} m\right) ((-1)^{\frac{L}{2}}-(-1)^{\frac{K}{2}}) \biggr) \Biggr] \right\} \\
    &= \sum_{L \in \Delta_I} \sum_{K \in \Delta_I} \Phi_{MK}^{I} \left( \Omega \right) \mathrm{e}^{-\mu^2 \beta^2/2}\, \beta^n\, \\
    & \left\{
    \begin{array}{lc}
        \cos(m\gamma)\left[ \delta_{KL} 
        + \frac{4 d_{KL}^{I} \left( \frac{\pi}{2} \right)}{\sqrt{(1+\delta_{L0})(1+\delta_{K0})}}   \cos \left(\frac{2\pi}{3} m\right) \right]
        & \text{$\frac{L}{2}$ even, $\frac{K}{2}$even}  \\
       \cos(m\gamma)\left[ \delta_{KL} 
        + \frac{4d_{KL}^{I} \left( \frac{\pi}{2} \right)}{\sqrt{(1+\delta_{L0})(1+\delta_{K0})}}   \sin \left(\frac{2\pi}{3} m\right) \right]
        & \text{$\frac{L}{2}$ odd, $\frac{K}{2}$even}  \\
        \sin(m\gamma)\left[ \delta_{KL} 
        + \frac{4d_{KL}^{I} \left( \frac{\pi}{2} \right)}{\sqrt{(1+\delta_{L0})(1+\delta_{K0})}}   \sin \left(\frac{2\pi}{3} m\right) \right]
        & \text{$\frac{L}{2}$ even, $\frac{K}{2}$ odd}  \\
        \sin(m\gamma)\left[ \delta_{KL}
        - \frac{4 d_{KL}^{I} \left( \frac{\pi}{2} \right) }{\sqrt{(1+\delta_{L0})(1+\delta_{K0})}}  \cos \left(\frac{2\pi}{3} m\right) \right]
        & \text{$\frac{L}{2}$ odd, $\frac{K}{2}$ odd}  \\
    \end{array}
    \right. \\
    \\
    &= \sum_{L \in \Delta_I} \sum_{K \in \Delta_I}  \Phi_{MK}^{I} \left( \Omega \right) \mathrm{e}^{-\mu^2 \beta^2/2}\, \beta^n\, c^I_{LmK}\,p_{Km}(\gamma)  \, .
 \end{split}
 \end{equation}}

where $c^I_{LmK}$ and $p_{Km}$ are given in Eqs.\eqref{cILmK:def}-\eqref{pKm:def}.


\section{Matrix elements}\label{sec:mel}

We provide the explicit matrix elements of the BH given in Eq.~\eqref{H:general} between two basis states $i,j$ in a given $I,M$ block.
We start with the vibrational sector, Eq.~\eqref{H:general:vib}. Since the Hamiltonian does not depend on the Euler angles, the integration over these angles can be performed analytically.
See Eq.\eqref{eq:basis:final:TOT} for the definition of $\Tilde{F}^{I}_{Kj}$.
Performing an integration by parts, we obtain the following result.
\begin{eqnarray}
 \langle \varphi^{IM}_{j} \vert \hat{H}_\text{vib} \vert \varphi^{IM}_{i}  \rangle
 &=& \frac{\hbar^2}{2}  \iint_{\beta,\gamma}  \sum_{K \in \Delta_I} \sum_{\mu,\nu=\beta,\gamma} \partial_\mu \Tilde{F}_{Kj}^{I} \; (B_\text{vib}^{-1})_{\mu,\nu} \; \partial_\nu \Tilde{F}_{Ki}^{I} \sqrt{G} d\tau_0 \, .
\end{eqnarray}

\noindent Using the identity $\partial_\mu \ln G  = \frac{1}{G} \partial_\mu G $, we get to the final expression of the vibrational matrix elements
\begin{eqnarray}
     \langle \varphi^{IM}_{j} \vert \hat{H}_\text{vib} \vert \varphi^{IM}_{i}  \rangle &=& \frac{\hbar^2}{2} \iint_{\beta,\gamma}  \sum_{K \in \Delta_I} \sum_{\mu,\nu=\beta,\gamma} (B_\text{vib}^{-1})_{\mu,\nu} \left\{ (\partial_\mu F_{Kj}^{I})   \, (\partial_\nu  F_{Ki}^{I}) \right.+ \frac{1}{16} F_{Kj}^{I}  \, F_{Ki}^{I}( \partial_\mu \ln G)  \,( \partial_\nu  \, \ln G ) \nonumber\\
    && \left.  - \frac{1}{4} \left[ F_{Kj}^{I}  \, (\partial_\mu \ln G )\, (\partial_\nu F_{Ki}^{I}) +  F_{Ki}^{I}  \, (\partial_\mu F_{Kj}^{I} )\, (\partial_\nu \ln G )   \right]\right\} \text{d}\tau_0.
\end{eqnarray}

\noindent The derivatives of the basis functions with respect to $\beta$ and $\gamma$ can both be evaluated analytically, while the integrals are calculated numerically using Gauss–Laguerre quadrature for $\beta$ and Gauss–Legendre quadrature for $\gamma$.

To calculate the rotational part, Eq.~\eqref{H:general:rot}, we first write
\begin{equation}
    \hat{H}_\text{rot}  
    =  a \hat{L}^{\mathrm{(in)}2} + b \hat{L}_z^{\mathrm{(in)}2} + c \left( \hat{L}_+^{\mathrm{(in)}2} + \hat{L}_-^{\mathrm{(in)}2} \right) ,
\end{equation}
where 
\begin{eqnarray}
    a&=&\frac{1}{4} \left(  \frac{1}{\mathcal{I}_x}  +   \frac{1}{\mathcal{I}_y}\right),\\
     b &=&   \frac{1}{2\mathcal{I}_z} - a,\\
      c &=&  \frac{1}{8} \left(  \frac{1}{\mathcal{I}_x}  -  \frac{1}{\mathcal{I}_y} \right).
\end{eqnarray}
\noindent The action of the intrinsic angular momentum operator  on the basis states is given in Eq.~\eqref{eq:L_in}. After some manipulations, the rotational matrix elements can therefore be written as

\begin{eqnarray}
       \langle \varphi^{IM}_{j} \vert \hat{H}_\text{rot} \vert \varphi^{IM}_{i}  \rangle &=&   \sum_{K \in \Delta_I} \hbar^2 \iint_{\beta,\gamma}\left( a I(I+1) + b K^2 \right)   F^{I}_{Kj} F^{I}_{Ki} \text{d}\tau_0 \\
       &+& \sum_{\substack{K\in\Delta_I \\ K\le I-2}} \sqrt{(1+\delta_{K0}) (I+K+1) (I - K) (I-K-1) (I+K+2)} \hbar^2\iint_{\beta,\gamma} c  \left( F^{I}_{K+2\,j} F^{I}_{Ki} +  F^{I}_{Kj} F^{I}_{K+2\,i} \right) \text{d}\tau_0 \nonumber  
\end{eqnarray}

\noindent Finally, one has to evaluate the potential matrix elements, which read
\begin{equation}
    \langle \varphi^{IM}_{j} \vert \hat{V}\left( \beta, \gamma \right) \vert \varphi^{IM}_{i} \rangle =  \iint_{\beta,\gamma}   \sum_{K \in \Delta_I} F^{I}_{Kj} F^{I}_{Ki} \hat{V}\left( \beta, \gamma \right)\text{d}\tau_0 \, .
\end{equation}
In the most general case, this is done numerically.

Diagonalising the Hamiltonian provides the eigenvalues $E_\alpha$ and their associated wave functions.
\begin{equation}\label{eq:diago:psi}
    \Psi^{I}_{M\alpha} \left( \beta, \gamma, \Omega \right) =\sum_{K \in \Delta_I} \psi^{I}_{ K \alpha}\left( \beta, \gamma \right) \Phi^{I}_{MK} (\Omega) \,.
\end{equation}

\noindent Having access to the wave functions of the excited states allows us to calculate other quantities, such as electromagnetic (see Appendix \ref{sec:electro}) transitions and invariants (see Eq.\eqref{eq:invariants}).

 \section{Electromagnetic moments}\label{sec:electro}

To calculate electromagnetic transitions between states with angular momenta $I_1$ and $I_2$, we first need to evaluate the following matrix elements \cite{Kumar_1967,Libert_1999}\footnote{Here, $\langle || \cdot || \rangle$ denotes the reduced matrix element, as obtained using the Wigner–Eckart theorem \cite{Varshalovich_1988}}.

\begin{equation}
\begin{split}
   \text{B}\left(\text{E}2;\alpha^\prime I_2 \to \alpha I_1\right) =& \frac{1}{2 I_2 + 1} \vert \langle \alpha I_1 \vert \vert \hat{\mathcal{M}}(\text{E}2) \vert \vert \alpha^\prime I_2 \rangle \vert^2 \\
    =& \sum_{M_1,\mu} \vert \langle \alpha I_1 M_1 \vert \hat{\mathcal{M}}(\text{E}2,\mu) \vert \alpha^\prime I_2 M_2 \rangle \vert^2 
\end{split}
\end{equation}
\noindent where $\hat{\mathcal{M}}(\text{E}2)$ denotes the electric quadrupole moment and $\alpha,\alpha^\prime=0,1,2,\dots$ enumerate the states. In this case, it can be expressed in the laboratory frame as
$\hat{Q}^\mathrm{(lab)-el}_{2 \mu} = \sum_\nu \hat{Q}^\mathrm{(in)-el}_{2 \nu} \mathcal{D}^2_{\mu \nu}$, where

\begin{equation}
\begin{split}
    \hat{{Q}}^\mathrm{(in)-el}_{20} = \frac{1}{2}\sqrt{\frac{5}{4\pi}} e \sum_p( 2z_p^2 -x_p^2 -y_p^2 ) , & \quad\quad \hat{{Q}}^\mathrm{(in)-el}_{22} = \frac{1}{2}\sqrt{\frac{5}{4\pi}}e \sqrt{3} \sum_p(  x_p^2 -y_p^2 ) \, ,
\end{split}
\end{equation}

\noindent where the sum is restricted only to protons. We rewrite $\langle \alpha I_1 M_1 \vert \hat{\mathcal{M}}(\text{E}2,\mu) \vert \alpha^\prime I_2 M_2 \rangle$ as

\begin{equation}
\begin{split}
     \langle \Psi^{I_1}_{M_1 \, \alpha} \vert \hat{Q}^{\mathrm{(lab)-el}}_{2\mu} \vert  \Psi^{I_2}_{M_2 \, \alpha^\prime} \rangle
     &=  \sum_{\nu,K_1,K_2} \langle \psi^{I_1}_{ K_1 \, \alpha}  \vert  \hat{Q}^{\text{(in)-el}}_{2\nu} \vert \psi^{I_2}_{K_2 \, \alpha^\prime} \rangle   \langle \Phi^{I_1}_{M_1K_1}  \vert \mathcal{D}^2_{\mu\nu} \vert \Phi^{I_2}_{M_2K_2} \rangle .
\end{split}
\label{eq:BE_sum_n}
\end{equation}

\noindent Using the relation
$\int_{\Omega} \mathcal{D}^{I_1*}_{M_1 K_1} \mathcal{D}^{2}_{\mu\nu} \mathcal{D}^{I_2}_{M_2 K_2} \text{d} \Omega = \frac{8 \pi^2}{2 I_1 + 1 } C_{I_2 M_2 2\mu}^{I_1 M_1}C_{I_2 K_2 2\nu}^{ I_1 K_1 }$, where $C_{I_2 M_2 2\mu}^{I_1 M_1}$ is the Clebsch–Gordan coefficient\footnote{Here, we use the Clebsch–Gordan notation adopted in Ref.~\cite{Varshalovich_1988}.}, we obtain
 
\begin{eqnarray}
      \langle \Phi^{I_1}_{M_1K_1}  \vert \mathcal{D}^2_{\mu \nu} \vert \Phi^{I_2}_{M_2K_2} \rangle 
      &= &\frac{ A_{I_1 I_2} }{2} b_{K_1} b_{K_2}    \left[ C_{I_2 K_2 2 \nu}^{ I_1 K_1 } \right.  + (-1)^{I_1} C_{I_2 K_2 2 \nu}^{ I_1 -K_1 } + (-1)^{I_2} C_{I_2 -K_2 2 \nu}^{I_1 K_1 } \left. + (-1)^{I_1+I_2} C_{I_2 -K_2 2 \nu}^{I_1 -K_1 } \right]\,,
\end{eqnarray}

\noindent where $ b_{K} = (1+\delta_{K0})^{-1/2} $ and $ A_{I_1 I_2} = C_{I_2 M_2 2 \mu}^{I_1 M_1 } \sqrt{\frac{2I_2+1}{2 I_1+1}}$.
In the case $\nu=0$, the various terms appearing in the previous equation can be further simplified as
\begin{eqnarray}
    \sum_{K_1,K_2} \langle \psi^{I_1}_{K_1 \, \alpha} \vert \hat{Q}^{\text{(in)-el}}_{20} \vert \psi^{I_2}_{K_2 \, \alpha^\prime} \rangle \langle \Phi^{I_1}_{M_1 \, K_1} \vert \mathcal{D}^2_{\mu 0} \vert  \Phi^{I_2}_{M_2 \, K_2} \rangle = A_{I_1 I_2} \sum_{K_2 \ge 0} \langle \psi^{I_1}_{K_2 \, \alpha} \vert \hat{Q}^{\text{(in)-el}}_{20} \vert \psi^{I_2}_{K_2 \, \alpha^\prime} \rangle C_{I_2 K_2 2 0}^{I_1 K_2} .
\end{eqnarray}

\noindent For $\nu=\pm2$ we have
\begin{equation}
\begin{split}
    &\sum_{K_1,K_2} \langle \psi^{I_1}_{K_1 \, \alpha} \vert \hat{Q}^{\text{(in)-el}}_{22} \vert \psi^{I_2}_{K_2 \, \alpha^\prime} \rangle \left( \langle \Phi^{I_1}_{M_1K_1} \vert \mathcal{D}^2_{\mu2} \vert \Phi^{I_2}_{M_2K_2} \rangle + \langle \Phi^{I_1}_{M_1K_1} \vert \mathcal{D}^2_{\mu-2} \vert \Phi^{I_2}_{M_2K_2} \rangle \right)=\\
    &  A_{I_1 I_2}\sum_{K_2 \ge 0}  \left[\frac{(1+\delta_{K_2 0})}{2}\right]^{\frac{1}{2}} \left( C_{I_2 K_2 2 2}^{ I_1 K_2+2 } \langle \psi^{I_1}_{K_2+2 \, \alpha}  \vert  \hat{Q}^{\text{(in)-el}}_{22} \vert \psi^{I_2}_{ K_2 \, \alpha^\prime} \rangle +  C_{I_2 K_2+2, 2 -2}^{I_1 K_2 }\langle \psi^{I_1}_{ K_2 \, \alpha}  \vert \hat{Q}^{\text{(in)-el}}_{22} \vert \psi^{I_2}_{ K_2+2 \, \alpha^\prime} \rangle \right) .
\end{split}
\end{equation}

\noindent Putting together the above equations, we get

\begin{equation}\label{q:transition}
\begin{split}
    &\text{B}(\text{E}2; \alpha^\prime I_2 \to \alpha  I_1 )  =  \left| \sum_{K_2 \ge 0 } \left[C_{I_2 K_2 2 0}^{I_1 K_2} \langle \psi^{I_1}_{K_2 \, \alpha}  \vert  \hat{Q}^{\text{(in)-el}}_{20} \vert \psi^{I_2}_{K_2 \, \alpha^\prime} \rangle + \right. \right.\\ 
   & \left.\left. \left[\frac{(1+\delta_{K_2 0})}{2}\right]^{\frac{1}{2}}\left( C_{I_2 K_2 2 2}^{I_1 K_2+2 } \langle \psi^{I_1}_{K_2+2 \, \alpha}  \vert  \hat{Q}^{\text{(in)-el}}_{22} \vert \psi^{I_2}_{K_2 \, \alpha^\prime} \rangle +  C_{I_2 K_2+2, 2 -2}^{ I_1 K_2 }\langle \psi^{I_1}_{K_2 \, \alpha}  \vert \hat{Q}^{\text{(in)-el}}_{22} \vert \psi^{I_2}_{ K_2+2 \, \alpha^\prime} \rangle \right) \right] \right|^2 .
\end{split}
\end{equation}

The spectroscopic quadrupole moment is easily obtained as \cite{BohrMotelson_1953}

\begin{equation}
\begin{split}
    Q_{spec, I\alpha} &= \sqrt{\frac{16\pi}{5}} \frac{C_{II20}^{ II} }{ \sqrt{2I+1} } \langle \alpha I \vert \vert \mathcal{M}(\text{E}2) \vert \vert \alpha I \rangle.
\end{split} 
\end{equation}


\section{Curvilinear coordinates }\label{app:curv}
\medskip 

The transformation relating the laboratory coordinates $\{\alpha_{\lambda\mu}\}=\{ \alpha_{2-2},\alpha_{2-1},\alpha_{20},\alpha_{21},\alpha_{22}\}$ to the intrinsic ones $\left\{ q_\nu \right\} = \{ a_{0}, a_{2}, \Omega \}$ is given in Eq.~\eqref{eq:coord_combined}. 

The metric tensor associated with this transformation can be expressed in the basis in the intrinsic frame $\{\frac{\partial}{\partial q_{\mu} }\} = \{ \frac{\partial}{\partial a_0}, \frac{\partial}{\partial a_2}, \frac{i}{\hbar}\hat{L}^{\mathrm{(in)}}_x, \frac{i}{\hbar}\hat{L}^{\mathrm{(in)}}_y, \frac{i}{\hbar}\hat{L}^{\mathrm{(in)}}_z \}$, as \cite{rowe}
\begin{equation}\label{eq:curv_g}
    g_{ij} = \sum_\mu \frac{\partial \alpha_{2\mu}^*}{\partial q_i} \frac{\partial \alpha_{2\mu}}{\partial q_j} \, . 
\end{equation}

\noindent The metric tensor is symmetric, and its components can be computed explicitly using the following relations
\begin{eqnarray}
    \frac{\partial \alpha_{2\mu}}{\partial a_0} &=&  \mathcal{D}^{2(+)}_{\mu0}, \\
    \frac{\partial \alpha_{2\mu}}{\partial a_2} &=&  \mathcal{D}^{2(+)}_{\mu2}, \\
   \hat{L}_{k}^{\mathrm{(in)}} \alpha_{2\mu} &=&  \hat{L}_{k}^{\mathrm{(in)}} \mathcal{D}^{2(+)}_{\mu0} a_0 + \hat{L}_{k}^{\mathrm{(in)}} \mathcal{D}^{2(+)}_{\mu2} a_2  ,
\end{eqnarray}
The explicit expressions of $\hat{L}_{k}^{\mathrm{(in)}}$ in terms of the Euler angles can be found in Ref.~\cite{Varshalovich_1988}. Here, we provide only their action on the Wigner functions.

\begin{equation}\label{eq:L_in}
\begin{aligned}
    &\hat{L}^{\mathrm{(in)}}_{x} \mathcal{D}^\lambda_{\mu\nu} = \frac{\hbar}{2} \left( \sqrt{\lambda(\lambda+1)-\nu(\nu-1)} \mathcal{D}^\lambda_{\mu\nu-1} + \sqrt{\lambda(\lambda+1)-\nu(\nu+1)} \mathcal{D}^\lambda_{\mu\nu+1}  \right),  \\
    &\hat{L}^{\mathrm{(in)}}_{y} \mathcal{D}^\lambda_{\mu\nu}  = \frac{-i\hbar}{2} \left( \sqrt{\lambda(\lambda+1)-\nu(\nu-1)} \mathcal{D}^\lambda_{\mu\nu-1}  -\sqrt{\lambda(\lambda+1)-\nu(\nu+1)} \mathcal{D}^\lambda_{\mu\nu+1}  \right),  \\
    &\hat{L}^{\mathrm{(in)}}_{z}  \mathcal{D}^\lambda_{\mu\nu}  = \hbar \nu \mathcal{D}^\lambda_{\mu\nu} ,\\
    &\hat{L}^{2\mathrm{(in)}}  \mathcal{D}^\lambda_{\mu\nu}  = \hbar^2 \lambda (\lambda+1) \mathcal{D}^\lambda_{\mu\nu}, \\ 
    &\hat{L}^{\mathrm{(in)}}_{\pm} \mathcal{D}^\lambda_{\mu\nu} = \hbar \sqrt{\lambda(\lambda+1)-\nu(\nu\mp1)} \mathcal{D}^\lambda_{\mu\nu\mp1} .
\end{aligned} 
\end{equation}

We recall that the operators satisfy the relation $[\hat{L}^{\mathrm{(in)}}_x,\hat{L}^{\mathrm{(in)}}_y] =-i\hbar \hat{L}^{\mathrm{(in)}}_z$ \cite{BohrMottelson_1975}, in the main text we dropped the (in) index, to make the notation lighter.
After some calculations, the metric tensor can be written in the following compact form
\begin{eqnarray}
    g&=& 
    \begin{pmatrix}
        1&0&0&0 &0 \\
        0&1&0&0 &0 \\
        0&0& 4a_x^2 &0&0\\
        0&0&0 & 4a_y^2&0 \\
        0&0&0&0 & 4a_z^2 \\
    \end{pmatrix} ,
\end{eqnarray} 
\noindent where 
\begin{equation}\label{eq:axayaz}
    \begin{array}{llll}
         a_x = & -\frac{1}{2} \left( \sqrt{3} a_0 + a_2 \right) & = -\frac{1}{2} \left(\sqrt{3} \beta \cos \gamma + \beta  \sin \gamma \right) & = \beta \sin (\gamma-\frac{2 \pi}{3}), \\
         a_y = &  \frac{1}{2} \left(\sqrt{3} a_0 - a_2 \right) & = \frac{1}{2} \left(\sqrt{3} \beta \cos \gamma - \beta  \sin \gamma  \right) & = \beta \sin (\gamma + \frac{2 \pi}{3}), \\
         a_z = & a_2  & = \beta \sin \gamma  &.
    \end{array} 
\end{equation}
From the metric tensor, we calculate the Jacobian of the transformation as $ \sqrt{\vert\text{det}(g)\vert} = \vert 2a_2(3a_0^2-a_2^2) \vert =2\beta^3 \vert \sin(3\gamma) \vert$. Note that the factor of 2 is usually omitted, since it is simply absorbed into the normalisation factor of the wave function. In this work, we therefore follow the usual convention and discard this factor. The transformation to the coordinates $(\beta,\gamma)$ introduces an additional factor of $\beta$ in the Jacobian, since $\text{d}a_0 \text{d}a_2=\beta \text{d}\beta \text{d}\gamma$, while the transformation to Euler angles introduces a factor of $\sin\theta$. Thus, the final measure is given by 

\begin{eqnarray}
   \text{d} \tau&=& \beta^4 \vert\sin(3\gamma)\vert \, \sin \theta \, \text{d}\beta \, \text{d}\gamma \, \text{d}\phi \, \text{d} \theta \, \text{d} \psi\,,
\end{eqnarray}

since the integration over the Euler angles is carried out analytically, it is convenient to introduce the following volume elements

\begin{eqnarray}
   \text{d} \Omega&=&  \sin \theta \, \text{d}\phi \, \text{d} \theta \, \text{d} \psi\ \label{eq:DOmega]}\,,\\
   \text{d} \tau_0&=&\beta^4 \vert\sin(3\gamma)\vert \, \text{d}\beta \, \text{d}\gamma \label{eq:dtau0}\,.
\end{eqnarray}

\section{Numerical details}\label{app:num:det}
It is well known that the harmonic oscillator (HO) basis exhibits very slow convergence for absolute quantities, such as the total nuclear binding energy \cite{sch12}. Of course, increasing the number of basis states leads to a progressively higher computational cost, which quickly becomes prohibitive for triaxial calculations \cite{sch12b}, especially when dealing with finite-range interactions.
Since most BH results depend more on differences and relative changes than on absolute values, we can expect a faster rate of convergence.

\begin{figure}[h!]
    \centering
          \includegraphics[width=0.6\textwidth]{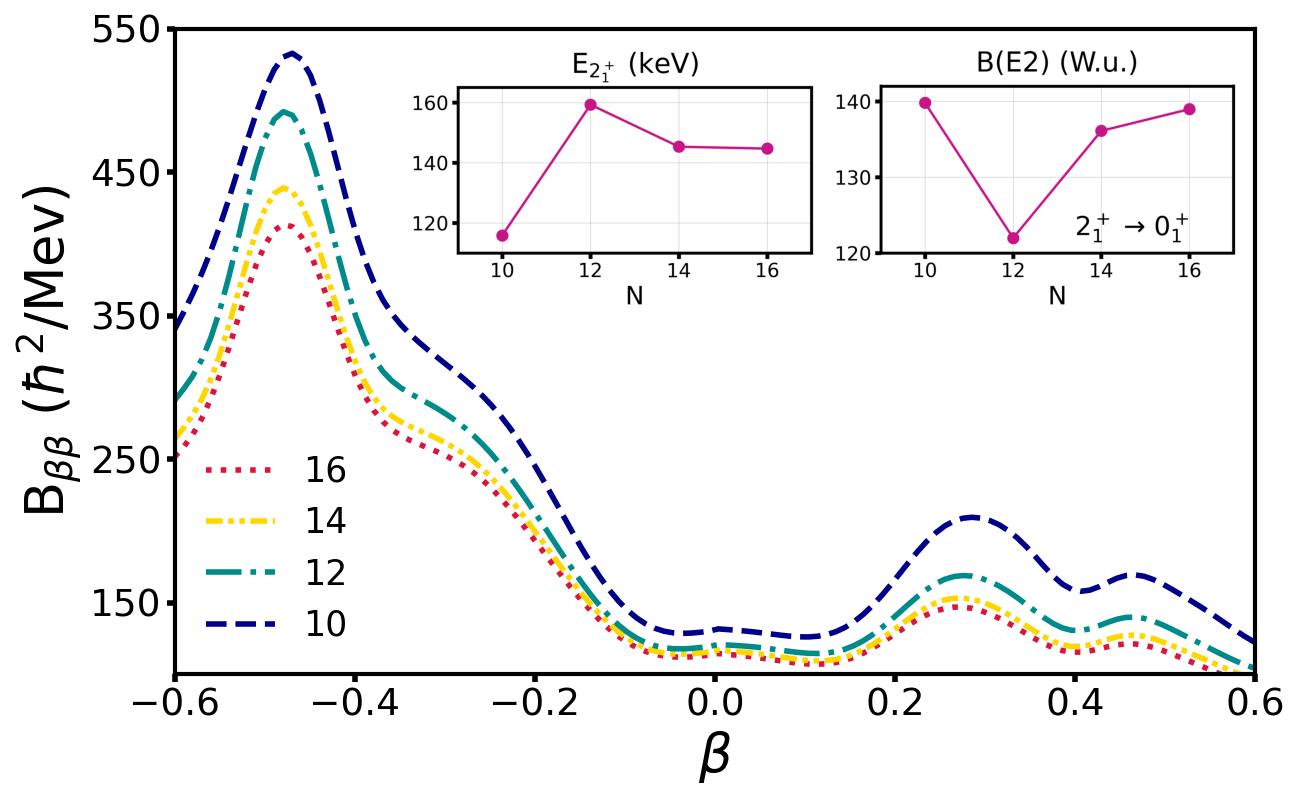}
          \caption{Evolution of the axial mass parameter $B_{\beta\beta}$ in $^{154}$Gd calculated using the Gogny D3G3M interaction and various cutoffs in the size of the HO basis. In the inset, we show the evolution of the first $2^+$ state and its electromagnetic transition to the ground state as a function of the number of shells. See text for details.}
    \label{fig:convergence}
\end{figure}

In Ref.~\cite{nikvsic2009beyond}, the authors showed that the structure of the PES converges for $N_{shell}=14$.
In Fig.~\ref{fig:convergence}, we examine the convergence of the mass parameter $B_{\beta\beta}$ in $^{154}$Gd, calculated using the D3G3M interaction in the axial representation.
We see that the major peak at $\beta\approx-0.5$ changes significantly when going from 10 to 14 shells, but then remains almost unchanged when going from 14 to 16 shells.

In the inset of the same figure, we show the evolution of the first $2^+$ state and its corresponding transition to the ground state as a function of the number of shells. We observe that the convergence path is not monotonic, and thus it is important to check the stability of the calculations; otherwise, we could overestimate or underestimate the results.
We conclude that $N_{shell}=14$ represents a good compromise between the accuracy of the results and the computational cost.

In order to perform the BH calculations, we first determine all relevant quantities, such as collective masses, moments of inertia, and potential energy surfaces, in the first sextant of the $(\beta,\gamma)$ plane on a uniform grid. For practical applications, we employ a mesh spacing of $0.05$ in $\beta$, with $\beta_\mathrm{max}=0.95$, and $6^\circ$ in $\gamma$.
Using the symmetries of the mass parameters and moments of inertia defined in Tab.~\ref{tab:B_symmetries}, we can then easily reconstruct these quantities over the full integration domain.
\begin{table}[h]
    \centering
    \caption{Transformations of the mass parameters into the different sextants.}
    \label{tab:B_symmetries}
    \begin{tabular}{c | c c c }
        $\gamma$ & $\gamma+\frac{\pi}{3}$ & $\gamma+\frac{2\pi}{3}$ & $-\gamma$ \\ \hline\hline
        $B_{\beta\beta}\left(\beta,\gamma\right)$   & $B_{\beta\beta}\left(\beta,\frac{\pi}{3}-\gamma\right)$   & $B_{\beta\beta}\left(\beta,\gamma\right)$  & $B_{\beta\beta}\left(\beta,\gamma\right)$ \\
        $B_{\beta\gamma}\left(\beta,\gamma\right)$  & $-B_{\beta\gamma}\left(\beta,\frac{\pi}{3}-\gamma\right)$  & $B_{\beta\gamma}\left(\beta,\gamma\right)$ & $-B_{\beta\gamma}\left(\beta,\gamma\right)$ \\
        $B_{\gamma\gamma}\left(\beta,\gamma\right)$ & $B_{\gamma\gamma}\left(\beta,\frac{\pi}{3}-\gamma\right)$ & $B_{\gamma\gamma}\left(\beta,\gamma\right)$  & $B_{\gamma\gamma}\left(\beta,\gamma\right)$  \\
        $B_x\left(\beta,\gamma\right)$ & $B_z\left(\beta,\frac{\pi}{3}-\gamma\right)$ & $B_z\left(\beta,\gamma\right)$ & $B_y\left(\beta,\gamma\right)$\\
        $B_y\left(\beta,\gamma\right)$ & $B_y\left(\beta,\frac{\pi}{3}-\gamma\right)$ & $B_x\left(\beta,\gamma\right)$ & $B_x\left(\beta,\gamma\right)$\\
        $B_z\left(\beta,\gamma\right)$ & $B_x\left(\beta,\frac{\pi}{3}-\gamma\right)$ & $B_y\left(\beta,\gamma\right)$ & $B_z\left(\beta,\gamma\right)$\\
    \end{tabular}
\end{table}

To compute the matrix elements discussed in Appendix~\ref{sec:mel}, we use Gauss-Laguerre quadrature in the $\beta$ direction and Gauss-Legendre quadrature in the $\gamma$ direction, with 30 and 40 integration points, respectively. Reducing these numbers to 20 and 30, respectively, changes the energy of the first $2^+$ state in $^{154}$Gd by less than 2 keV, thus demonstrating that our integration methods do not affect the accuracy of the results.

Since the quadrature points do not generally coincide with the grid points at which the HFB calculations are performed, we employ radial basis function (RBF) interpolation \cite{buhmann2000radial,virtanen2020scipy}. RBF interpolation reconstructs a smooth function from scattered data by placing a radially symmetric kernel at each sample location. A key advantage of this method is that the interpolation is performed globally over the entire dataset, without imposing a predefined functional form. This provides greater flexibility and allows for the exclusion of non-converged HFB points, thereby improving the robustness of the procedure.

Once the RBF kernel has been trained, the resulting function can be efficiently evaluated at arbitrary points in the $(\beta,\gamma)$ plane. Particular care must be taken in the choice of the quadrature points in the $\beta$ direction. In practice, we always verify that the zeros of the generalised Laguerre polynomials lie within the region covered by the available HFB data in order to avoid unreliable extrapolations.

To further assess the robustness of the method, we studied the sensitivity of the results to the random removal of HFB training points. The mean and standard deviation of the $E(2^+_1)$ and $B(E2; 2^+_1 \to 0^+_1)$ values obtained from 100 calculations for different percentages of randomly removed data points are given in Tab.~\ref{tab:point}. Using the full dataset for $^{154}$Gd with D3G3M (\emph{i.e.}, 220 points), we obtain, for the first $2^+$ state (after rescaling), $E_{2^+}=143.5$~keV and $B(E2)=136.80$~W.u. Randomly removing $1\%$ of the training points yields $E_{2^+}=143.1$~keV and $B(E2)=137.0$~W.u. The mean  remains essentially unchanged even when up to $5\%$ of the points are removed, provided that the discarded points are sparsely distributed over the sextant rather than clustered in a localised region. When $10\%$ of the points are removed, we obtain $E_{2^+}=141.8$~keV and $B(E2)=137.8$~W.u which remain quite close to the complete result although we observe an increase in the standard deviation on both quantities. These tests indicate that the RBF interpolation procedure is robust and, compared to a simple local interpolation scheme, provides sufficient flexibility to accommodate missing HFB points arising from non-converged calculations.

\begin{table}
    \centering
        \caption{Mean ($\bar{\mu}$) and standard deviation ($\sigma$) of $E(2^+_1)$ and $B(E2; 2^+_1 \to 0^+_1)$ obtained from 100 calculations using the D3G3M parametrization for $^{154}\mathrm{Gd}$, for different percentages of randomly removed data points.
}
    \label{tab:point}
    \begin{tabular}{c|ccc}
        & $1 \%$ & $5 \%$ & $10 \%$ \\ \hline
        $\bar{\mu}_{E_{2^+}} (keV) $ & $143.1$ & $142.4$ & $141.8$ \\ 
        $\sigma_{E_{2^+}} (keV)$ & $1.4$ & $3.2 $ & $5.4$ \\  
        $\bar{\mu}_{B(E2)}$ (W.u.) & $137.0$ & $137.6$ & $137.8$ \\ 
        $\sigma_{B(E2)}$ (W.u.) & $1.1$ & $3.6$ & $5.70$ \\  
    \end{tabular}

\end{table}

\bibliography{biblio.bib}

@book{hamermesh2012group,
  title={Group theory and its application to physical problems},
  author={Hamermesh, Morton},
  year={2012},
  publisher={Courier Corporation}
}

@article{nik11,
  title={Relativistic nuclear energy density functionals: Mean-field and beyond},
  author={Nik{\v{s}}i{\'c}, Tamara and Vretenar, Dario and Ring, Peter},
  journal={Progress in Particle and Nuclear Physics},
  volume={66},
  number={3},
  pages={519--548},
  year={2011},
  publisher={Elsevier}
}

@article{moller2012new,
  title={New finite-range droplet mass model and equation-of-state parameters},
  author={M{\"o}ller, Peter and Myers, William D and Sagawa, Hiroyuki and Yoshida, Satoshi},
  journal={Physical Review Letters},
  volume={108},
  number={5},
  pages={052501},
  year={2012},
  publisher={APS}
}

@article{pro04,
  title={A self-consistent approach to the quadrupole dynamics of medium heavy nuclei},
  author={Pr{\'o}chniak, L and Quentin, P and Samsoen, D and Libert, J},
  journal={Nuclear Physics A},
  volume={730},
  number={1-2},
  pages={59--79},
  year={2004},
  publisher={Elsevier}
}

@article{suz24,
  title = {Shape transition and coexistence in Te isotopes studied with the quadrupole collective Hamiltonian based on a relativistic energy density functional},
  author = {Suzuki, K. and Nomura, K.},
  journal = {Physical Review C},
  volume = {110},
  issue = {6},
  pages = {064304},
  numpages = {16},
  year = {2024},
  month = {Dec},
  publisher = {American Physical Society},
  doi = {10.1103/PhysRevC.110.064304},
}

@article{pro09,
doi = {10.1088/0954-3899/36/12/123101},
year = {2009},
month = {oct},
publisher = {},
volume = {36},
number = {12},
pages = {123101},
author = {L Próchniak and S G Rohoziński},
title = {Quadrupole collective states within the Bohr collective Hamiltonian},
journal = {Journal of Physics G}
}

@book{Bohr_1952,
  title={The coupling of nuclear surface oscillations to the motion of individual nucleons},
  author={Bohr, Aage},
  year={1952},
  publisher={Munksgaard}
}

@article{iachello2000dynamic,
  title={Dynamic symmetries at the critical point},
  author={Iachello, F},
  journal={Physical Review Letters},
  volume={85},
  number={17},
  pages={3580},
  year={2000},
  publisher={APS}
}

@article{hin13,
  title={Low-energy collective modes of deformed superfluid nuclei within the finite-amplitude method},
  author={Hinohara, Nobuo and Kortelainen, Markus and Nazarewicz, Witold},
  journal={Physical Review C},
  volume={87},
  number={6},
  pages={064309},
  year={2013},
  publisher={APS}
}

@article{avo13,
  title={Efficient calculation for the quasiparticle random-phase approximation matrix},
  author={Avogadro, Paolo and Nakatsukasa, Takashi},
  journal={Physical Review C},
  volume={87},
  number={1},
  pages={014331},
  year={2013},
  publisher={APS}
}

@article{car12,
  title={Collective vibrational states within the fast iterative quasiparticle random-phase approximation method},
  author={Carlsson, BG and Toivanen, Jussi and Pastore, Alessandro},
  journal={Physical Review C},
  volume={86},
  number={1},
  pages={014307},
  year={2012},
  publisher={APS}
}

@article{cho24,
  title={Note sur une m{\'e}thode de r{\'e}solution des {\'e}quations normales provenant de l’application de la m{\'e}thode des moindres carr{\'e}sa un syteme d’{\'e}quations lin{\'e}aires en nombre inf{\'e}rieura celui des inconnues (Published six years after Cholesky’s death by Benoit)},
  author={Cholesky, Andr{\'e}-Louis},
  journal={Bull. G{\'e}od{\'e}sique},
  volume={2},
  pages={67--77},
  year={1924}
}

@book{Varshalovich_1988,
  author    = {D. A. Varshalovich and A. N. Moskalev and V. K. Khersonskii},
  title     = {Quantum Theory of Angular Momentum},
  year      = {1988},
  publisher = {World Scientific}
}

@phdthesis{fro21,
  title={Ab initio description of doubly-open shell nuclei via a novel resolution method of the quantum many-body problem},
  author={Frosini, Mikael},
  year={2021},
  school={Universit{\'e} Paris-Saclay}
}

@article{virtanen2020scipy,
  title={SciPy 1.0: fundamental algorithms for scientific computing in Python},
  author={Virtanen, Pauli and Gommers, Ralf and Oliphant, Travis E and Haberland, Matt and Reddy, Tyler and Cournapeau, David and Burovski, Evgeni and Peterson, Pearu and Weckesser, Warren and Bright, Jonathan and others},
  journal={Nature methods},
  volume={17},
  number={3},
  pages={261--272},
  year={2020},
  publisher={Nature Publishing Group US New York}
}

@article{buhmann2000radial,
  title={Radial basis functions},
  author={Buhmann, Martin Dietrich},
  journal={Acta numerica},
  volume={9},
  pages={1--38},
  year={2000},
  publisher={Cambridge university press}
}

@article{batail2025three,
  title={A three-ranged Gogny-HFB nuclear mass model suited for astrophysical applications},
  author={Batail, Lysandra and Goriely, St{\'e}phane and P{\'e}ru, S and Hilaire, St{\'e}phane and Davesne, Dany and Pastore, Alessandro},
  journal={Physics Letters B},
  pages={139719},
  volume = {868},
  issn = {0370-2693},
  year={2025},
  publisher={Elsevier}
}

@article{sun25,
  title={Moments of inertia of rare-earth nuclei and the nuclear time-odd mean fields within exact solutions of the adiabatic theory},
  author={Sun, Xuwei and Dobaczewski, Jacek and Kortelainen, Markus and Sadhukhan, Jhilam and S{\'a}nchez-Fern{\'a}ndez, Adri{\'a}n and Wibowo, Herlik},
  journal={Physics Letters B},
  volume={868},
  pages={139685},
  year={2025},
  publisher={Elsevier}
}

@article{bel61,
  title={Concerning the calculation of the nuclear moment of inertia},
  author={Beliaev, ST},
  journal={Nuclear Physics},
  volume={24},
  number={2},
  pages={322--325},
  year={1961},
  publisher={Elsevier}
}

@article{ing56,
  title={Nuclear moments of inertia due to nucleon motion in a rotating well},
  author={Inglis, DR},
  journal={Physical Review},
  volume={103},
  number={6},
  pages={1786},
  year={1956},
  publisher={APS}
}

@article{li12,
  title={Efficient method for computing the Thouless-Valatin inertia parameters},
  author={Li, ZP and Nik{\v{s}}i{\'c}, Tamara and Ring, Peter and Vretenar, Dario and Yao, JM and Meng, Jie},
  journal={Physical Review C},
  volume={86},
  number={3},
  pages={034334},
  year={2012},
  publisher={APS}
}

@article{davesne2025isovector,
  title={Isovector properties of effective finite-range nuclear interactions},
  author={Davesne, D and Navarro, J and Pastore, A and Lallouet, Y},
  journal={The European Physical Journal A},
  volume={61},
  number={2},
  pages={1--23},
  year={2025},
  publisher={Springer}
}

@article{sellahewa2014isovector,
  title={Isovector properties of the Gogny interaction},
  author={Sellahewa, Roshan and Rios, Arnau},
  journal={Physical Review C},
  volume={90},
  number={5},
  pages={054327},
  year={2014},
  publisher={APS}
}

@incollection{sat13,
  title={Microscopic description of large-amplitude collective motions with local QRPA inertial masses},
  author={Sato, K and Hinohara, N and Nakatsukasa, T and Matsuo, M and Matsuyanagi, K},
  booktitle={Nuclear Structure Problems},
  pages={126--129},
  year={2013},
  publisher={World Scientific}
}

@article{Sad13,
  title = {Spontaneous fission lifetimes from the minimization of self-consistent collective action},
  author = {Sadhukhan, Jhilam and Mazurek, K. and Baran, A. and Dobaczewski, J. and Nazarewicz, W. and Sheikh, J. A.},
  journal = {Physical Review C},
  volume = {88},
  issue = {6},
  pages = {064314},
  numpages = {5},
  year = {2013},
  month = {Dec},
  publisher = {American Physical Society},
  doi = {10.1103/PhysRevC.88.064314},
}

@article{bar78,
  title={An adiabatic time-dependent Hartree-Fock theory of collective motion in finite systems},
  author={Baranger, M and Veneroni, M},
  journal={Annals of Physics},
  volume={114},
  number={1-2},
  pages={123--200},
  year={1978},
  publisher={Elsevier}
}

@misc{ENSDF,
  howpublished = {\url{https://www.nndc.bnl.gov/ensdf/}},
}

@article{llewellyn2020establishing,
  title={Establishing the maximum collectivity in highly deformed N= Z nuclei},
  author={Llewellyn, RDO and Bentley, MA and Wadsworth, R and Iwasaki, H and Dobaczewski, J and de Angelis, G and Ash, J and Bazin, D and Bender, PC and Cederwall, Bo and others},
  journal={Physical review letters},
  volume={124},
  number={15},
  pages={152501},
  year={2020},
  publisher={APS}
}

@phdthesis{mui21,
  title={Microscopic modelling of collective quadrupole excitations of nuclei},
  author={Muir, David},
  year={2021},
  school={University of York}
}

@misc{del26,
  author       = {Delaroche, J.-P. and Girod, M. and Libert, J. and Goutte, H. and Hilaire, S. and P{\'e}ru, S. and Pillet, N. and Bertsch, G. F.},
  title        = {{HFB+5DCH} results: tabulated nuclear properties},
  howpublished = {\url{https://www-phynu.cea.fr/science_en_ligne/carte_potentiels_microscopiques/tables/HFB-5DCH-table_eng.htm}},
  note         = {Accessed November 26, 2026}
}

@article{pro99b,
title = {Collective quadrupole excitations in the 50 < Z, N < 82 nuclei with the general Bohr Hamiltonian},
journal = {Nuclear Physics A},
volume = {648},
number = {3},
pages = {181-202},
year = {1999},
issn = {0375-9474},
doi = {https://doi.org/10.1016/S0375-9474(99)00023-8},
author = {L. Prochniak and K. Zajac and K. Pomorski and S.G. Rohozinski and J. Srebrny},
}

@article{Corrigan_1976,
  title={Exact solution of the quadrupole surface vibration Hamiltonian in body-fixed coordinates},
  author={Corrigan, TM and Margetan, FJ and Williams, SA},
  journal={Physical Review C},
  volume={14},
  number={6},
  pages={2279},
  year={1976},
  publisher={APS}
}

@article{wil56,
  title={Surface oscillations in even-even nuclei},
  author={Wilets, Lawrence and Jean, Maurice},
  journal={Physical Review},
  volume={102},
  number={3},
  pages={788},
  year={1956},
  publisher={APS}
}

@article{smi76,
  title={Solution of the Bohr Hamiltonian for J≤ 20},
  author={Smith, BC and Ko, C-M},
  journal={Canadian Journal of Physics},
  volume={54},
  number={18},
  pages={1862--1865},
  year={1976},
  publisher={NRC Research Press Ottawa, Canada}
}

@article{roh77,
  title={Microscopic dynamic calculations of collective states in Xenon and Barium isotopes},
  author={Rohozi{\'n}ski, SG and Dobaczewski, J and Nerlo-Pomorska, B and Pomorski, K and Srebrny, J},
  journal={Nuclear Physics A},
  volume={292},
  number={1-2},
  pages={66--87},
  year={1977},
  publisher={Elsevier}
}

@article{Kumar_1967,
title = {Complete numerical solution of Bohr's collective Hamiltonian},
journal = {Nuclear Physics A},
volume = {92},
number = {3},
pages = {608-652},
year = {1967},
issn = {0375-9474},
doi = {https://doi.org/10.1016/0375-9474(67)90636-7},
author = {Krishna Kumar and Michel Baranger}
}

@article{per24,
doi = {10.1088/1361-6471/adb50b},
year = {2025},
month = {mar},
publisher = {IOP Publishing},
volume = {52},
number = {3},
pages = {033001},
author = {Coello Pérez, E A and Papenbrock, T},
title = {Effective field theories for collective excitations of atomic nuclei},
journal = {Journal of Physics G},
}

@article{was24,
  title={Five-dimensional collective Hamiltonian with improved inertial functions},
  author={Washiyama, Kouhei and Hinohara, Nobuo and Nakatsukasa, Takashi},
  journal={Physical Review C},
  volume={109},
  number={5},
  pages={L051301},
  year={2024},
  publisher={APS}
}

@article{bar11,
  title={Quadrupole collective inertia in nuclear fission: cranking approximation},
  author={Baran, Andrzej and Sheikh, JA and Dobaczewski, J and Nazarewicz, Witold and Staszczak, Andrzej},
  journal={Physical Review C},
  volume={84},
  number={5},
  pages={054321},
  year={2011},
  publisher={APS}
}

@article{pil93,
  title={Coupling of nuclear shape oscillations with pairing vibrations},
  author={Pi{\l}at, Stanis{\l}aw and Pomorski, Krzysztof},
  journal={Nuclear Physics A},
  volume={554},
  number={3},
  pages={413--420},
  year={1993},
  publisher={Elsevier}
}

@article{xia20,
  title={Coupling of shape and pairing vibrations in a collective Hamiltonian based on nuclear energy density functionals},
  author={Xiang, J and Li, ZP and Nik{\v{s}}i{\'c}, Tamara and Vretenar, Dario and Long, WH},
  journal={Physical Review C},
  volume={101},
  number={6},
  pages={064301},
  year={2020},
  publisher={APS}
}

@article{xia2024,
  title={Coupling of shape and pairing vibrations in a collective Hamiltonian based on nuclear energy density functionals. II. Low-energy excitation spectra of triaxial nuclei},
  author={Xiang, J and Li, ZP and Nik{\v{s}}i{\'c}, T and Vretenar, D and Long, WH and Wu, XY},
  journal={Physical Review C},
  volume={109},
  number={4},
  pages={044319},
  year={2024},
  publisher={APS}
}

@article{nom21,
  title={Coupling of pairing and triaxial shape vibrations in collective states of $\gamma$-soft nuclei},
  author={Nomura, Kosuke and Vretenar, Dario and Li, ZP and Xiang, J},
  journal={Physical Review C},
  volume={103},
  number={5},
  pages={054322},
  year={2021},
  publisher={APS}
}

@article{goz85,
  title={Collective pairing Hamiltonian in the GCM approximation},
  author={G{\'o}{\'z}d{\'z}, A and Pomorski, K and Brack, Matthias and Werner, E},
  journal={Nuclear Physics A},
  volume={442},
  number={1},
  pages={50--67},
  year={1985},
  publisher={Elsevier}
}

@article{goriely2009first,
  title={First gogny-hartree-fock-bogoliubov nuclear mass model},
  author={Goriely, Stanislas and Hilaire, St{\'e}phane and Girod, Michel and P{\'e}ru, S},
  journal={Physical review letters},
  volume={102},
  number={24},
  pages={242501},
  year={2009},
  publisher={APS}
}

@article{pomorski2003nuclear,
  title={Nuclear liquid-drop model and surface-curvature effects},
  author={Pomorski, K and Dudek, Jerzy},
  journal={Physical Review C},
  volume={67},
  number={4},
  pages={044316},
  year={2003},
  publisher={APS}
}

@article{chappert2008towards,
  title={Towards a new Gogny force parameterization: Impact of the neutron matter equation of state},
  author={Chappert, F and Girod, M and Hilaire, S},
  journal={Physics Letters B},
  volume={668},
  number={5},
  pages={420--424},
  year={2008},
  publisher={Elsevier}
}

@article{ryssens2022skyrme,
  title={Skyrme--Hartree--Fock--Bogoliubov mass models on a 3D mesh: II. Time-reversal symmetry breaking},
  author={Ryssens, Wouter and Scamps, Guillaume and Goriely, Stephane and Bender, Michael},
  journal={The European Physical Journal A},
  volume={58},
  number={12},
  pages={246},
  year={2022},
  publisher={Springer}
}

@article{baran1996gd,
  title={Gd isotope systematics with Skyrme and $\delta$-pairing forces},
  author={Baran, Andrzej and H{\"o}henberger, Walter},
  journal={Physical Review C},
  volume={53},
  number={4},
  pages={1571},
  year={1996},
  publisher={APS}
}

@article{boh76,
  title={Rotational motion in nuclei},
  author={Bohr, Aage},
  journal={Reviews of Modern Physics},
  volume={48},
  number={3},
  pages={365},
  year={1976},
  publisher={APS}
}

@article{sch12b,
  title={Solution of the Skyrme--Hartree--Fock--Bogolyubov equations in the Cartesian deformed harmonic-oscillator basis.:(VII) hfodd (v2. 49t): A new version of the program},
  author={Schunck, Nicolas and Dobaczewski, J and McDonnell, J and Satu{\l}a, W and Sheikh, JA and Staszczak, Andrzej and Stoitsov, M and Toivanen, P},
  journal={Computer Physics Communications},
  volume={183},
  number={1},
  pages={166--192},
  year={2012},
  publisher={Elsevier}
}

@article{sch12,
  title={Error analysis in nuclear density functional theory},
  author={Schunck, Nicolas and McDonnell, Jordan D and Sarich, Jason and Wild, Stefan M and Higdon, Dave},
  journal={Journal of Physics G},
  volume={42},
  number={3},
  pages={034024},
  year={2015},
  publisher={IOP Publishing}
}

@article{nikvsic2009beyond,
  title={Beyond the relativistic mean-field approximation. III. Collective Hamiltonian in five dimensions},
  author={Nik{\v{s}}i{\'c}, Tamara and Li, ZP and Vretenar, Dario and Pr{\'o}chniak, L and Meng, Jie and Ring, Peter},
  journal={Physical Review C},
  volume={79},
  number={3},
  pages={034303},
  year={2009},
  publisher={APS}
}

@article{bloch1954theory,
  title={Theory of nuclear level density},
  author={Bloch, Claude},
  journal={Physical Review},
  volume={93},
  number={5},
  pages={1094},
  year={1954},
  publisher={APS}
}

@article{zhao20,
  title={Microscopic model for the collective enhancement of nuclear level densities},
  author={Zhao, Jie and Nik{\v{s}}i{\'c}, Tamara and Vretenar, Dario},
  journal={Physical Review C},
  volume={102},
  number={5},
  pages={054606},
  year={2020},
  publisher={APS}
}

@article{dob16,
  title={Consistent quadrupole-octupole collective model},
  author={Dobrowolski, Artur and Mazurek, Kamil and G{\'o}{\'z}d{\'z}, Andrzej},
  journal={Physical Review C},
  volume={94},
  number={5},
  pages={054322},
  year={2016},
  publisher={APS}
}

@article{thou62,
  title={Time-dependent Hartree-Fock equations and rotational states of nuclei},
  author={Thouless, DJ and Valatin, JG},
  journal={Nuclear Physics},
  volume={31},
  pages={211--230},
  year={1962},
  publisher={Elsevier}
}

@article{bonatsos2004sequence,
  title={Sequence of potentials interpolating between the U (5) and E (5) symmetries},
  author={Bonatsos, Dennis and Lenis, D and Minkov, N and Raychev, PP and Terziev, PA},
  journal={Physical Review C},
  volume={69},
  number={4},
  pages={044316},
  year={2004},
  publisher={APS}
}

@article{Fortunato_2005,
  title={Solutions of the Bohr Hamiltonian, a compendium},
  author={Fortunato, Lorenzo},
  journal={The European Physical Journal A},
  volume={26},
  number={Suppl 1},
  pages={1--30},
  year={2005},
  publisher={Springer}
}

@article{delaroche2024investigations,
  title={Investigations of electric monopole transitions in medium-mass to heavy nuclei: Beyond mean field calculations with the Gogny force},
  author={Delaroche, J-P and Libert, J and Girod, M and Deloncle, I and Dupuis, M},
  journal={Physical Review C},
  volume={109},
  number={1},
  pages={014320},
  year={2024},
  publisher={APS}
}

@article{yul99,
  title={Mass parameters for large amplitude collective motion: A perturbative microscopic approach},
  author={Yuldashbaeva, E Kh and Libert, J and Quentin, P and Girod, M},
  journal={Physics Letters B},
  volume={461},
  number={1-2},
  pages={1--8},
  year={1999},
  publisher={Elsevier}
}

@article{lib16,
  title={Five-dimensional collective Hamiltonian with the Gogny force: An ongoing saga},
  author={Libert, J and Delaroche, J-P and Girod, M},
  journal={The European Physical Journal A},
  volume={52},
  number={7},
  pages={197},
  year={2016},
  publisher={Springer}
}

@article{gir09,
  title={The role of triaxiality for the coexistence and evolution of shapes in light krypton isotopes},
  author={Girod, M and Delaroche, J-P and G{\"o}rgen, A and Obertelli, A},
  journal={Physics Letters B},
  volume={676},
  number={1-3},
  pages={39--43},
  year={2009},
  publisher={Elsevier}
}

@article{Gar08,
  title = {Relation between $E(5)$ models and the interacting boson model},
  author = {Garc{\'{\i}}a-Ramos, Jos{\'e} Enrique and Arias, Jos{\'e} M.},
  journal = {Physical Review C},
  volume = {77},
  issue = {5},
  pages = {054307},
  numpages = {9},
  year = {2008},
  month = {May},
  publisher = {American Physical Society},
  doi = {10.1103/PhysRevC.77.054307},
}

@article{cap02,
  title={Finite well solution for the E (5) Hamiltonian},
  author={Caprio, MA},
  journal={Physical Review C},
  volume={65},
  number={3},
  pages={031304},
  year={2002},
  publisher={APS}
}

@article{delaroche2010structure,
  title={Structure of even-even nuclei using a mapped collective Hamiltonian and the D1S Gogny interaction},
  author={Delaroche, J-P and Girod, M and Libert, J and Goutte, H and Hilaire, S and P{\'e}ru, S and Pillet, N and Bertsch, GF},
  journal={Physical Review C},
  volume={81},
  number={1},
  pages={014303},
  year={2010},
  publisher={APS}
}

@book{abr68,
  title={Handbook of mathematical functions with formulas, graphs, and mathematical tables},
  author={Abramowitz, Milton and Stegun, Irene A},
  volume={55},
  year={1968},
  publisher={US Government printing office}
}

@book{cornwell1997group,
  title={Group theory in physics: An introduction},
  author={Cornwell, John F},
  volume={1},
  year={1997},
  publisher={Academic press}
}

@article{Libert_1999,
  title={Microscopic descriptions of superdeformed bands with the Gogny force: Configuration mixing calculations in the A~ 190 mass region},
  author={Libert, J and Girod, M and Delaroche, J-P},
  journal={Physical Review C},
  volume={60},
  number={5},
  pages={054301},
  year={1999},
  publisher={APS}
}

@article{Libert_Quentin_1982,
  title={Self-consistent description of heavy nuclei. II. Spectroscopic properties of some odd nuclei},
  author={Libert, J and Meyer, M and Quentin, Ph},
  journal={Physical Review C},
  volume={25},
  number={1},
  pages={586},
  year={1982},
  publisher={APS}
}

@book{bortignon2019giant,
  title={Giant Resonances: Nuclear structure at finite temperature},
  author={Bortignon, Pier Francesco and Bracco, Angela and Broglia, Ricardo A},
  year={2019},
  publisher={CRC Press}
}

@article{nakada2008mean,
  title={Mean-field approach to nuclear structure with semi-realistic nucleon-nucleon interactions},
  author={Nakada, H},
  journal={Physical Review C},
  volume={78},
  number={5},
  pages={054301},
  year={2008},
  publisher={APS}
}

@article{decharge1980hartree,
  title={Hartree-Fock-Bogolyubov calculations with the D 1 effective interaction on spherical nuclei},
  author={Decharg{\'e}, J and Gogny, D},
  journal={Physical Review C},
  volume={21},
  number={4},
  pages={1568},
  year={1980},
  publisher={APS}
}

@article{sheikh2021symmetry,
  title={Symmetry restoration in mean-field approaches},
  author={Sheikh, Javid A and Dobaczewski, J and Ring, Peter and Robledo, Luis Miguel and Yannouleas, Constantine},
  journal={Journal of Physics G},
  volume={48},
  number={12},
  pages={123001},
  year={2021},
  publisher={IOP Publishing}
}

@article{peru2014mean,
  title={Mean field based calculations with the Gogny force: Some theoretical tools to explore the nuclear structure},
  author={P\'eru, Sophie and Martini, Marco},
  journal={The European Physical Journal A},
  volume={50},
  pages={1--35},
  year={2014},
  publisher={Springer}
}

@inproceedings{mui20,
  title={Structure of Krypton isotopes using the generalised Bohr Hamiltonian method},
  author={Muir, David and Pr{\'o}chniak, Leszek and Pastore, Alessandro and Dobaczewski, Jacek},
  booktitle={Journal of Physics: Conference Series},
  volume={1643},
  pages={012147},
  year={2020},
  organization={IOP Publishing}
}

@article{bender2006global,
  title={Global study of quadrupole correlation effects},
  author={Bender, Micha{\"e}l and Bertsch, GF and Heenen, P-H},
  journal={Physical Review C},
  volume={73},
  number={3},
  pages={034322},
  year={2006},
  publisher={APS}
}

@article{pod28,
  title={Quantum-mechanically correct form of Hamiltonian function for conservative systems},
  author={Podolsky, Boris},
  journal={Physical Review},
  volume={32},
  number={5},
  pages={812},
  year={1928},
  publisher={APS}
}

@article{sch16,
  title={Microscopic theory of nuclear fission: a review},
  author={Schunck, N and Robledo, LM},
  journal={Reports on Progress in Physics},
  volume={79},
  number={11},
  pages={116301},
  year={2016},
  publisher={IOP Publishing}
}

@book{ring2004nuclear,
  title={The nuclear many-body problem},
  author={Ring, Peter and Schuck, Peter},
  year={2004},
  publisher={Springer Science \& Business Media}
}

@article{hoo72,
  title={A technique for numerical solution of the Schroedinger equation with non-local potentials},
  author={Hooverman, RH},
  journal={Nuclear Physics A},
  volume={189},
  number={1},
  pages={155--160},
  year={1972},
  publisher={Elsevier}
}

@article{xia25,
  title   = {Microscopic triaxial quadrupole-octupole collective Hamiltonian for low-energy nuclear excitations},
  author  = {Xiang, J. and Zhao, J. and Li, Z. P. and Vretenar, D.},
  journal = {Physical Review C},
  volume  = {113},
  number  = {4},
  pages   = {044328},
  year    = {2026},
}

@article{girod1979zero,
  title={The zero-point energy correction and its effect on nuclear dynamics},
  author={Girod, M and Grammaticos, B},
  journal={Nuclear Physics A},
  volume={330},
  number={1},
  pages={40--52},
  year={1979},
  publisher={Elsevier}
}

@article{rei75,
  title={Zero-point energies in potential-energy surfaces},
  author={Reinhard, PG},
  journal={Nuclear Physics A},
  volume={252},
  number={1},
  pages={120--132},
  year={1975},
  publisher={Elsevier}
}

@article{lib82,
  title={A general solution of the Bohr collective Hamiltonian},
  author={Libert, J and Quentin, P},
  journal={Zeitschrift f{\"u}r Physik A Atoms and Nuclei},
  volume={306},
  number={4},
  pages={315--322},
  year={1982},
  publisher={Springer}
}

@article{aza26,
  title={Error analysis of the finite-range liquid-drop model},
  author={Azam, C and Blondeau-Patissier, D and Tamagno, P and Frosini, M and Pastore, A},
  journal={Physical Review C},
  volume={113},
  number={2},
  pages={024327},
  year={2026},
  publisher={APS}
}

@inproceedings{iverson1962programming,
  title={A programming language},
  author={Iverson, Kenneth E},
  booktitle={Proceedings of the May 1-3, 1962, spring joint computer conference},
  pages={345--351},
  year={1962}
}

@book{pauli,
  title={General principles of quantum mechanics},
  author={Pauli, Wolfgang},
  year={2012},
  publisher={Springer Science \& Business Media}
}

@book{BohrMotelson_1953,
    author = {Bohr, A and Mottelson, B R},
    title = {Collective and individual-particle aspects or nuclear structure},
    publisher = {K. Danske Vidensk. Selsk., Mat.-Fys. Medd. 27 No 16},
    year = {1953}
}

@book{BohrMottelson_1975,
  author    = {Bohr, Aage and Mottelson, Ben R.},
  title     = {Nuclear Structure, Vol. II: Nuclear Deformations},
  publisher = {W. A. Benjamin},
  address   = {Reading, MA},
  year      = {1975}
}

@book{rowe,
  title={Fundamentals of nuclear models: foundational models},
  author={Rowe, David J and Wood, John L},
  year={2010},
  publisher={World Scientific}
}

\end{document}